\documentclass[a4paper,11pt]{article}
\usepackage{jcappub}
\usepackage{tikz, xcolor,hyperref}
\usepackage{amsmath, amssymb, amsthm, graphicx, epsfig, fancyhdr,epsfig, slashed}
\usepackage[normalem]{ulem}
\usepackage{natbib}
\usepackage{float}
\usepackage{tikzsymbols}
\usepackage{makecell}
\usepackage{comment}
\usepackage{cleveref}
\usepackage{cancel}

\newcommand{\aend}{a_\text{end}}

\def\be{\begin{equation}}

\def\ee{\end{equation}}
\def\bea{\begin{eqnarray}}
\def\eea{\end{eqnarray}}
\def\bi{\begin{itemize}}
\def\ei{\end{itemize}}
\def\beq{\begin{equation}\begin{aligned}}
\def\eeq{\end{aligned}\end{equation}}

\numberwithin{equation}{section}
\usepackage[compat=1.1.0]{tikz-feynman}

\begin{document}
%%%%%%%%%%%%%%%%%%%%
\title{Asymmetric Reheating of Dark QED}
%%%%%%%%%%%%%%%%%%%%%%%%%%%%%%%

\author[a]{Simon Cl\'ery,}
\author[b]{Jean Kimus}
\author[b]{\& Michel H.G. Tytgat}

\affiliation[a]{Technical University of Munich, TUM School of Natural Sciences, Physics Department, 85748
Garching, Germany}
\affiliation[b]{\,Service de Physique Th\'eorique, Universit\'e Libre de Bruxelles,
Boulevard du Triomphe, CP225, 1050 Brussels, Belgium}

\emailAdd{simon.clery@tum.de}
\emailAdd{jean.kimus@ulb.be}
\emailAdd{michel.tytgat@ulb.be}

%%%%%%%%%%%%%%%%%
\abstract{We study in detail a scenario in which the inflaton scalar field couples to both a visible sector (VS) and a hidden sector (HS). The VS is assumed to contain the Standard Model (SM), while the HS contains a dark matter (DM) candidate. We are in particular interested in a scenario in which the inflaton decays dominantly into the HS degrees of freedom. The DM candidate is taken to be a dark Dirac fermion $\chi$, coupled to a massive dark photon $\gamma'$, a popular model for a HS also known as Dark QED. The inflaton decays into particles of both sectors generate an initial asymmetry between the SM and HS fermion abundances, which we model as being proportional to the ratio of effective Yukawa couplings, $y$ and $y'$. We pay particular attention to the process of thermalisation of the HS, with temperature $T'$, as a function of $y'$ and $\alpha'$, the HS fine structure constant. We investigate the several possible ways of producing the observed DM relic abundance, and their interplay with the reheating of the HS and the transfer of energy between the HS and the VS. Key results, beyond the systematic character of our analysis, include: a new mechanism for DM production, which occurs when DM particles annihilate while still being produced by the inflaton decay; a study of the temperature ratio $\xi = T'/T$
 and its relation with the initial energy asymmetry between the HS and VS, as parameterized by $\xi_i = \sqrt{y'/y}$; a reassessment of the domain of viable DM candidates, taking into account the constraints set by unitarity and the thermalisation of the HS, accounting for the LPM effect; and, in cases where the HS does not reach thermal equilibrium, an analysis of how non-thermal DM production fits within the domain of thermal DM candidates.}
%%%%%%%%%%%%%%%%%%
\maketitle
%%%%%%%%%%%
\section{Introduction}
\label{sec:intro}

Inflation followed by reheating plays a central role in setting the initial conditions for all the history of our universe. In the most conventional scenario, the inflaton decays into what will eventually be the Standard Model (SM) degrees of freedom --- the visible sector (VS) in the sequel. However, the cosmological evolution prior to Big Bang nucleosynthesis (BBN) remains largely unconstrained \cite{Allahverdi:2002pu}. Also, many well-motivated extensions of the SM predict the existence of hidden sectors (HS), consisting of particles that interact only feebly with the VS \cite{Strassler:2006im,Feng:2008mu,Curtin:2017quu}. If the inflaton decays differently into visible and hidden particles, leading to unequal amounts of energy deposited into each sectors, reheating of the universe can be asymmetric. This is a rather natural expectation in theories with multiple Yukawa couplings or portal interactions \cite{Berezhiani:1995am,Adshead:2016xxj,Tenkanen:2016jic,Hardy:2017wkr}. This can give rise to a rich set of alternative cosmological histories, with non-trivial temperature ratios, $T'/T \neq 1$ --provided the HS thermalises--, entropy transfer from one sector to the other, which can have significant implications for relics from the early universe, such as gravitational waves (GW) \cite{Schwaller:2015tja,Ertas:2021xeh}, the baryon \& lepton asymmetries \cite{Kamada2019Gravitational,Alonso-Alvarez:2023bat,Coy:2024itg}, or  dark matter (DM) \cite{Berlin:2016vnh,Hardy:2017wkr,Coy_2021,Bernal:2022wck,Coy:2024itg,Bernal:2025fdr} to mention a few. 

In this work, we consider a HS that consists of dark QED, a simple yet instructive and widely studied framework \cite{Ackerman:2008kmp,Feng:2008mu}. In our approach, DM is made of a Dirac dark fermion $\chi$, charged under a gauge $U(1)'$ symmetry, along with a massive dark photon, a companion particle, $m_{\gamma'} < m_\chi$ that is unstable due to a kinetic mixing portal between the VS and dark QED \cite{Holdom:1985ag}. Our motivation to study this scenario is manifold. Typically, asymmetric reheating is considered in the framework of DM production through freeze-in \cite{Hall:2009bx,Chu:2011be}. Indeed, a central issue is that one must typically assume that DM abundance is negligible prior to freeze-in. Whether this is the case can be  addressed given an asymmetric reheating framework \cite{Adshead:2016xxj,Hardy:2017wkr}. Instead, we will consider an inflaton that decays predominantly into a HS, here dark QED. If that sector thermalises, then 
$$\xi \,\hat =\, T'/T \gg 1\,$$ 
and the HS is hot (rather than cold or chilly \cite{Hardy:2017wkr}) compared to the VS, at least prior to Big Bang Nucleosynthesis (BBN). Such scenario is not new, see e.g. \cite{Berlin:2016vnh,Tenkanen:2016jic}, but has been more  extensively addressed in \cite{Coy_2021,Coy:2024itg} to delineate the possible range (aka the \emph{domain}) of DM candidates that were once in thermal equilibrium. In the latter works, the temperature ratio 
$\xi$ has been treated as a mere initial condition. Our goal here is to study in details how it can arise in a dynamical setup. This includes the  possibility that the temperature ratio evolves along the process of reheating of the universe after inflation. 

Doing so, we will be confronted to the problem of dark QED thermalisation. Surprisingly, thermalisation of a HS with only abelian interactions is not much discussed in the literature \footnote{The two exceptions we are aware of are the works \cite{Garny:2018grs,Arvanitaki:2021qlj} which rely on gravitational production during and after inflation of the HS and so are  different from the scenario we consider. The relevance of the LPM effect is also covered in \cite{Garny:2018grs}. } This stands in contrast to the well-studied problem of thermalisation of the SM sector after inflation \cite{Davidson:2000er,Kurkela:2011ti,Harigaya:2013vwa}, and more generally of a  HS with non-abelian interactions, see e.g. \cite{Garani:2021zrr}. Here we will derive a simple condition, valid for weak coupling and an initially under-occupied abundance of dark fermions (see section \ref{sec:Thermalisation_HS} for more concrete definitions) and, for the sake of comparison, confront the situation of dark QED to that of a non-abelian HS. {A key aspect of our analysis is the identification of significant differences in the thermalisation process and timescales for abelian and non-abelian HS, both during and after reheating. In particular, we will show that thermalisation in an abelian HS is parametrically slower than in the non-abelian case, as it is controlled by soft scatterings and LPM-suppressed bremsstrahlung, leading to a distinct equilibration process during reheating}. This study of the thermalisation process will allow us to track the evolution of the temperature ratio $\xi = T'/T$, {from its initial value set by the inflaton decay rates into the visible and hidden sectors, through the subsequent stages of energy transfer between the two sectors, controlled by the feeble kinetic mixing.  We will consider different DM production { regimes from standard freeze-out in the radiation-dominated era, to freeze-in sourced by inflaton decays, including a not yet studied scenario in which freeze-out occurs while inflaton is still sourcing both sectors during reheating. After highlighting some possible benchmark DM candidates, we finally discuss the implications for the domain of all thermal and non-thermal dark QED dark matter candidates, taking into account the constraints coming from unitarity, the thermalisation of the HS, and the evolution of the temperature ratio $\xi$.} 

Our work is organised as follows. In section \ref{sec:asymmetric_reheating_HS}, we introduce the framework of dark QED and asymmetric reheating, describe the relevant interactions and derive the set of Boltzmann equations governing the evolution of both sectors. In section \ref{sec:Thermalisation_HS}, we perform a detailed analysis of HS thermalisation in the perturbative limit, highlighting the specific features of abelian gauge interactions contrasting them with the non-abelian case, and identifying the relevant timescales. In section \ref{sec:DMabundance}, we determine the DM relic abundance across the different regimes, including non-thermal production, freeze-out after reheating, and freeze-out during reheating. A summary of the timeline for each regime is illustrated in figure \ref{fig:timeline}. Finally, in section \ref{sec:Constraints_DM}, we discuss the phenomenological constraints on this model and characterise the viable parameter space for thermal DM (the domain). We conclude about the main results of this work in section \ref{sec:conclusions}. Some more technical aspects are relegated to appendices, including a discussion of the physics underlying the LPM effect.

\section{Asymmetric production of a HS from inflation}
\label{sec:asymmetric_reheating_HS}
\subsection{Setup and definitions}
\label{sec:framework}
%%%%%%%%%%%
Following \cite{Ackerman:2008kmp,Feng:2009mn}, we consider dark QED, a dark sector made of a Dirac fermion $\chi$ charged under a group $U(1)'$ with charge $e'$. Interactions with a massive dark photon $\gamma'$ is characterized by $\alpha' ={e'^2}/{4\pi}$. We  do not address the origin of the dark photon mass $m_{\gamma'}$, which we take as a free parameter \cite{Cheung:2007ut,Feldman:2007wj}. The dark fermions and dark
photons interact with the SM through kinetic mixing  with mixing parameter $\varepsilon \ll 1$ \cite{Holdom:1985ag}. 

For inflation, we assume a single scalar inflaton field $\phi$ which, like the HS, is singlet of the SM. The inflaton dynamics depends on the shape of its potential $V(\phi)$. As a benchmark, we consider the Starobinsky model \cite{Starobinsky:1980te},
\be
V(\phi) = \frac{3}{4}m_\phi^2 M_P^2\left( 1- e^{-\sqrt{\frac{2}{3}}\frac{\phi}{M_P}}\right)^2
\label{eq:staro}
\ee
where $M_P = 1/\sqrt{8 \pi G} = 2.4 \times 10^{18} \,\text{GeV}$ is the reduced Planck mass. The potential is quadratic around the origin and the inflaton mass, $m_\phi$, is fixed by the amplitude of the scalar perturbations inferred from CMB measurements \cite{Planck:2018jri}, 
\be
m_\phi \simeq 3 \times 10^{13}~{\rm GeV}.
\ee
The end of inflation, $\ddot a =0$ with $a$ the scale factor, corresponds to \cite{Starobinsky:1980te}
\be
\rho_{\rm end} = \frac{3}{2}V(\phi_{\rm end})\simeq 0.175 m_\phi^2 M_P^2 \simeq \left(5.5\times10^{15}~\rm GeV\right)^4\,. 
\ee 

The problem of reheating of the universe after inflation is complex, see e.g. \cite{Barman:2025lvk}. In the Starobinsky model,  the inflaton field starts to oscillate about the minimum of its potential after the end of slow-roll, corresponding to $H_{\rm end} \sim m_\phi$, marking the onset of reheating. 
In what follows, we assume  $\phi\ll M_P$. Near the minimum, $V(\phi)\approx \frac{1}{2}m_\phi^2 \phi^2$, so that the inflaton equation of state is effectively $w_\phi = {P_\phi}/{\rho_\phi} \simeq 0$ and the universe is matter dominated. We suppose a setup in which the inflaton is weakly coupled to VS and HS degrees of freedom\footnote{{In particular, we restrict ourselves to very feeble effective inflaton couplings to both VS and HS, and do not consider possible non-perturbative effects of preheating occurring for larger couplings of the inflaton, see  for instance \cite{Garcia:2021iag,Bhusal:2025oqg}}}. Reheating proceeds through the inflaton decay into VS and HS particles, with total rate $\Gamma_\phi = \Gamma_\phi^{\rm vs} + \Gamma_\phi^{\rm hs}$. Matter domination by the inflaton field lasts until $H \lesssim \Gamma_\phi$ and the inflaton abundance is exponentially suppressed, marking the end of reheating \cite{Kolb:1990vq}. In the sequel, we call \emph{reheating} the phase during which the expansion is dominated by the inflaton. 

We consider  asymmetric reheating scenarios,  
\begin{equation}
    \Gamma_\phi  = \Gamma_\phi^{\rm vs} + \Gamma_\phi^{\rm hs}
\end{equation} with $\Gamma_\phi^{\rm hs} \ll \Gamma_\phi^{\rm vs}$ (cold or subdominant HS) or $\Gamma_\phi^{\rm hs} \gg \Gamma_\phi^{\rm vs}$ (hot or dominant HS). Our focus will be on  DM and so on the HS. Provided this sector thermalises, we will use $T'$ to denote its temperature. Regardless of the HS evolution, we assume that the VS degrees of freedom thermalise on a short timescale (i.e. instantaneous thermalisation), and so is fully characterized by its temperature $T$. In this approximation, the VS temperature evolves as \cite{Kolb:1990vq,Giudice:1999am} (see also \cite{Coy:2024itg})
\begin{equation}
\label{eq:Tmaxrh}
    T \sim (\Gamma_\phi^{\rm vs} H M^2_P)^{1/4}
\end{equation}
rapidly reaching a maximum temperature $T_{\rm max}$ at the end of inflation, $H \sim H_{\rm end}$, and then decreases down to the temperature $T_{\rm rh}$ at the end of reheating, which is characterized by the {\em total} decay rate, $H \sim \Gamma_\phi$.  

\subsection{Basic processes}
\label{sec:relevant_processes}

We model inflaton decay through Yukawa-like couplings to VS and HS fermions.
Our model for reheating is thus
\bea
\nonumber
\mathcal{L} &\supset& \frac{1}{2}\partial_\mu\phi\partial^\mu\phi -{1\over 2} m_\phi^2 \phi^2 -y\phi \bar{f}f -y'\phi\bar{\chi}\chi \\
&+& \bar{\chi}\left(i\slashed D - m_\chi\right)\chi -\frac{1}{4}F^{'\mu\nu}F'_{\mu\nu} +\frac{1}{2}m_{\gamma '}^2A'_\mu A^{'\mu} -\frac{\varepsilon}{2}B_{\mu\nu}F^{\mu\nu} \, .
\label{eq:lagrangian}
\eea
where $B_{\mu\nu}$ is the SM hypercharge field strength. As usual, the kinetic terms can be made diagonal through a redefinition of the gauge bosons fields, $B_\mu \rightarrow B_\mu -\varepsilon A_\mu'$ and $A_\mu' \rightarrow A_\mu'$, valid for $\varepsilon \ll 1$ and $m_{\gamma'} \neq m_Z$ \cite{Holdom:1985ag}; this induces a coupling of the dark photon to SM fermions proportional to $\varepsilon$. In \eqref{eq:lagrangian}, $f$ is meant to represent a generic SM fermion. Strictly speaking, the coupling of the inflaton to SM (or VS) fields should be given in a gauge-invariant form. As we assume that the VS thermalises instantaneously, the parameter $y$  merely serves to quantify the fraction of inflaton energy that is transferred into the VS,  
\begin{equation}
	\scalebox{1.}{
		\begin{tikzpicture}[baseline={-0.1cm}]
			\begin{feynman}[every blob={/tikz/fill=gray!30,/tikz/inner sep=2pt}]
				\vertex (f1) at (0,0.75) {\(f\)} ;
				\vertex (f2) at (0,-0.75) {\(\bar{f}\)};
				\vertex (f3) at (-1,0);
				\vertex (e1) at (-2.,0) {\(\phi\)} ;
				\diagram* {
					(f3) -- [fermion] (f1),
					(f2) -- [fermion] (f3),
					(e1) -- [scalar] (f3) }; \end{feynman}
	\end{tikzpicture}} \qquad \mbox{\rm with} \qquad  \Gamma_\phi^{\rm vs} = \dfrac{y^2}{8 \pi} m_\phi\,.
    \label{fig:phi_f_f_diagram}
\end{equation}
Similarly, 
\begin{equation}
\scalebox{1.}{
		\begin{tikzpicture}[baseline={-0.1cm}]
			\begin{feynman}[every blob={/tikz/fill=gray!30,/tikz/inner sep=2pt}]
				\vertex (f1) at (0,0.75) {\(\chi\)} ;
				\vertex (f2) at (0,-0.75) {\(\bar{\chi}\)};
				\vertex (f3) at (-1,0);
				\vertex (e1) at (-2.,0) {\(\phi\)} ;
				\diagram* {
					(f3) -- [fermion] (f1),
					(f2) -- [fermion] (f3),
					(e1) -- [scalar] (f3) }; \end{feynman}
	\end{tikzpicture}}\qquad \mbox{\rm with} \qquad  \Gamma_\phi^{\rm \chi} = \dfrac{y'^2}{8 \pi} m_\phi\,.
    \label{eq:phi_X_X_diagram}
\end{equation}
neglecting the fermions masses. The inflaton could also decay into two dark photons. To limit the number of free parameters, we assume it to occur at one-loop, 
\begin{equation}
	\scalebox{0.8}{
		\begin{tikzpicture}[baseline={-0.1cm}]
			\begin{feynman}[every blob={/tikz/fill=gray!30,/tikz/inner sep=2pt}]
				\vertex (i1) at (0.75, 1.2) {\(\gamma^\prime\)};
				\vertex (i2) at (0.75,-1.2) {\(\gamma^\prime\)};
				\vertex (f1) at (0,0.75);
				\vertex (f2) at (0,-0.75);
				\vertex (f3) at (-1.25,0);
				\vertex (e1) at (-2.5,0) {\(\phi\)} ;
				\diagram* {
					(f1) -- [boson, style=black] (i1),
					(f1) -- [fermion] (f2),
					(i2) -- [boson, style=black] (f2),
					(f3) -- [fermion, edge label=\(\chi\)] (f1),
					(f2) -- [fermion] (f3),
					(e1) -- [scalar] (f3) }; \end{feynman}
	\end{tikzpicture}}
    \qquad \mbox{\rm with} \quad \Gamma_\phi^{\gamma'} =\frac{\alpha'^2y'^2}{256\pi^3}\left(\frac{m_\phi^3}{m_\chi^2}\right)|F(\beta_\chi)|^2
    \label{eq:loop_decay}
\end{equation}
where $F$ is a function of $\beta_\chi = {4m_\chi^2}/{m_\phi^2}\ll 1$ such that the rate vanishes as $\beta_\chi \rightarrow 0$ \cite{Shifman:1979eb,Marciano:2011gm} (see Appendix \ref{app:average_cross_sections_and_decays}). The total inflaton decay rate into HS particles is thus\footnote{The inflaton radiative decay to dark photons is typically very suppressed compared to the tree-level decay into dark electrons, and so neglect the corresponding decay rate in our analytical estimates; it is however included in our numerical calculations.} given by\begin{equation}
    \Gamma_\phi^{\rm hs} = \Gamma_\phi^{\chi}+  \Gamma_\phi^{\gamma'} \approx \Gamma_\phi^{\chi}
\end{equation}
The Yukawa couplings also induce an irreducible interaction between HS and SM, mediated by the inflaton \cite{Heurtier:2017nwl,Heurtier:2019eou},
\begin{equation}
\scalebox{0.9}{
\begin{tikzpicture}[baseline={-0.1cm}]
  \begin{feynman}[every blob={/tikz/fill=gray!30,/tikz/inner sep=2pt}]
    \vertex (i1) at (0, 0.75) {\(\chi\)} ;
    \vertex (i2) at (0,-0.75) {\(\bar{\chi}\)};
    \vertex (c1) at (1., 0.) ;
    \vertex (c2) at (2., 0.) ;
    \vertex (f1) at (3.,0.75) {\(f\)};
    \vertex (f2) at (3.,-0.75)  {\(\bar{f}\)};;
    \diagram* {
      (i1) -- [fermion] (c1),
      (c1) -- [fermion] (i2),
      (c1) -- [scalar, style=black, edge label=\(\phi\)] (c2),
      (c2) -- [fermion] (f1),
      (f2) -- [fermion] (c2),      
      }; \end{feynman}
\end{tikzpicture}}\qquad \mbox{\rm with} \quad  \sigma_{\chi f}^\phi\sim \dfrac{(y y')^2 }{16 \pi}\, {s\over m_\phi^4} 
\label{Fig:inflaton_portal}
\end{equation}
for $s \approx 4 m_\chi^2 \lesssim m_\phi^2$. Since we  assume weak Yukawa couplings, we will neglect this and the corresponding t-channel scattering process. More important for our purpose are gauge interactions within the HS. Through the kinetic mixing, dark fermions can also annihilate into (or scatter with) VS fermions through
\begin{equation}\scalebox{0.9}{
\begin{tikzpicture}[baseline={-0.1cm}]
  \begin{feynman}[every blob={/tikz/fill=gray!30,/tikz/inner sep=2pt}]
    \vertex (i1) at (0, 0.75) {\(\chi\)} ;
    \vertex (i2) at (0,-0.75) {\(\bar{\chi}\)};
    \vertex (c1) at (1., 0.) ;
    \vertex (c2) at (1.5, 0.) {\(\varepsilon\)};
    \vertex (c3) at (2., 0.);
    \vertex (f1) at (3.,0.75) {\(f\)};
    \vertex (f2) at (3.,-0.75)  {\(\bar{f}\)};;
    \diagram* {
      (i1) -- [fermion] (c1),
      (c1) -- [fermion] (i2),
      (c1) -- [boson, style=black, edge label=\(\gamma^\prime\)] (c2),
      (c2) -- [boson, style=black, edge label=\(\gamma/Z\)] (c3),
      (c3) -- [fermion] (f1),
      (f2) -- [fermion] (c3),      
      }; \end{feynman}
\end{tikzpicture}}\qquad \mbox{\rm with (see Appendix)}\quad \sigma_{\chi f}^{\varepsilon} \sim  {4 \pi \alpha \alpha' \varepsilon^2\over s}
\label{Fig:DE_interactions2}
\end{equation}
We consider feeble coupling between the HS and VS, $\varepsilon \ll 1$, so such process too will be subdominant. Thus, to first approximation, both sectors evolve independently from each others. 

The relevant processes are then the following. First, the annihilation process 
\begin{equation}
\scalebox{1.}{
\begin{tikzpicture}[baseline={-0.1cm}]
  \begin{feynman}[every blob={/tikz/fill=gray!30,/tikz/inner sep=2pt}]
    \vertex (i1) at (-0.75, 1.2) {\(\chi\)};
    \vertex (i2) at (-0.75,-1.2) {\(\bar{\chi}\)};
    \vertex (c1) at (0,0.5);
    \vertex (c2) at (0,-0.5);
    \vertex (f1) at (0.75,1.2) {\(\gamma^\prime\)};
    \vertex (f2) at (0.75,-1.2) {\(\gamma^\prime\)};
    \diagram* {
      (i1) -- [fermion] (c1),
      (c2) -- [fermion] (i2),
      (c1) -- [fermion] (c2),
      (c1) -- [boson, style=black] (f1),
      (c2) -- [boson, style= black] (f2) }; \end{feynman}
\end{tikzpicture}}\qquad \mbox{\rm which scales as} \quad
\sigma_{\chi  \gamma'} \sim \dfrac{\pi \alpha'^2}{s} \,,
\label{Fig:DE_interactions}
\end{equation}
(neglecting the dark photon mass) will be relevant to set the DM abundance through secluded freeze-out \cite{Pospelov:2007mp}. Before DM freeze-out, such processes will also lead to the production of dark photons within the HS. Dark photons will eventually decay into VS  fermions through kinetic mixing,
\begin{equation}
\begin{tikzpicture}[baseline={-0.1cm}]
  \begin{feynman}[every blob={/tikz/fill=gray!30,/tikz/inner sep=2pt}]
    \vertex (f1) at (0,0.75) {\(f\)} ;
    \vertex (f2) at (0,-0.75) {\(\bar{f}\)};
    \vertex (f3) at (-1,0);
    \vertex (e1) at (-2,0) {\(\varepsilon\)} ;
    \vertex (e2) at (-3,0) {\(\gamma^\prime\)} ;
    \diagram* {
      (f3) -- [fermion] (f1),
      (f2) -- [fermion] (f3),
      (e2) -- [boson, style=black] (e1), 
      (f3) -- [boson, style=black, edge label=\(\gamma/Z\)] (e1)}; \end{feynman}
\end{tikzpicture} \qquad \mbox{\rm with  rate (see Appendix)}\quad \Gamma_{\gamma'} \sim \sum_f \alpha Q_f^2 \varepsilon^2 \, m_{\gamma'}
\label{Fig:DP_decay}
\end{equation}
where $Q_f$ is the electric charge of SM fermions.

An important question is whether the HS can reach thermal equilibrium. This depends crucially on 
t-channel scattering processes via dark photon exchange, such as
\begin{equation}
\begin{tikzpicture}[baseline={-0.1cm}]
  \begin{feynman}[
      every blob={fill=gray!30, inner sep=2pt},
      every fermion/.style={draw=black!70, line width=0.4pt, postaction={decorate},
                            decoration={markings,mark=at position 0.55 with {\arrow{>}}}},
      every boson/.style={draw=black!40, line width=0.4pt}
    ]
    \vertex (i1) at (-1.5, 0.5) {\(\chi\)};
    \vertex (i2) at (-1.5,-0.5) {\(\chi\)};
    \vertex (c1) at (0,0.5);
    \vertex (c2) at (0,-0.5);
    \vertex (f3) at (1.8, 0.5) {\(\chi\)}; 
    \vertex (f2) at (1.8,-0.5) {\(\chi\)};

    \diagram* {
      (i1) -- [fermion] (c1),
      (i2) -- [fermion] (c2),
      (c1) -- [boson, style=black, edge label=\(\gamma'\)] (c2),
      (c1) -- [fermion] (f3),
      (c2) -- [fermion] (f2)
    };
  \end{feynman}
\end{tikzpicture}
 \qquad \mbox{\rm with}\quad \sigma_{\chi\chi} \sim {\alpha'^2\over t - m_{\gamma'}^2}
\end{equation}
We will address this more extensively in section \ref{sec:Thermalisation_HS}. There, we will distinguish between hard and soft t-channel processes.\\ 

Figure ~\ref{fig:Schema_inflaton_HS_VS} gives a summary of the different components involved and their interactions, which gives rise to the rich dynamics to be discussed in the following sections. Detailed expressions for the cross sections are provided in Appendix \ref{app:average_cross_sections_and_decays}.
\begin{figure}[h!]
	\centering
	\includegraphics[width=0.5\textwidth]{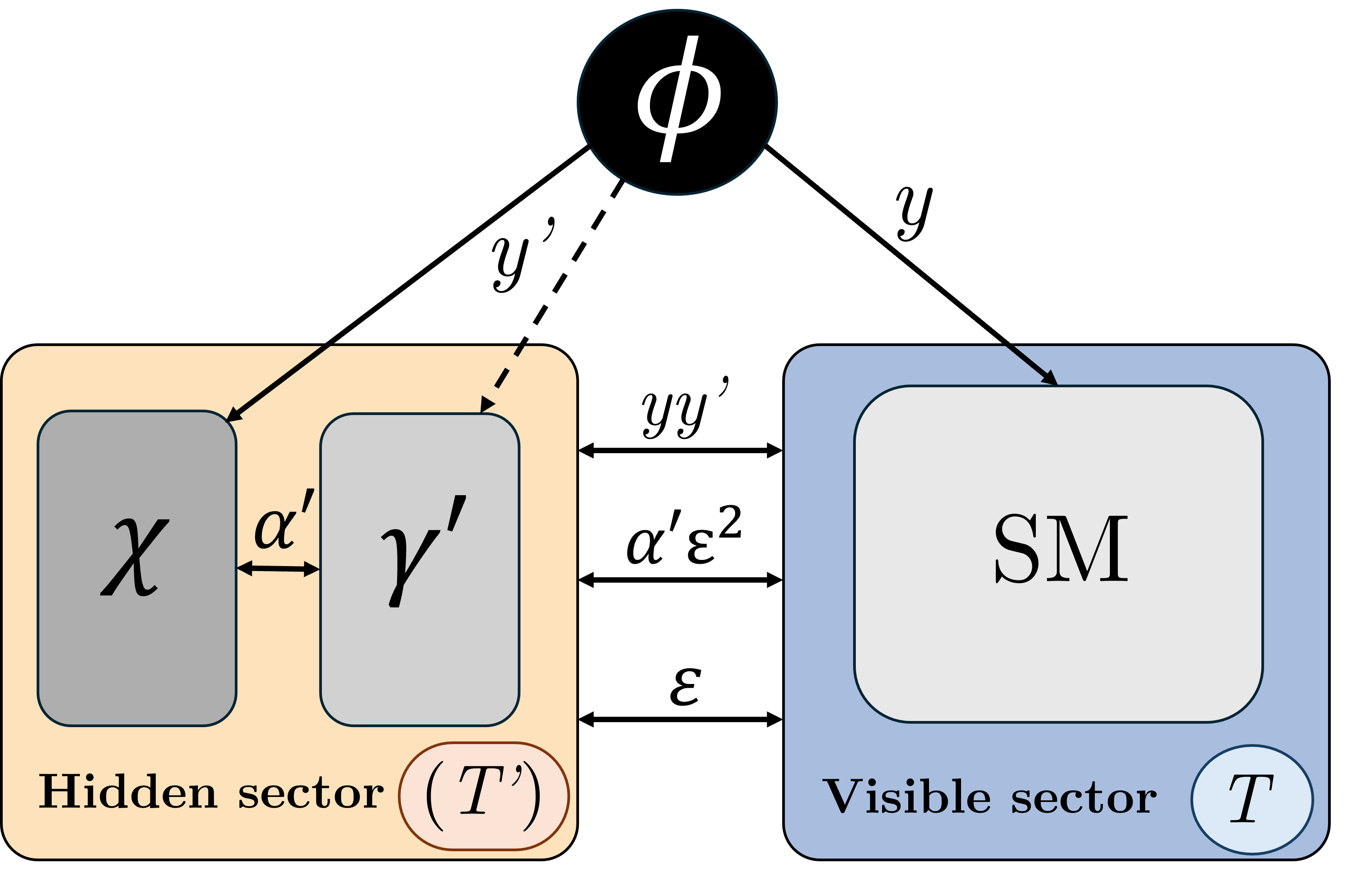}
	\caption{The inflaton $\phi$ transfers energy to the HS and VS through the Yukawa couplings $y'$ and $y${, via inflaton decays \eqref{eq:phi_X_X_diagram} and \eqref{fig:phi_f_f_diagram}}. The HS temperature $T'$ is between brackets as thermalisation of the HS is hypothetical (see section \ref{sec:Thermalisation_HS}). {HS dark fermions and dark photons is controlled by the HS fine structure constant $\alpha'$ \eqref{Fig:DE_interactions}. Interactions between hidden and visible sectors are dominated by inflaton-mediated dark fermions annihilation \eqref{Fig:inflaton_portal}, dark photon-mediated dark fermions annihilation \eqref{Fig:DE_interactions2} and dark photon decay \eqref{Fig:DP_decay}.} }
	\label{fig:Schema_inflaton_HS_VS}
\end{figure}

\subsection{Boltzmann equations}
\label{subsec:boltzmann}
We now write down the Boltzmann equations governing the evolution of the different species during and after reheating. The shape of the inflaton potential near the minimum determines the average equation of state $w_\phi$ for the homogeneous inflaton condensate. In the present case of a quadratic potential near the minimum, $\phi$ will behave like matter and its energy density $\rho_\phi$ follows
\begin{equation}
	\dot{\rho}_\phi + 3 H \rho_\phi = - \Gamma_\phi \rho_\phi
	\label{eq:Energy_conservation_inflaton}
\end{equation}
where $H$ is the Hubble rate and $\Gamma_\phi$ is the total {inflaton decay rate}. As the inflaton dominates the expansion throughout reheating, until $H_{\rm rh} \sim \Gamma_\phi$, its energy density evolution can be expressed as
\begin{equation}
    \rho_\phi(a) = \rho_{\rm end } \left(\dfrac{a_{\rm end }}{a} \right)^3 \exp \left\{-\dfrac{2 \Gamma_\phi}{3 H_{\rm end }} \left( \dfrac{a}{a_{\rm end }} \right)^{3/2} \right\}
\end{equation}
with $a$ the scale factor and $H_{\rm end}$ the expansion rate at the end of inflation. We will rely on numerical solutions in the sequel, but this analytic solution remains an excellent approximation for all scenarios considered below. Reheating ends  when the inflaton energy density falls exponentially fast,  corresponding to $H_{\rm rh}\simeq \Gamma_\phi$, at which point the scale factor is of the order of 
\be
        a_{\rm RH} \simeq \aend \left( \dfrac{3 H_{\rm end}}{2\Gamma_\phi} \right)^{\frac{2}{3}}
        \label{eq:a_RH}\, .
\ee
Conservation of the total energy implies that, in the absence of any interaction other than inflaton decay, the energy densities of the other species considered, $\rho_\chi$, $\rho_{\gamma'}$, and $\rho_{\text{vs}}$, will evolve according to the following simple Boltzmann equations \cite{Chung:1998rq}
\begin{eqnarray}
	&&\dot{\rho}_\chi + 3(1+w_\chi) H \rho_\chi = \Gamma_\phi^{\chi} \,\rho_\phi \nonumber\\
	&&\dot{\rho}_{\gamma'} + 3 (1+w_{\gamma'}) H \rho_{\gamma'} = \Gamma_\phi^{\gamma'}\, \rho_\phi \\
	&&\dot{\rho}_{\text{vs}} + 4 H \rho_{\text{vs}} = \Gamma_\phi^{\rm vs}\, \rho_\phi\nonumber
\label{eq:Energy_conservation_rest}
\end{eqnarray}
where $w_\chi$ and $w_{\gamma'}$ are the equation of state (EoS) of respectively, the dark electrons and the dark photons, with  $w_{\gamma'} = w_\chi = 1/3\, (0)$ for relativistic (non-relativistic) particles. We use the subscript vs for the VS and hs for the HS, $\rho_{\rm hs} = \rho_\chi + \rho_{\gamma'}$. 

For a HS that remains non-thermal throughout its evolution, we can largely rely on solving this simple system of Boltzmann equations up to the end of reheating to determine the DM number density and its relic abundance. We return to this non-thermal scenario in Section \ref{sec:non_thermal}.\\

In the sequel, we will be mostly interested in scenarios along which both the VS and HS reach thermal equilibrium, with temperatures $T$ and $T'$ respectively. Throughout, we will assume that the VS is reaching thermal equilibrium almost instantaneously,  with $T \sim \rho_{\rm vs}^{1/4}$, where $\rho_{\rm vs}$ is dominated by relativistic degrees of freedom. For the HS, the temperature will be defined with respect to the dark photons, which in most cases are assumed to be much lighter than the dark fermions, yielding $T' \sim \rho_{\gamma'}^{1/4}$. The  set of Boltzmann equations for energy transfer to the HS and VS, including now all relevant interaction channels, is more involved
{\small
\begin{eqnarray}
		\dot{\rho}_\phi + 3 H \rho_\phi &=& -\Gamma_\phi \rho_\phi \nonumber\\
		\dot{\rho}_\chi + 3(1+w_\chi) H \rho_\chi &=& \Gamma_\phi^\chi \rho_\phi - \langle \sigma_{\chi\gamma'} v E \rangle_{T'} \left( n_\chi^2 - \left(n _\chi^{\text{eq}}(T') \right)^2 \right) \nonumber \\
        &&- \langle \sigma_{\chi f} v E \rangle_{T'} n_\chi^2 + \langle \sigma_{\chi-f} v\, E \rangle_T \left(n_\chi^{ \text{eq}}(T) \right)^2 \nonumber\\
		\dot{\rho}_{\gamma'} + 3(1+w_{\gamma'}) H \rho_{\gamma'} &=& \Gamma_\phi^{\gamma'} \rho_\phi + \langle \sigma_{\chi\gamma'} v E \rangle_{T'} \left( n_{\chi}^2 - \left( n_\chi^{\text{eq}}(T') \right)^2 \right)\\
		&&- \,m_{\gamma'} \Gamma_{\gamma'} \left( n_{\gamma'} - n_{\gamma'}^{\text{eq}}(T) \right) \nonumber\\
		\dot{\rho}_{\text{vs}} + 4 H \rho_{\text{vs}} &=& \Gamma_\phi^{\text{vs}} \rho_\phi + \langle \sigma_{\chi f} v\, E \rangle_{T'} n_\chi^2 - \langle \sigma_{\chi f} v\, E \rangle_T \left( n_\chi^{ \text{eq}}(T) \right)^2 \nonumber\\
        &&+ \,m_{\gamma'} \Gamma_{\gamma'} \left( n_{\gamma'} - n_{\gamma'}^{\text{eq}}(T) \right)\nonumber
\label{eq:Full_Boltzmann_system}
\end{eqnarray}
}
In these equations, $n_i^{\rm eq}(T)$ refers to equilibrium number densities at a temperature $T$ (VS) or $T'$ (HS). Similarly, thermal averages are taken over Maxwell-Boltzmann distribution at $T$ or $T'$, depending on the sector,  see Appendix \ref{app:average_cross_sections_and_decays}. 

Throughout, we focus on the HS and assume that the VS is dominantly made of thermalised relativistic particles, so we define the temperatures $T$ through $\rho_{\rm vs} \sim g_\ast T^4$, with $g_\ast$ the effective number of relativistic degrees of freedom, as usual. As it is, the above system is not closed since it involves both energy and particle number densities. It also involves the pressures, through the equation of states $w_i = p_i/\rho_i$. So, in principle, more Boltzmann equations  are necessary. To solve these equations, we will make the following assumptions.

First and if relevant, we will assume that the HS is thermalised at a temperature $T'$. The condition for this is subtle but is discussed at length in section \ref{sec:Thermalisation_HS}, see in particular table \ref{tab:thermalisation_abelian_NA}. Then, the temperature $T'$ will be related to the energy of dark photons which, by assumption, are relativistic around dark electron freeze-out, 
\begin{equation}
	\rho_{\gamma'} \sim   T'^4 
	\label{eq:definition_Tp_T_part_2}
\end{equation} 
Next, the dark electron number density is not evaluated via its Boltzmann equation but instead is determined from their energy density. In practice, for numerical resolution, this is implemented through 
\begin{equation}
    n_\chi(T') \approx \frac{\rho_\chi(T')}{m_\chi + 3T'}
\end{equation}
which allows to track the number of dark electrons from their energy density both in the relativistic and non-relativistic regimes, together with a consistent equation of state $w_\chi(T’)$, using Maxwell-Boltzmann statistics, e.g. $p_\chi = n_\chi T'$, see \cite{Coy:2024itg}.
While we assume a fully thermalised VS, the relation is valid as long as dark photons are relativistic and at thermal and chemical equilibrium. Strictly speaking, this is true only before their decoupling from the dark electrons, which is essentially at chemical freeze-out. After that, they are simply free-streaming, gradually moving away from an equilibrium distribution when they become non-relativistic \cite{Coy:2024itg}. In this work, we ignore such effects in the determination of the averaged cross sections. This potentially leaves several loopholes; however, under these assumptions the system of equations is closed and can be solved. Specific solutions will be discussed in Section~\ref{sec:DMabundance}. \\

Finally, following  \eqref{eq:definition_Tp_T_part_2}, we  define the initial temperature ratio $\xi_i = T^\prime_i/T_i$ as 
\begin{equation}
	\xi_i \equiv \left.\left( \dfrac{g_\ast}{g_{\gamma'}} \dfrac{\rho_{\gamma'}}{\rho_{\text{vs}}} \right)^{1/4}\right\vert_i
	\label{eq:xi_i_definition_part_2}
\end{equation}
Barring significant interactions between the hidden and visible sectors and assuming, for the sake of the argument, instantanous thermalisation of both sectors,  the initial ratio can be written as
\begin{equation}
	 \xi_i \simeq \left( \dfrac{g_\ast}{g_\chi} \dfrac{\Gamma_\phi^\chi}{\Gamma_\phi^{\text{vs}}} \right)^{1/4} \sim \sqrt{\dfrac{y'}{y}}
	\label{eq:xi_i_approx}
\end{equation}
In the sequel, we will track the evolution of the temperature ratio, which may significantly depart from this naive estimate, depending on the details of the process of reheating (see section \ref{sec:xi_evolution}). This requires to have a handle on the process of thermalisation of the HS, thus of dark QED.

\section{Thermalisation of dark QED}
\label{sec:Thermalisation_HS}

The initial situation we consider involves production of dark fermion pairs through inflaton decay. The initial distribution of dark fermions {(before any additional interactions)} takes the form
\begin{equation}
    f_\chi \simeq 24 \pi^2 \frac{n_\chi}{m_\phi^3} \left( \dfrac{m_\phi}{2 p} \right)^{3/2} \Theta(m_\phi/2 - p) \, ,
    \label{eq:distribution_nonthermal}
\end{equation}
with $\Theta$ the step function and $p \propto a^{-1}$, see for instance \cite{Garcia:2018wtq}. From this non-thermal distribution, we can track the averaged rates involving dark electrons at early times, for instance $\chi\bar \chi \rightarrow \gamma' \gamma'$. Starting from this, we want to determine whether the HS, here dark QED, thermalises during reheating. Thermalisation is a complex problem but it has been much studied in the literature, in particular  in the framework of QCD or heavy ions physics \cite{Baier:2000sb, Arnold:2002zm,Kurkela:2011ti}.  There is also a quite large literature that focuses on the problem of thermalisation of a sector produced by inflaton decay \cite{Kofman:1997yn,Davidson:2000er, Kurkela:2011ti, Mazumdar:2013gya, Harigaya:2013vwa, Harigaya:2014waa, Mukaida:2015ria, Drees:2022vvn, Mukaida:2024jiz, Bernal:2020gzm}. Most of these works are concerned with thermalisation through non-abelian interactions, which is relevant for the SM. Our HS however has only abelian interactions. Surprisingly, this seemingly simpler framework has not been much considered in the cosmological literature\footnote{Noticeable exceptions are \cite{Garny:2018grs,Arvanitaki:2021qlj}, but the production mechanism of the HS in both studies is different than the one we consider here. In \cite{Garny:2018grs}, they consider freeze-in production of a completely secluded HS sourced by Planck-suppressed effective couplings to VS, and in the limit of instantaneous reheating. In \cite{Arvanitaki:2021qlj} instead, inflation leads to the production of coherent massive dark electric fields that transfer their energy to DM particles, through non-perturbative effects.}. In the following sections, we review the essential aspects of thermalisation through gauge interactions. Our goal will be to derive a criterion for thermalisation of the HS in the form of a characteristic time-scale, see  \eqref{eq:abelianth_here} in section \ref{sec:thermalisation}.\footnote{See \cite{Chang:2022gcs} for a similar approach in the case of thermalisation of dark electrons in stars.} For the sake of comparison, we treat cases where gauge interaction is abelian and non-abelian in parallel, following \cite{Kurkela:2011ti} and \cite{Mukaida:2015ria}. Our discussion of thermalisation of dark QED may be deemed to be long. The reader mainly interested with the implications on the parameter space of our model, \eqref{eq:lagrangian}, may directly go to section \ref{sec:DMabundance}. \\

\subsection{Kinetic vs thermal equilibrium}  
\label{subsec:kinetic}

By thermalisation we mean thermodynamic (thermal, for short) equilibrium, thus both kinetic equilibrium  --characterised by a temperature $T$-- and chemical equilibrium --characterised by chemical potentials. Strictly speaking, chemical equilibrium only imposes relations among chemical potentials, here those of dark electrons and dark photons. In this work, we consider equilibrium to be reached when the chemical potentials of dark electrons and photons have relaxed to zero \cite{Kolb:1990vq} (we do not consider the possibility of asymmetric DM). In this sense, a nonzero chemical potential provides a measure of departure from thermal equilibrium. For instance, if $\mu < 0$,  the particle distribution is under-occupied compared to that in thermal equilibrium, which typically corresponds to an under-abundance relative to equilibrium number density. In the sequel, we will be somewhat sloppy and will refer to under-abundance rather than under-occupancy. 

To be concrete, consider the number density of dark electrons produced along inflaton decay during reheating, which have energy $E \sim m_\phi$,
\begin{equation}
	\dot{n}_\chi + 3 H n_\chi = \Gamma_\phi^\chi {\frac{\rho_\phi}{m_\phi}} \quad \mbox{\rm and so} \quad n_\chi \sim \frac{\Gamma_\phi^\chi}{H} \frac{\rho_\phi}{m_\phi} \sim \frac{\Gamma_\phi H M_{P}^2}{m_\phi}
    \label{eq:nchifromphi}
\end{equation}
with $\Gamma_\phi^\chi \lesssim H$. We can compare this to the number density dark fermions would have under the assumption of instantaneous thermal (th) equilibrium, 
\be
n_\chi^{\rm th}\sim T^{\prime 3}  \sim \left(\frac{\Gamma^\chi_\phi \rho_\phi}{H}\right)^{3/4} \sim (\Gamma^\chi_\phi H M_P^2)^{3/4}.
\ee
where we used the equivalent of Eq.\eqref{eq:Tmaxrh} but for $T'$, the temperature of the HS.
The ratio of these two quantities is
\be
\frac{n_\chi}{n_\chi^{\rm th}} \sim \frac{(\Gamma_\phi^\chi H M_P^2)^{1/4}}{m_\phi} \sim \frac{T^\prime}{m_\phi} \lesssim 1
\label{eq:ratio_kinetic_thermal_density}
\ee
Just like the VS, see Eq.\eqref{eq:Tmaxrh}, if the HS were to thermalize,\footnote{We assume for concreteness that the inflaton decay into dark photons is subdominant compare to its decay into dark fermions. }
\begin{equation}
    T_{\rm rh}'\lesssim T'\lesssim T'_{\rm max}
\end{equation}
with maximal temperature 
\begin{equation}
    T'_{\rm max} \sim (\Gamma^\chi_\phi H_{\rm end} M_P^2)^{1/4}
    \label{eq:Tpmax}
\end{equation} and temperature at reheating 
\begin{equation}
    T'_{\rm rh} \sim (\Gamma^\chi_\phi M_P)^{1/2}.
    \label{eq:Tprh}
\end{equation}
\cite{Kolb:1990vq,Giudice:1999am} (see also \cite{Coy:2024itg})
The inequality in Eq.\eqref{eq:ratio_kinetic_thermal_density} stems from energy conservation as the temperature cannot be higher than the initial energy of dark fermions. This implies that dark fermions produced from inflaton decay tend to be under-abundant. Now, an under-abundant initial condition ensures a clear separation between particle production and equilibration, allowing thermalisation to be studied in a regime which dominated by inelastic processes, like bremsstrahlung. This sharply contrasts with the case of $n \sim n_{\rm th}$, where the medium is dense and energy redistribution rather than particle multiplication governs the approach to equilibrium. In the sequel, we will assume that the dark sector is dilute, so that a perturbative approach to thermalisation applies. This also corresponds to a regime in which the particles are weakly coupled \cite{Kurkela:2011ti}.

The ratio \eqref{eq:ratio_kinetic_thermal_density} can be expressed in terms of the inflaton Yukawa couplings using \eqref{eq:phi_X_X_diagram}, so that $\Gamma_\phi^\chi \sim y'^2 m_\phi$.  At the end of inflation, 
\begin{equation}
    \left. \frac{n_\chi}{n_\chi^{\rm th}} \right|_{\rm end} \sim \frac{(\Gamma_\phi^\chi H_{\rm end}M_P^2)^{1/4}}{m_\phi} \sim y'^{1/2}\left(\frac{M_P}{m_\phi}\right)^{1/2}
\end{equation}
In the Starobinsky model, $H_{\rm end}\sim m_\phi \sim 10^{13}~{\rm GeV}$,
and so the $\chi$ are under-occupied at the end of inflation provided
\begin{equation}
    y' \lesssim \frac{m_\phi}{M_P} \simeq 10^{-5} \qquad {\text{(under-occupancy at inflation)}}
\end{equation}
The condition is however time-dependent. For instance, at the end of reheating, $H \sim \Gamma_\phi^\chi$ (assuming a hot HS scenario), then 
\begin{equation}
    \left. \frac{n_\chi}{n_\chi^{\rm th}} \right|_{\rm rh} \sim \frac{(\Gamma_\phi^\chi M_P)^{1/2}}{m_\phi} \sim y'\left(\frac{M_P}{m_\phi}\right)^{1/2}
\end{equation}
corresponding to 
\begin{equation}
    y' \lesssim \sqrt{\frac{m_\phi}{M_P}} \simeq 3\times 10^{-3} \qquad {\text{(under-occupancy at the end of reheating)}}
\end{equation}
We will come back to this in section \ref{sec:hubble_th} {(see item 3)} where we will correlate this condition with the condition for thermalisation, see figure \ref{fig:thermalisation_domain}.

\subsection{Hard $2-2$ collisions}
\label{sec:hard}

Starting from an initial abundance of dark fermions only, we may expect the leading processes to be $2\to2$ scatterings, $\chi\bar{\chi}\to\gamma'\gamma'$ and $\chi\chi\to\chi\chi$. We refer to these as hard (h) collisions. The condition for their equilibrium reads, as usual,
\begin{equation}
  \Gamma^{h}_{2\to2} = \langle \sigma^{h}_{2\to2} v \rangle n_\chi \gtrsim H \,,
  \label{eq:eqhard}
\end{equation}
where the average is taken over the out-of-equilibrium distribution of Eq.~\eqref{eq:distribution_nonthermal} (see Appendix~\ref{app:average_cross_sections_and_decays}), and $n_\chi$ is given by Eq.~\eqref{eq:nchifromphi}. For instance,
\begin{equation}
  \langle \sigma_{\chi\gamma'} v \rangle_{\rm NT} \simeq \frac{72\pi\,\alpha'^2}{m_\phi^2}\,,
\end{equation}
which is suppressed by $1/m_\phi^2$, reflecting the large energy transfer $\sim m_\phi$ involved in collisions between dark fermions and dark photons. A similar scaling holds for hard Compton scattering. Such processes can, at best, lead to kinetic equilibrium, as they essentially do not change the total particle number \cite{PhysRevLett.99.125003}. In this case, kinetic equilibrium only enforces equality of chemical potentials, $\mu_\chi = \mu_{\gamma'}$, while the corresponding kinetic temperature $T'_{\rm kin}\sim m_\phi$ is set by the mean energy per particle and is parametrically  { larger than the equilibrium temperature, $T'_{\rm kin} \gg T'$}.

Using Eq.~\eqref{eq:nchifromphi}, the equilibrium condition~\eqref{eq:eqhard} for annihilation of hard dark fermion into dark photons reads
\begin{equation}
  \langle \sigma^{h}_{2\to2} v \rangle n_\chi 
  \sim \frac{\alpha'^2}{m_\phi^2}\,
       \frac{\Gamma_\phi^\chi H M_P^2}{m_\phi}
  \gtrsim H \,,
  \label{eq:hard2-2}
\end{equation}
and is therefore time independent. 
Expressing the inflaton decay rate in terms of the Yukawa coupling $y'$, this condition becomes
\begin{equation}
  y'\,\alpha' \gtrsim \left( \frac{m_\phi}{M_P} \right)
  \qquad
  (\text{equilibrium of hard $2\to2$ collisions}) \,.
  \label{eq:kinetic_2_2}
\end{equation}
This condition is depicted in Fig.~\ref{fig:thermalisation_domain} (orange dashed line), using the value of $m_\phi$ corresponding to the Starobinsky model.

To the left of this line, hard $2-2$ collisions are out of equilibrium. This, however, does not imply that the HS cannot thermalise. As we shall see, thermalisation does in fact occur even for weaker couplings, because the relevant processes are controlled by soft rather than hard scatterings. Nevertheless, in practice, the condition~\eqref{eq:kinetic_2_2} also delineates the boundary for under-abundant dark fermions. To see this, we now derive the conditions under which thermalisation occurs in dark QED.

\subsection{Thermalisation}
\label{sec:thermalisation}

Achieving thermalisation requires taking into account particle-number–changing processes, beginning with the bremsstrahlung of dark photons from the initial distribution of dark fermions with energy $E \sim m_\phi$,
\begin{center}
\[
\scalebox{1.}{
\begin{tikzpicture}[baseline={-0.1cm}]
  \begin{feynman}[
      every blob={fill=gray!30, inner sep=2pt},
      every fermion/.style={draw=black!70, line width=0.4pt, postaction={decorate},
                            decoration={markings,mark=at position 0.55 with {\arrow{>}}}},
      every boson/.style={draw=black!40, line width=0.4pt}
    ]
    \vertex (i1) at (-1.5, 0.5) {\(\chi\)};
    \vertex (i2) at (-1.5,-0.5) {\(\chi\)};
    \vertex (c1) at (0,0.5);
    \vertex (c2) at (0,-0.5);
    \vertex (f1) at (0.6,0.5);
    \vertex (f3) at (1.8, 0.5) {\(\chi\)}; 
    \vertex (f4) at (1.8, 1.2) {\(\gamma'\)};
    \vertex (f2) at (1.8,-0.5) {\(\chi\)};

    \diagram* {
      (i1) -- [fermion] (c1),
      (i2) -- [fermion] (c2),
      (c1) -- [boson] (c2),
      (c1) -- [fermion] (f1),
      (f1) -- [fermion] (f3),
      (f1) -- [boson] (f4),
      (c2) -- [fermion] (f2)
    };
  \end{feynman}
\end{tikzpicture}
}
\]\label{Fig:softemission}
\end{center}
Naively, such processes are  $\sigma_{2-3} \sim \alpha^\prime \sigma^h_{2-2} \propto \alpha^{\prime 3}/m_\phi^2$, suppressed by the hard scale $m_\phi$. This overlooks the fact that it is only necessary to put slightly off-shell the parent particle to emit a gauge boson which can carry away a substantial fraction of the initial momentum. Number changing processes are thus driven by soft collisions  between hard particles \cite{Davidson:2000er,Kurkela:2011ti,Mukaida:2015ria} with a cross section that behaves as
\begin{equation}
 \sigma_{2\rightarrow 3}
\sim \alpha' \sigma_{2-2}^s \quad {\rm with} \quad \sigma_{2-2}^s   \sim {\alpha'^2\over \bar m^2}
\end{equation}
for exchange of a dark photon in a t-channel. 
For $\bar m \ll m_\phi$, the cross section is much larger than in the case of hard scatterings. 
In a medium, the cut-off $\bar m^2$ is typically set by the Debye mass  $\bar m = m_D$ \cite{Kapusta:2006pm}. Focusing on a medium that is initially dominated by hard fermions of mean energy $m_\phi$, 
\begin{equation}
    m_D^2 \sim \alpha' \int {d^3 p\over \omega_p} f(p) \sim \alpha' {n_\chi/m_\phi}.
\end{equation}
More generally, for a dark photon of mass $m_{\gamma'}$, 
\begin{equation}
    \bar m = \mbox{\rm max}(m_D,m_{\gamma'}).
 \end{equation}
The outcome is that the rate of bremsstrahlung is proportional to the rate of soft $2-2$ elastic collisions between hard particles, 
\begin{equation}
\label{eq:soft2-2_abelian}
    \Gamma^s_{2-2} \approx \sigma_{2-2}^s n_\chi \sim \alpha' m_\phi 
\end{equation}
 which is thus independent of $n_\chi$ as long as $m_D \gtrsim m_{\gamma'}$.
 
A further complication  is that, in a medium,  bremsstrahlung may be affected by the Landau, Pomeranchuk \& Migdal or LPM effect  \cite{Baier:2000sb,Arnold:2002zm,Kurkela:2011ti}. 
 A summary of this effect is given in Appendix \ref{app:LPM}, to which we also refer for intermediate results. The end result is that the medium leads to a suppression of the rate for bremsstrahlung.  In the case of abelian interactions, the suppression is such that
\begin{equation}
\label{eq:23abelian_here}
  \Gamma_{2-3}^A  \sim \left\{\begin{array}{cc} \alpha' \, \Gamma_{2-2}^s\sqrt{k\over k^A_{\rm LPM}} & \quad k \lesssim k^A_{\rm LPM}\\
 \\ \alpha'\, \Gamma_{2-2}^s & \quad k \gtrsim k^A_{\rm LPM}
  \end{array}\right. 
\end{equation}
see Eq. \eqref{eq:23abelian}.
Here, $k$ is the momentum of the dark photon and the pivot momentum $k^A_{\rm LPM}$ is given by $k^A_{\rm LPM}\sim {p^2 m^2_\phi/n_\chi}$ for $m_D \gtrsim m_{\gamma'}$ (see Appendix \ref{app:LPM} for $m_D \lesssim m_{\gamma'}$); $p$ is the momentum of the emitting dark electron. For $p\sim m_\phi$, $k_{\rm LPM}^{\rm A} \sim m_\phi^4/n_\chi \gg m_\phi$. Thus, the LPM suppression affects all bremsstrahlung processes, but is less effective for hard emissions with  $k\sim p$. As discussed in Appendix~\ref{app:LPM}, this implies that thermalisation proceeds via a {\it top-down} mechanism, in which energy cascades from hard modes to progressively softer ones. In this regime, bremsstrahlung dominates the energy redistribution, and therefore the {LPM-suppressed} bremsstrahlung rate parametrically controls the thermalisation timescale,
\begin{equation}
    t^A_{\rm th} \sim 1/\Gamma^A_{2-3}\sim {1\over \alpha'^2 m_\phi} \left({m_\phi^3\over n_\chi}\right)^{1/2} \qquad \mbox{\rm (thermalisation, abelian interactions),}
    \label{eq:abelianth_here}
\end{equation}
see Eq.\eqref{eq:abelianth}.

While our focus is on a HS with abelian interactions only, it is instructive to compare  this result to the case of non-abelian (NA) gauge interactions. As we recap in Appendix \ref{app:LPM}, thermalisation is a far more complex in the case of non-abelian interactions. The timescale for thermalisation is estimated to be  of order
\begin{equation}
\label{eq:nonabelianth_here}
t_{\rm th}^{\rm NA}\sim {1\over \alpha'^2 m_\phi} \left(\frac{m_\phi^3}{n_\chi}\right)^{3/8}\qquad \mbox{\rm (thermalisation, non-abelian)}
\end{equation}
using the same notation for $\alpha'$ as for the abelian case. 
All other things being kept the same,  a non-abelian HS thermalises more rapidly than an abelian one, as we shall see in the next section.

\subsection{Hubble time at thermalisation}
\label{sec:hubble_th}

\begin{table}[htb]
    \centering
    \begin{tabular}{|c|c|c|}
    \hline
         &  abelian & non-abelian  \\ 
         \hline
         &&\\
         LPM & $k\gtrsim k_{\rm LPM}$& $k\lesssim k_{\rm LPM}$ \\
         &&\\
         \hline
         &&\\
        $k_{\rm LPM}$ & $\dfrac{m_\phi^4}{n_\chi} \gg m_\phi$  &   $\dfrac{n_\chi}{m_\phi^2} \ll m_\phi$\\
        &&\\
         \hline
         &&\\
         $\Gamma_{2\rightarrow3}^{\rm LPM}$ & $\alpha'\Gamma^s_{2-2}\sqrt{\dfrac{k}{k_{\rm LPM}}}$ & $\alpha'\Gamma^s_{2-2}\sqrt{\dfrac{k_{\rm LPM}}{k}}$ \\
         &&\\
         \hline
         &&\\
        $t_{\rm th}$  & $\dfrac{1}{\alpha'^{2} m_\phi} \left(\dfrac{m_\phi^3}{n_\chi}\right)^{1/2}$ &  $\dfrac{1}{\alpha'^{2} m_\phi} \left(\dfrac{m_\phi^3}{n_\chi}\right)^{3/8}$\\
        &&\\
        \hline
        &&\\
        $H_{\rm th}$& $ \alpha'^{4}\, m_\phi\,\left( \dfrac{\Gamma_\phi^\chi M_P^2}{m_\phi^3} \right)$ &  $ \alpha'^{16/5}\,m_\phi\, \left( \dfrac{\Gamma_\phi^\chi M_P^2}{m_\phi^3} \right)^{3/5}$\\
        &&\\
        \hline
    \end{tabular}
 \caption{Comparison of thermalisation via  $2\rightarrow3$ processes for abelian and non-abelian interactions. We refer to a situation in which high-energy fermions (hard sector) are generated by inflaton decays with a rate $\Gamma_\phi^{\chi}$. They initially have hard (h) momenta $p\sim m_\phi$ and a {\it low occupancy} initial number density $n_h\ll m_\phi^3$ compared to a thermalised system. The momentum $k$ is that of the emitted gauge boson. The quantities $k_{\rm LPM}$ represent pivot scale {above (below)} which one has to take into account LPM suppression in determining $2\rightarrow3$ splitting rate, $\Gamma_{2\rightarrow3}^{\rm LPM}$ {in the abelian (non-abelian) case}. Without suppression, the rate are given by $\alpha' \Gamma_{2-2}^s$, with $s$ referring to soft $2-2$ elastic scatterings. Results for a non-abelian plasma were first derived in \cite{Harigaya:2013vwa}. All are estimates.}
    \label{tab:thermalisation_abelian_NA}
\end{table}

From the timescales \eqref{eq:abelianth_here} and  \eqref{eq:nonabelianth_here}, we  can determine if and when thermalisation occurs after inflation. From Eq.\eqref{eq:nchifromphi},
\be
n_\chi\sim \dfrac{\Gamma_\phi^\chi}{H} {\rho_\phi\over m_\phi}\sim \frac{\Gamma_\phi^\chi M_P^2}{m_\phi \,t}
\ee
Combining with Eq.\eqref{eq:abelianth_here}, an abelian HS thermalises when
\begin{equation}
    t_{\rm th}^{\rm A} \sim H_{\text{th}}^{-1} \sim {1\over m_\phi \,\alpha'^{4}} \left( \dfrac{m_\phi^3}{\Gamma_\phi^\chi M_P^2} \right) \qquad \mbox{\rm (Hubble time, abelian)}
    \label{eq:th_time_A_here}
\end{equation}
For a non-abelian HS, Eq.\eqref{eq:nonabelianth_here} gives \cite{Harigaya:2013vwa,Harigaya:2014waa}
\begin{equation}
    t_{\rm th}^{\rm NA} \sim H_{\text{th}}^{-1} \sim {1\over m_\phi \,\alpha'^{16/5}} \left( \dfrac{m_\phi^3}{\Gamma_\phi^\chi M_P^2} \right)^{3/5}\qquad \mbox{\rm (Hubble time, non-abelian)}
    \label{eq:th_time_NA_here}
\end{equation}
The ratio of the these two times is thus
\begin{equation}
    {t^{\rm A}_{\rm th}\over t^{\rm NA}_{\rm th} } \sim {1\over \alpha'^{4/5}} \left({m_\phi^3 \over \Gamma_\phi^\chi M_P^2}\right)^{2/5}
\end{equation}
For the Starobinsky model we use as a benchmark (see section \ref{sec:framework}), 
\begin{equation}
    {t^{\rm A}_{\rm th}\over t^{\rm NA}_{\rm th} } \sim  10^{-4} \, (\alpha' y')^{-4/5} 
\end{equation}
which is typically larger than one in the limit of weak couplings; {\em mutatis mutandis}, this means that thermalisation takes more time for a HS with only abelian interactions. 
\\

The key results discussed so far in this section are summarized in Table~\ref{tab:thermalisation_abelian_NA}. From there, we can distinguish different scenarios, depending on whether the HS thermalises (if at all) at the beginning of, during, or after reheating. {Following the results of the section above, we use expressions \eqref{eq:th_time_A_here} and \eqref{eq:th_time_NA_here} to evaluate the Hubble rate at which thermalisation occurs.} To recap, we defined reheating as the period of domination by the inflaton, see section~\ref{sec:framework}. The different HS thermalisation scenarios then correspond to distinct regions in the $(\alpha', y')$ plane, as shown in Fig.~\ref{fig:thermalisation_domain} both for an abelian and for a non-abelian HS for comparison. 
We fix the value of the effective Yukawa coupling of the VS to $y = 10^{-8}$. Then, along the horizontal axis, $y' \gtrsim 10^{-8}$ thus corresponds to a dominant HS (hot HS scenario), while  $y' \lesssim 10^{-8}$ corresponds to the subdominant case (cold HS). Also, when necessary, we adopt the values  $H_{\rm end} \sim m_\phi$ and $m_\phi \sim 10^{13}$ GeV of the Starobinsky model, see section \ref{sec:framework}. Finally,  we also report the {couplings} for which the dark fermions are initially under-occupied, see section \ref{subsec:kinetic}. The different limits depicted are  the  following:
\begin{figure}[hbtp]
    \centering
    \includegraphics[width=0.67\linewidth]{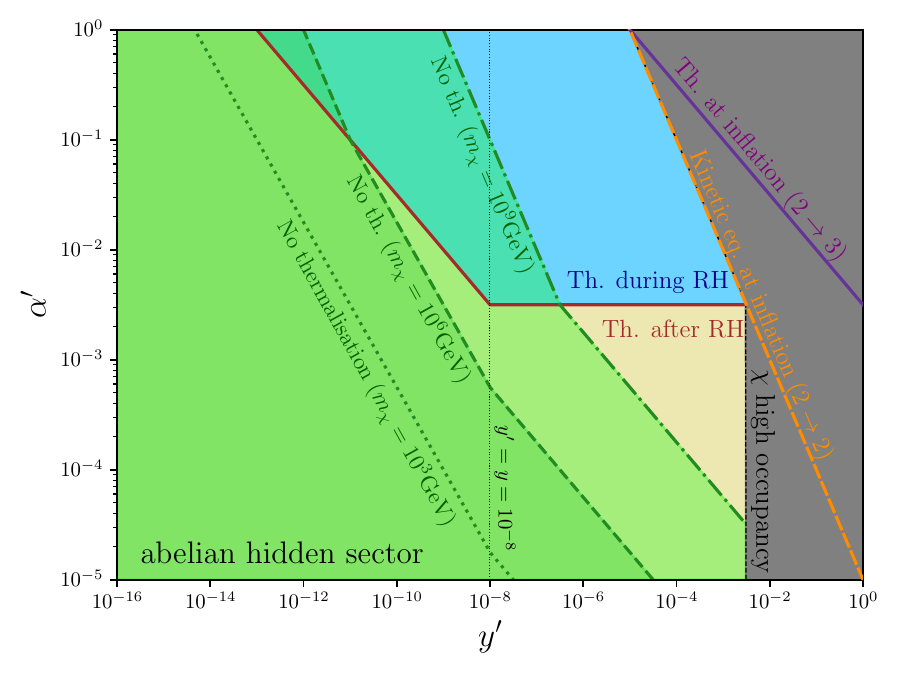}
    \vspace{0.5cm}
    \includegraphics[width=0.67\linewidth]{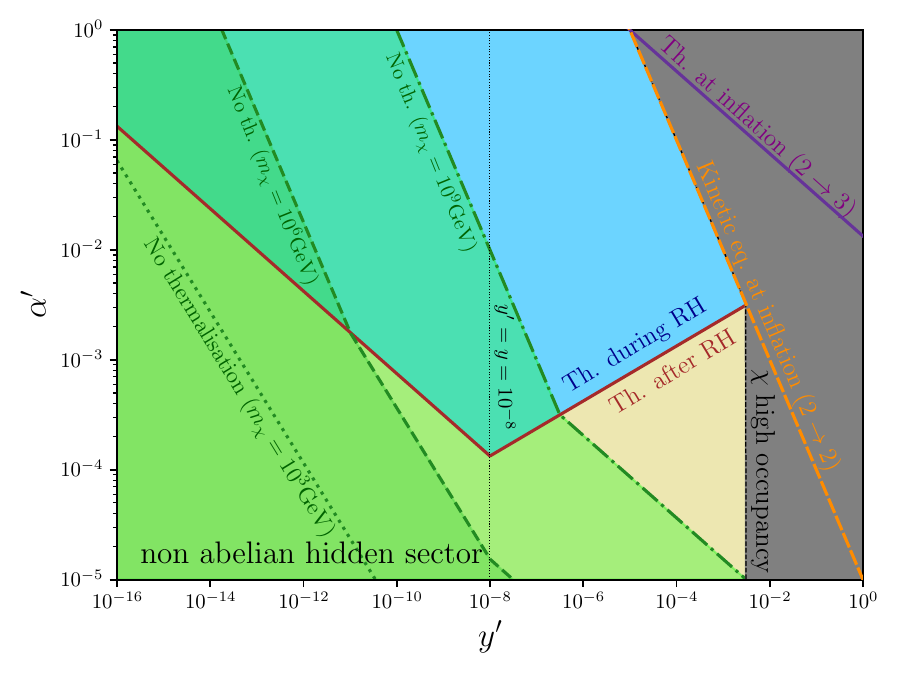}
    \caption{Summary of the constraints related to the thermalisation of a HS. The upper panel corresponds to an abelian case, the lower panel to a non-abelian one (see section~\ref{sec:hubble_th}). The constraints are illustrated for a fixed value of $y = 10^{-8}$; for $y' \gtrsim y$, the HS is hot, while for $y' \lesssim y$, the HS is cold. The blue regions indicate thermalisation occurring during reheating (see item~2 in section~\ref{sec:hubble_th}), whereas the yellow regions correspond to thermalisation after reheating. The grey shaded region lies beyond the validity of our perturbative approach; its boundary corresponds to dark fermions whose initial abundance is comparable to the thermal equilibrium abundance (limit of under-occupancy, item 3). This region partially overlaps with the orange dashed line, which represents the condition for equilibrium of hard $2 \to 2$ collisions (item~1). {The purple line indicates, for comparison, the condition for instantaneous thermalisation at the end of inflation. As it lies beyond the validity of our perturbative approach, it should be considered as indicative.} The green shaded regions, corresponding to various DM mass choices, are excluded because we require $m_\chi \lesssim T'(a_{\rm th})$, where $T'(a_{\rm th})$ is the HS temperature at thermalisation.}
    \label{fig:thermalisation_domain}
\end{figure}
\begin{description}
\item{\textbf{1. Equilibrium of hard collisions:} }

If couplings are large enough, hard collisions may be in equilibrium,  see section \ref{sec:hard} and Eq.\eqref{eq:kinetic_2_2}. The condition is $\alpha' y' \gtrsim m_\phi/M_P$ and is time-independent. The boundary of this region is shown as the orange dashed curve {in figure \ref{fig:thermalisation_domain}}.

\item{\textbf{2. Thermalisation during reheating:}} 

{This scenario occurs if $H_{\rm th} \gtrsim \Gamma_\phi$.} We must distinguish the hot and cold HS scenarios: 
\begin{equation}
	H_{\text{th}} \gtrsim \Gamma_\phi \sim \begin{cases}
		\Gamma_\phi^\chi \qquad &(y' \gtrsim y,  \text{hot HS}) \\
		\Gamma_\phi^{\rm vs} \qquad &(y' \lesssim y,  \text{cold HS})
	\end{cases}
\end{equation}
For a hot HS, Eqs.\eqref{eq:th_time_A_here} and \eqref{eq:th_time_NA_here} lead to
\begin{equation}
	H_{\text{th}} \gtrsim \Gamma_\phi^\chi \quad \mbox{\rm and so} \quad  \alpha' \gtrsim \begin{cases}
		 \left( \dfrac{m_\phi}{M_P} \right)^{1/2} \qquad &(\text{hot abelian HS})\\
		y'^{1/4} \left( \dfrac{m_\phi}{M_P} \right)^{3/8} \qquad &(\text{hot non-abelian HS})
	\end{cases}
\end{equation}
In {figure} \ref{fig:thermalisation_domain}, these conditions separate the  blue region --thermalisation during reheating-- from the yellow one --thermalisation  after reheating. 

The case of a cold HS corresponds to $y' \lesssim 10^{-8}$, in which case the condition becomes
\begin{equation}
	H_{\text{th}} \gtrsim \Gamma_\phi^{\text{vs}} \quad \mbox{\rm and thus} \quad \alpha' \gtrsim \begin{cases}
		 y^{1/2} y'^{-1/2} \left( \dfrac{m_\phi}{M_P} \right)^{1/2} \ &(\text{cold  abelian HS})\\
		y^{5/8} y'^{-3/8} \left( \dfrac{m_\phi}{M_P} \right)^{3/8}  &(\text{cold non-abelian HS})
	\end{cases}
\end{equation}
The main message is that  thermalisation is more efficient through non-abelian interactions than through abelian ones. 

At some point, hard collisions become as effective as the soft processes that lead to thermalisation.  Also, at least
in principle, we may ask under which {conditions} the HS thermalises right at the end of inflation, $H_{\rm th} \sim H_{\rm end}$. This corresponds to
\begin{equation}
	H_{\text{th}} \sim H_{\rm end} \quad \mbox{\rm so}\quad  \alpha' \gtrsim \begin{cases}
    y'^{-1/2} \left( \dfrac{H_{\rm end}}{m_\phi} \right)^{1/4} \left( \dfrac{m_\phi}{M_P} \right)^{1/2} \quad &(\text{abelian}) \\
	y'^{-3/8} \left( \dfrac{H_{\rm end}}{m_\phi} \right)^{5/16} \left( \dfrac{m_\phi}{M_P} \right)^{3/8} \quad &(\text{non-abelian})
\end{cases}
\end{equation}
This bound is depicted by the purple line in figure \ref{fig:thermalisation_domain}. We see that it corresponds to a situation in which the initial distribution would be over-occupied, beyond the validity of our approximations.

\item{\textbf{3. Dark fermions under-abundant:}}

Throughout we assumed that the dark fermions were initially under-abundant compare to their thermal equilibrium abundance. This basically amounts to assume that the HS is dilute, so that we can make a clear distinction between the different equilibration processes and time-scales, see section \ref{subsec:kinetic}. 
The condition for under-abundance is given by Eq.\eqref{eq:ratio_kinetic_thermal_density},
\begin{equation}
    \Gamma_\phi^\chi H M_P^2 \lesssim m_\phi^4   \label{eq:under_abundance_condition}
\end{equation}
which is time-dependent, as the dark fermions abundance builts up along reheating. Requiring  that the dark fermions are under-abundant by the time of  thermalisation, Eqs.\eqref{eq:th_time_A_here} and \eqref{eq:th_time_NA_here}, this condition becomes
\begin{equation}
    y' \alpha' \lesssim {m_\phi\over M_P} \quad \mbox{\rm (under-occupancy before thermalisation)}
\end{equation}
and is the same for both abelian and non-abelian interactions. This condition turns out to be the the same as the one for the equilibrium of hard collisions, provided thermalisation occurs before the end of reheating.  The HS remains under-occupied after the end of reheating if condition \eqref{eq:under_abundance_condition} is fulfilled at $H \sim \Gamma_\phi^\chi$. This gives the following bound
\begin{equation}
    y' \lesssim \sqrt{\dfrac{m_\phi}{M_P}} \simeq 3 \times 10^{-3} \quad \mbox{\rm (under-occupancy at the end of reheating)} \ ,
\end{equation}
which is independent of $\alpha'$ and corresponds to the vertical boundary of the grey-shaded regions of figure \ref{fig:thermalisation_domain}. 
We conclude that the dark fermion occupancy is of the order of the equilibrium abundance within the black shaded region. Again, what happens there is beyond the domain of validity of our approach to thermalisation, as it enters a non-perturbative regime.

\item{\textbf{4. DM particles relativistic at thermalisation:}}

For the consistency of our framework, we must require that the dark fermions have not frozen out before thermalisation of the HS. First, the derivation of the thermalisation timescale sketched above assumed that all particles are relativistic. Also, one of our  goals is to determine the initial conditions for thermal freeze-out of DM particles. Consequently,  we  impose that
\begin{equation}
\label{eq:rel_th}
	T'{(a_{\rm th})} \gtrsim m_\chi \quad \mbox{\rm with} \quad \rho_{\gamma'} \sim g_{\gamma'} T'^4  \quad \mbox{\rm or} \quad  T' \sim \rho_{\gamma'}^{1/4}
\end{equation}
We need to consider different situations. 
First, if thermalisation occurs before the end of reheating, then 
\begin{equation}
	 \rho_{\gamma'}\sim \rho_\chi  \sim \Gamma_\phi^\chi H M_P^2
\end{equation}
and the condition \eqref{eq:rel_th} reads 
\begin{equation}
	\Gamma_\phi^\chi H_{\text{th}} M_P^2 \gtrsim m_\chi^4 \quad \mbox{\rm or} \quad \alpha' \gtrsim \begin{cases}
		y'^{-1} \left( \dfrac{m_\chi}{M_P} \right) \quad &(\text{abelian}) \\
		y'^{-1} \left( \dfrac{m_\chi^5}{m_\phi M_P^4} \right)^{1/4} \quad &(\text{non-abelian})
	\end{cases}
	\label{eq:R_thermalisation_condition_pre_RH_th}
\end{equation}
Next, if $H_{\text{th}}$ is reached after reheating, we need to treat the hot and cold HS cases separately. If $y' \gg y$ (hot HS), the Hubble rate after reheating is driven by the HS, implying that $\rho_\chi \sim H^2 M_P^2$. From this, the condition \eqref{eq:rel_th} can be written as
\begin{equation}
	H_{\text{th}}^2 M_P^2 \gtrsim m_\chi^4 \rightarrow \alpha' \gtrsim \begin{cases}
		y'^{-1/2} \left( \dfrac{m_\chi^2 m_\phi}{M_P^3} \right)^{1/4} \quad &(\text{abelian})\\
		y'^{-3/8} \left( \dfrac{m_\chi^{10} m_\phi}{M_P^{11}} \right)^{1/16} \quad &(\text{non-abelian}) 
        \label{eq:R_thermalisation_condition_post_RH_hot}
	\end{cases}
	\quad \text{(hot HS)}
\end{equation}
Conversely, when $y' \lesssim y$, from solutions of Boltzmann equations (\ref{eq:Full_Boltzmann_system}) (to be discussed in the next section)  and assuming negligible interactions between the HS and VS,  we got that $\rho_\chi \sim (\Gamma_\phi^\chi/\Gamma_\phi^{\text{vs}}) \rho_{\text{vs}} \sim (\Gamma_\phi^\chi/\Gamma_\phi^{\text{vs}}) H^2 M_P^2$ and the  condition \eqref{eq:rel_th} becomes
\begin{equation}
	\dfrac{\Gamma_\phi^\chi}{\Gamma_\phi^{\text{vs}}} H_{\text{th}}^2 M_P^2 \gtrsim m_\chi^4 \rightarrow \alpha' \gtrsim \begin{cases}
		y^{1/4} y'^{-3/4} \left( \dfrac{m_\chi^2 m_\phi}{M_P^3} \right)^{1/4} \quad &(\text{abelian}) \\y^{5/16} y'^{-11/16} \left( \dfrac{m_\chi^{10} m_\phi}{M_P^{11}} \right)^{1/16} \quad &(\text{non-abelian})
	\end{cases}  \label{eq:R_thermalisation_condition_post_RH_cold}
	\quad \text{(cold HS)}
\end{equation}

These different conditions are depicted by the green-shaded areas. For a given DM mass, the green region is excluded; the case of $m_\chi = 10^6$ GeV covers all the 3 cases discussed here. 
\end{description}

\noindent Equipped with the above results, we now address the problem of DM abundance, taking into account that the HS could have thermalised. 

\section{Dark matter abundance}
\label{sec:DMabundance}

We remind that the Yukawa couplings $y'$ and $y$ set the initial energy density ratio, $\rho_{\rm hs}/\rho_{\rm vs} \sim (y'/y)^2$, see section \ref{subsec:boltzmann}. In turn, if both sectors thermalise, the temperature ratio is $\xi = {T'/T} \sim (\rho_{\rm hs}/\rho_{\rm hs})^{1/4}$. We will consider various regimes in the following, both with  $\rho_{\rm hs} \gtrsim \rho_{\rm vs}$ (hot HS) and with $\rho_{\rm hs} \lesssim \rho_{\rm vs}$ (cold HS) along reheating. Following the previous section, we will distinguish regimes for which the HS thermalises or remains out of equilibrium (non-thermal DM production). We begin with the latter (regime I). We then consider standard freeze-out, which can occur  after reheating, both for a hot or a cold HS (regime II). Between these two extremes, there is an intermediate regime along which the dark fermions are non-relavitivistic and annihilate while being sourced by inflaton decay (regime III). A summary of the timeline for each regime is illustrated in figure \ref{fig:timeline}, at the end of this section.

\subsection{Regime I: Non-thermal production}
\label{sec:non_thermal}

This may be obvious but, for the sake of the argument, let us say that a scenario in which the HS is dominant ($y' \gtrsim y$) but has not reached thermal equilibrium, is not possible. Indeed, the annihilation cross section between hard dark fermions, $E \sim m_\phi$, which scales as $\propto m_\phi^{-2}$ (see Eq.~\eqref{eq:hard2-2}), must be efficient  to deplete the otherwise over-abundant DM particles. As the orange line in Fig.~\ref{fig:thermalisation_domain} shows,  this is possible only for large couplings, for which the HS is thermalised. 

A non-thermal scenario is thus only possible for a subdominant HS, $y'<y$, which remains out of equilibrium. This is just a freeze-in scenario \cite{McDonald:2001vt,Hall:2009bx,Chu:2011be}.
The  dark electrons number density is given by Eq.\eqref{eq:nchifromphi}, $n_\chi \sim \Gamma_\phi^\chi \rho_\phi/H\, m_\phi$ with $\Gamma_\phi^\chi \sim y'^2 m_\phi$. Similarly, 
VS particles reach a (thermal, by assumption) number density given by 
\begin{equation}
 \rho_{\text{vs}} \sim  s_{\rm vs} \, T \sim \dfrac{\Gamma_\phi^{\text{vs}}}{H} \rho_\phi\,.
\end{equation}
with $T \sim \rho_{\rm vs}^{1/4}$.
The DM yield $Y_{\rm dm}$ at the end of reheating is then given by
\begin{equation}
	Y_\chi = \dfrac{n_\chi}{s} \sim \dfrac{\Gamma_\phi^\chi M_P^{1/2}}{m_\phi H_{\text{rh}}^{1/2}} \sim y'^2 y^{-1} \left( \dfrac{M_P}{m_\phi} \right)^{1/2}
\end{equation}
using $H_{\text{rh}} \sim \Gamma_\phi^{\text{vs}} \sim y^2 m_\phi$. Neglecting a possible contribution from dark photons, the observed DM density \cite{Planck:2018VI} gives 
\begin{equation}
	m_\chi Y_\chi^0 \simeq 4 \times 10^{-10} \text{GeV} \quad \mbox{\rm or} \quad m_\chi \sim \xi_i^{-4} \left( \dfrac{10^{-11}}{y} \right) \rm GeV
    \label{eq:Constraint_relic_abundance}
\end{equation}
Here, $\xi_i \sim \sqrt{y'/y} \lesssim 1$ is just a measure of the ratio of energy densities, as the HS has not thermalised.

\subsection{Regime II: Freeze‐out after reheating}
\label{sec:post_RH_FO}

We now turn to thermal scenarios. We consider couplings such that the HS has thermalised before the end of reheating, see section \ref{sec:hubble_th} and figure \ref{fig:thermalisation_domain}. The dark fermions may freeze-out during reheating (RH) or after reheating (post-RH). We begin with the latter, which occurs when $T' \gg m_\chi$ at the end of reheating. This is illustrated in figure \ref{fig:Boltzmann_illustrations_post_RH} both for hot ($y'>y$, top panel) and cold HS ($y'<y$, bottom panel), obtained solving numerically the system of Boltzmann equations \eqref{eq:Full_Boltzmann_system}, with the appropriate thermal averages for each sectors, as derived in Appendix \ref{app:average_cross_sections_and_decays}. During reheating, $\rho_\chi \approx \rho_{\gamma'}$ (we assume classical Maxwell Boltzmann statistics for simplicity, see \cite{Coy:2024itg}). The parameters are chosen for the sake of illustration; the scale factor is normalised to the one at the end of inflation ($\aend$). 
\begin{figure}[hbt]
	\centering
	\includegraphics[width=0.7\textwidth]{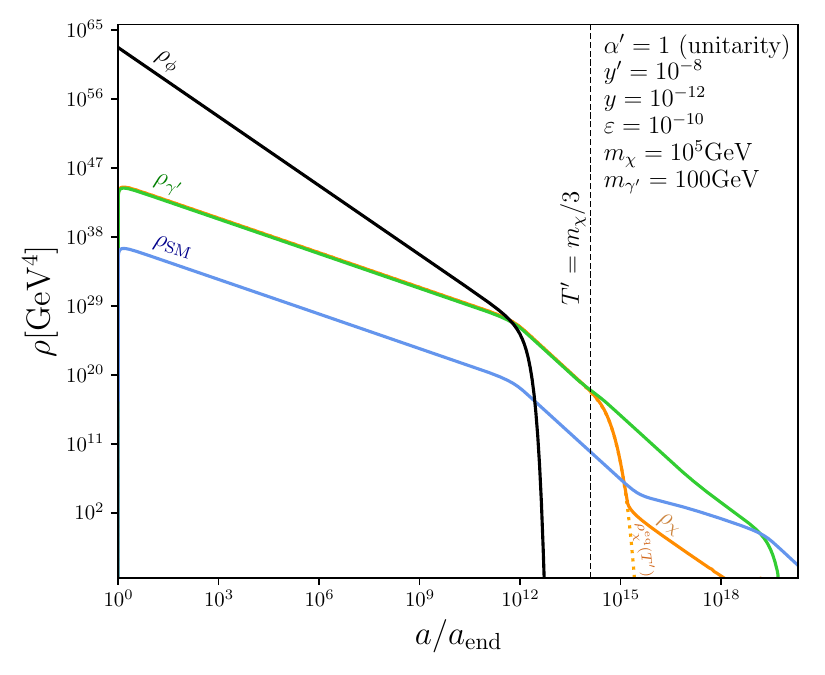}
	\includegraphics[width=0.7\textwidth]{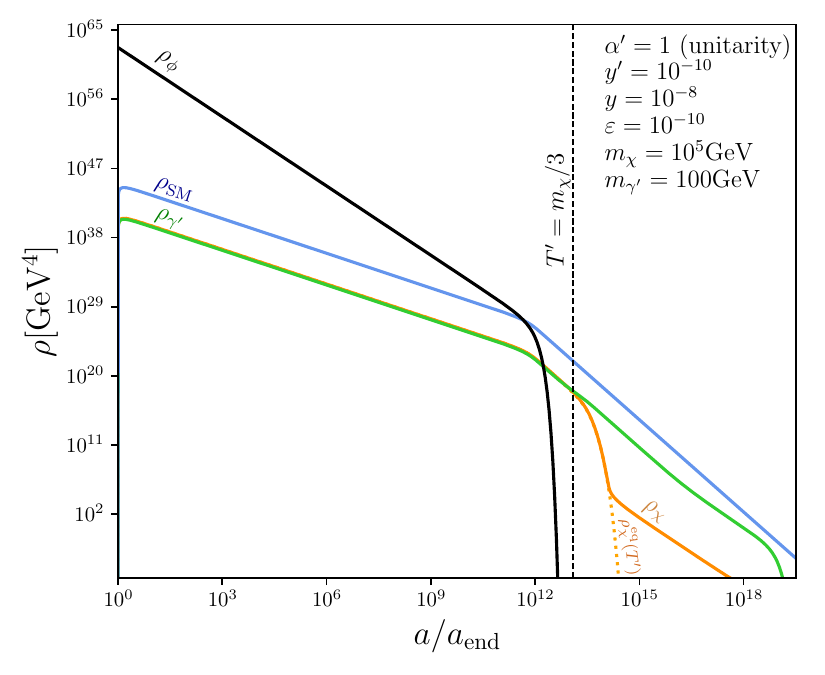}
	\caption{{\bf Freeze-out after reheating}. The top (bottom) panel corresponds to a hot, $y'\gtrsim y$ HS (respectively, cold $y'\lesssim y$). Along reheating, $\rho_{\gamma'} \sim \rho_\chi$. The end of reheating is marked by the exponential fall-off of the inflaton, after which the expansion becomes radiation dominated. In the top panel, dark photons (green curve) decay after DM freeze-out, transferring their entropy to the VS (blue curve), through a second phase of reheating. In the bottom panel, they become non-relativistic but never come to dominate the energy density of the universe, so the entropy transferred to the VS is marginal. The $\chi \bar{\chi} \rightarrow \gamma' \gamma'$ cross section is set to its maximal value, as allowed by unitarity (\emph{unitarity $\alpha'$}).}
	\label{fig:Boltzmann_illustrations_post_RH}
\end{figure}
Using the instantaneous freeze-out approximation, the particle density of dark fermions and the total entropy density are given by
\begin{equation}
	n_{\chi,\text{fo}} \approx \left.\dfrac{H}{\langle \sigma_{\chi\gamma'} v \rangle_{T'}} \right\vert_{\rm fo} \quad \text{and} \quad s_{\text{t,fo}} \sim g_{\ast} T_{\text{fo}}^3 + g_{\gamma'} T_{\text{fo}}'^3 \sim (H_{\rm fo} M_P)^{3/2}
\end{equation}
so that
\begin{equation}
	Y_{\chi,\text{fo}} = \left.\dfrac{n_{\chi}}{s_{\text t}}\right\vert_{\rm fo}  \quad \mbox{\rm gives} \quad \begin{cases}
		\dfrac{1}{\langle \sigma_{\chi\gamma'} v \rangle_{T'} M_P m_\chi} \quad & \quad (\mbox{\rm hot HS})  \\
		\dfrac{\xi_{\text{fo}}}{\langle \sigma_{\chi\gamma'} v \rangle_{T'} M_P m_\chi} \quad &\quad (\mbox{\rm cold HS}) 
        \label{eq:DM_yield_post_RH}
	\end{cases}
\end{equation}
where the temperature $T^{(\prime)}$ of the dominant sector is $\sim {H^{1/2} M_P}^{1/2}$ and the Hubble rate $H_{\text{fo}}$ at freeze-out is $\sim m_\chi^2/M_P$ or $m_\chi^2/\xi_{\text{fo}}^2 M_P$ depending on which sector is the dominant one \cite{Coy:2024itg}. We underline that the temperature ratio is that at DM freeze-out, $\xi_{\rm fo}$, to emphasize that this quantity may depart from the initial value $\xi_i$, see section \ref{sec:xi_evolution}.\\

In this work, we assume that dark photons are lighter than dark fermions, as it is the most generic way to realize DM freeze-out  \cite{Griest:1990kh}. After dark fermions freeze-out, the  number density of dark photons is relatively large,
\[
n_{\gamma'} \sim T_{\rm fo}'^{\,3} \gg n_{\chi,{\rm fo}},
\]
and they are free-streaming with, at least initially, a relativistic distribution that can be  parameterized by $T' \propto 1/a$. In the case of a hot HS, the dark photons must decay back into VS degrees of freedom before BBN. As discussed in Ref.~\cite{Coy:2024itg}, one may distinguish two possibilities,  depending on the mean dark photon decay rate in VS particles, which is controlled both by the kinetic mixing parameter (see Eq.~\eqref{Fig:DP_decay}) and the ratio $m_{\gamma'}/T'$. By assumption, initially $\langle \Gamma_{\gamma'} \rangle \ll H$. As $H$ decreases and  $\langle \Gamma_{\gamma'} \rangle \sim \Gamma_{\gamma'} m_{\gamma'}/{T'}$ 
increases, it may be that $\langle \Gamma_{\gamma'} \rangle \gtrsim H$
while the dark photons are still relativistic. In this situation, the dark photons  reach equilibrium with the VS with $T' = T$. As the  condition $\langle \Gamma_{\gamma'} \rangle \gtrsim H$ holds thereafter, the dark photons remain in thermal equilibrium, and their abundance simply becomes Boltzmann suppressed when they become non-relativistic \cite{Harvey:1981yk}. If their decay rate is slower, with $\Gamma' \lesssim H$ when the dark photons becomes non-relativistic, then the universe undergoes a period of matter domination, driven by the dark photons. When eventually $H \sim \Gamma_{\gamma'}$, the dark photons decay releases entropy, thereby reheating the VS. The summary of possibilities, depending on $\epsilon$ and $m_{\gamma'}$ is given in \cite{Coy:2024itg}, see in particular figure 5. In the case of a cold HS, there are also several possibilities. First, dark photons do not have to decay if they are light enough to constitute a hot but subdominant DM specie \cite{Hambye:2020lvy,Coy_2021,Coy:2024itg}. Otherwise, they must decay. Depending on their mass and abundance, they may come to dominate the expansion of the universe and, again, their decay will release entropy and reheat the VS \cite{Coy:2024itg}.
\\

If the dark photon produces entropy, we must multiply \eqref{eq:DM_yield_post_RH} by a dilution factor $D$ given by the ratio of comoving entropy before and after dark photon decay,
\begin{equation}
	D = \dfrac{s_{\text{t,fo}} a_{\text{fo}}^3}{s_{\text{t,f}} a_{\text{f}}^3} \sim \left( \dfrac{\rho_{\text{t,fo}} a_{\text{fo}}^4}{\rho_{\text{vs,f}} a_{\text{f}}^4} \right)^{3/4} \lesssim 1
    \label{eq:Entropy_dilution_factor_expression}
\end{equation}
where the subscript ($f$) refers to the end of entropy production from dark photon decay. With this factor, the final {(and thus current)} DM yield {$Y_\chi^0$} is given by  
\begin{equation}
    Y_{{\chi}}^{{0}} = D \cdot Y_{\chi,\rm fo}
\end{equation}
As explained above, the dilution factor depends on the scenario we consider. We begin with the case of a hot HS. As shown in \cite{Coy:2024itg}, the dilution factor depends in general on the {square root of} the ratio of two timescales, one being the lifetime of the dark photons $\tau_{\gamma'} = \Gamma_{\gamma'}^{-1}$ and the other one the moment when the dark photons become non-relativistic, $H_{\rm nr}^{-1}$. Indeed, 
\begin{equation}
	D = \dfrac{s_{\text{t,fo}} a_{\text{fo}}^3}{s_{\text{t,f}} a_{\text{f}}^3} \sim \left( \dfrac{\rho_{\gamma', \text{nr}} a_{\text{nr}}^4}{\rho_{\text{vs,f}}a_{\text{f}}^4}
\right)^{3/4}\sim \underbrace{\left( \dfrac{\rho_{\gamma', \text{nr}} a_{\text{nr}}^3}{\rho_{\text{vs,f}}a_{\text{f}}^3}
\right)^{3/4}}_{\sim 1} \left( \dfrac{a_{\text{nr}}}{a_{\text{f}}} \right)^{3/4} \sim \left( \dfrac{\Gamma_{\gamma'}}{H_{\text{nr}}} \right)^{1/2} \quad \mbox{\rm (hot HS)}
\label{eq:D_hot_II}
\end{equation}
with, by assumption, $\Gamma_{\gamma'} \ll H_{\rm nr}$.
In this relation, we used that the energy density of the dark photons scales as matter, ${\rho_{\gamma'}} \propto a^{-3}$, when they become non-relativistic and that, when they decay, their energy density is transferred to the VS, $\rho_{\rm vs,f} \sim \rho_{\gamma',\rm nr} (a_{\rm nr}/a_{\rm f})^3$, so the pre-factor is actually ${\cal O}(1)$; the dependence on the time scales stems simply from the fact that $H \propto a^{-3/2}$ during a matter dominated era, here driven by the non-relativistic dark photons \cite{Coy:2024itg}. If, instead, the HS is cold but the dark photons become non-relativistic, they may start dominating the expansion if $\rho_{\gamma'}$ and $\rho_{\text{vs}}$ become equal {before dark photons decay}, in which case the relevant initial time scale is the time $H_{\rm eq}^{-1}$ of equality between the HS and the VS energy densities \cite{Kolb:1990vq,Coy:2024itg},
\begin{equation}
	D = \dfrac{s_{\text{t,fo}} a_{\text{fo}}^3}{s_{\text{t,f}} a_{\text{f}}^3} \sim \left( \dfrac{\Gamma_{\gamma'}}{H_{\text{eq}}} \right)^{1/2}\quad \mbox{\rm (cold HS)}
    \label{eq:D_cold_II}
\end{equation}
where by assumption $\Gamma_{\gamma'} \ll H_{\rm eq}$.
In both cases, a more long-lived dark photon will lead to more entropy dilution.

\subsection{Regime III: Freeze-out during reheating}
\label{sec:pre_RH_FO}

We now discuss  dark fermions freeze-out during reheating. This possibility has also been discussed in \cite{Bernal:2022wck, Chowdhuryand:2024uvi} but here we consider a scenario in which dark fermions freeze-out while being still produced by inflaton decay. 
We consider model parameters such that the HS has thermalised during reheating. If this is not the case, DM production is non-thermal, as in regime I, section \ref{sec:non_thermal}, which is only viable for a cold HS. Freeze-out during reheating however may occur both for a cold or hot HS. We focus on the latter but the end result can be readily applied to the cold case, Eq.\eqref{eq:Y_X_RH_pre_RH_FO}. The processes that can significantly affect the dark electron abundance are inflaton decay and dark fermions annihilation. From the set of Boltzmann equations \eqref{eq:Full_Boltzmann_system},  consider 
\begin{equation}
	\dot{\rho}_\chi + 3(1+w_\chi) H \rho_\chi = \Gamma_\phi^\chi \rho_\phi - \langle \sigma_{\chi\gamma'} v E \rangle_{T'} \left( n_\chi^2 - \left( n_\chi^{\text{eq}}(T') \right)^2 \right) \,.
  \label{eq:bolt_simple1}
\end{equation}
As long as the dark electrons remain relativistic, their abundance is  $n_\chi \approx n_\chi^{\rm eq}(T') \sim T'^3$. The evolution of $T'$ is in turn driven by the source term from inflaton decay, 
$
    \rho_\chi \sim {\Gamma_\phi^\chi \rho_\phi/ H} \sim a^{-3/2}
$
so that $n_\chi \sim \rho_\chi^{3/4}\propto a^{-9/8}$ and $T' \propto a^{-3/8}$, see figure \ref{fig:Boltzmann_illustrations_pre_RH}. When $T' \sim m_\chi$, the dark electrons become non-relativistic but are still tracking the equilibrium abundance, which becomes Boltzmann suppressed. As  $\rho_\chi \approx m_\chi n_\chi^{\rm eq}$ the equation \eqref{eq:bolt_simple1}  simplifies to 
\begin{equation}
	\dot{n}_\chi + 3 H n_\chi \approx \Gamma_\phi^\chi {\rho_\phi\over m_\chi} - \langle \sigma_{\chi\gamma'} v \rangle_{T'} \left( n_\chi^2 - \left( n_\chi^{\text{eq}}(T') \right)^2 \right)  
    \label{eq:bolt_simple2}
\end{equation}
{once $T' \lesssim m_\chi$.} The appearance of $\rho_\phi/m_\chi$ in the source term (instead of $\rho_\phi/m_\phi$, see \eqref{eq:nchifromphi}) 
 reflects the fact that the inflaton decay products are assumed to be thermalised.

{From this time on, the dark electrons follow a Boltzmann suppressed equilibrium density $n_\chi^{\rm eq}$. } In absence of the inflaton source term, the dark electrons would eventually freeze-out at a temperature $T'_{\rm fo}$, as in the case of post-reheating freeze-out, Eq.\eqref{eq:DM_yield_post_RH}. Here, instead, the inflaton source term takes over at some $T' \gtrsim T'_{\rm fo}$, 
\begin{equation}
     {\Gamma_\phi^\chi\rho_\phi\over m_\chi}  \gtrsim \langle \sigma_{\chi\gamma'} v \rangle (n_\chi^{\text{eq}})^2 
    \label{eq:equality1}
\end{equation}
Indeed, using the instantaneous approximation, 
$
    n_{\chi,\rm fo} \sim {H_{\rm fo}}/{\langle \sigma_{\chi \gamma'} v\rangle}
$
and 
$
    \rho_\phi \sim H^2 M_P
$
show that the inflaton term would {remain} subdominant {until} freeze-out only if    
\begin{equation}
    {\Gamma_\phi^\chi \rho_{\phi,\rm fo}\over m_\chi} \lesssim \langle\sigma_{\chi \gamma'} v \rangle n_{\chi,\rm fo}^2 \implies y'\alpha' \lesssim \frac{m_\chi}{M_P}\left(\frac{m_\chi}{m_\phi}\right)^{1/2} 
\end{equation}
Comparing with Eq.\eqref{eq:R_thermalisation_condition_pre_RH_th}, we see that this condition is never satisfied if the HS is thermalised. Thus, once \eqref{eq:equality1} is satisfied, the Boltzmann equation  \eqref{eq:bolt_simple2} further simplifies to take the form 
\begin{equation}
	\dot n_\chi + 3 H n_\chi \approx \Gamma_\phi^\chi {\rho_\phi\over m_\chi} - \langle \sigma_{\chi\gamma'} v \rangle n_\chi^2
    \label{eq:reannihilation}
\end{equation}
 As long as the inflaton is still driving the expansion, $H \gtrsim \mbox{\rm max}(\Gamma_\phi^\chi,\Gamma_\phi^{\rm vs})$, the dark fermion density follows a quasi-equilibrium solution that stems from near cancellation of the rhs of Eq.\eqref{eq:reannihilation}, 
\begin{equation}
	n_\chi \approx \sqrt{\dfrac{\Gamma_\phi^\chi}{\langle \sigma_{\chi\gamma'} v \rangle} {\rho_\phi\over m_\chi}} \propto a^{-3/2} \quad (\text{inflaton resourcing})
	\label{eq:rho_X_resourcing}
\end{equation}
see figure \ref{fig:Boltzmann_illustrations_pre_RH}.
We dub this "inflaton resourcing" but, for a cold HS, this dynamics is akin to the so-called re-annihilation regime  familiar from other portal scenarios \cite{Hall:2009bx,Cheung:2010gj,Chu:2011be}. In these models, the production of the HS proceeds from the VS which competes with annihilation within the HS, provided it has thermalised.\footnote{Obviously, this is only possible for a cold HS.  Incidentally, for the curves depicted in the lower panel of Fig.~\ref{fig:Boltzmann_illustrations_pre_RH} we have assumed that other possible source terms, such as $f \bar{f} \rightarrow \chi \bar{\chi}$ can be neglected.}
\begin{figure}[hbt]
	\centering
	\includegraphics[width=0.7\textwidth]{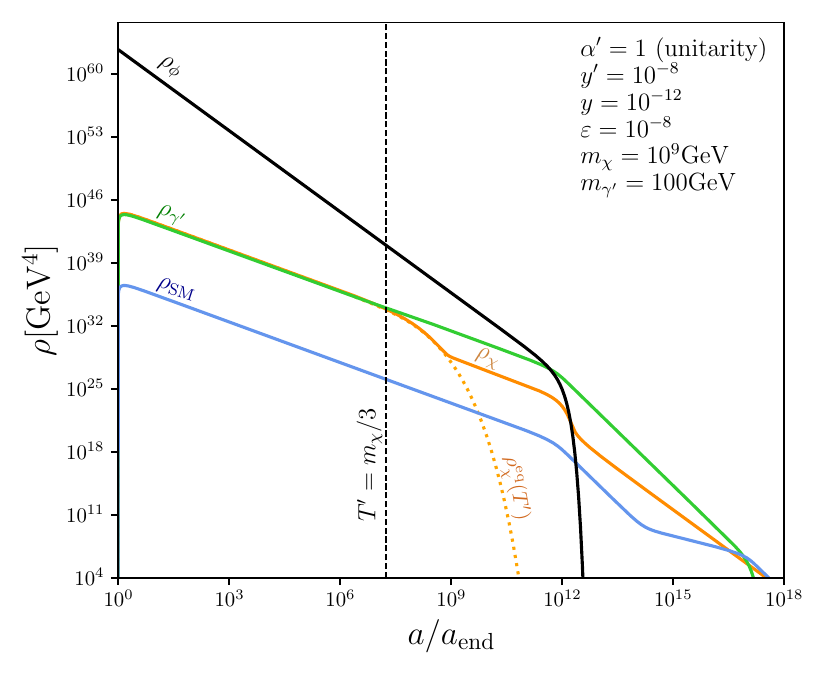}
	\includegraphics[width=0.7\textwidth]{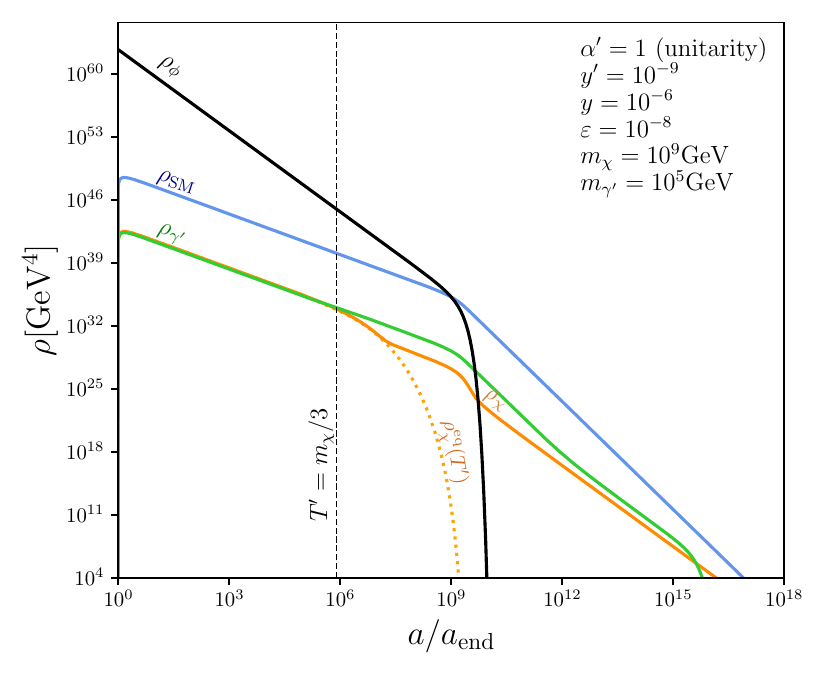}
	\caption{{\bf Freeze-out during reheating}. The top  panel corresponds to a hot HS ($y'\gtrsim y$) and bottom  for a cold one ($y'<y$). As in figure \ref{fig:Boltzmann_illustrations_post_RH}, we set $\alpha'$ to the maximal value allowed by unitarity. See the text for a discussion of behavior of the curve for the DM. }
	\label{fig:Boltzmann_illustrations_pre_RH}
\end{figure}

This ends at $H_{\rm rh} \sim \mbox{\rm max}(\Gamma_\phi^\chi,\Gamma_\phi^{\rm vs})$, at which point the DM relic abundance is driven only by the annihilation term so that the dark electron relic abundance is given by
\begin{equation}
	Y_{\chi,\text{fo}} = \left.\dfrac{n_\chi}{s_{\text{t}}}\right\vert_{\rm rh} \sim \dfrac{H_{\text{rh}}}{\langle \sigma_{\chi\gamma'} v \rangle \left( H_{\text{rh}} M_P \right)^{3/2}} \quad \mbox{\rm giving} \quad \begin{cases}
		\dfrac{1}{\langle \sigma_{\chi\gamma'} v \rangle (\Gamma_\phi^\chi)^{1/2} M_P^{3/2}} \quad &(\text{hot HS}) \\
		\dfrac{1}{\langle \sigma_{\chi\gamma'} v \rangle (\Gamma_\phi^{\text{vs}})^{1/2} M_P^{3/2}} \quad &(\text{cold HS})
	\end{cases}
	\label{eq:Y_X_RH_pre_RH_FO}
\end{equation}
 in the instantaneous freeze-out approximation. The dark fermion abundance is inversely proportional to the annihilation cross section, as for freeze-out after reheating, but the time of freeze-out is set by the inflaton decay rate. For fixed dark electron mass, this implies that the DM abundance is controlled by $\alpha'^2 y' \sim $ const for the case of a hot HS.

As for the regime of freeze-out after reheating (regime {II}, section \ref{sec:post_RH_FO}), the final dark fermion relic abundance depends on the fate of the dark photon, which may release entropy. Following the results of that section, the final DM abundance is set
by 
\begin{equation}
    Y_{{\chi}}^{{0}} = D \cdot Y_{\chi, \rm fo} 
\end{equation}
with $D$ the entropy dilution factor. 
As in the previous section, $D$ is relevant only if dark photons are non-relativistic and dominate or come to dominate the expansion; otherwise $D\sim 1$, see section \ref{sec:post_RH_FO}. 

In the regime of freeze-out during reheating,  the HS is thermalised, thus the dark photons track the energy density injected in the HS during reheating, $\rho_{\gamma'} \sim \Gamma_\phi^\chi \rho_\phi/H$. Hence, at the end of reheating,
\begin{equation}
\rho_{\gamma', \text{rh}} \sim \dfrac{\Gamma_\phi^\chi}{H_{\text{rh}}} \rho_{\phi, \text{rh}} \sim \Gamma_\phi^\chi H_{\text{rh}} M_P^2 \sim \begin{cases}
		\Gamma_\phi^{\chi 2} M_P^2 \quad &(\text{hot HS}) \\
		\Gamma_\phi^\chi \Gamma_\phi^{\text{vs}} M_P^2 \quad &(\text{cold HS})
	\end{cases}
	\label{eq:rho_A_rh}
\end{equation}
If the HS is hot,
\begin{equation}
    {D = \dfrac{s_{\text{t,rh}} a_{\text{rh}}^3}{s_{\text{t,f}} a_{\text{f}}^3} \sim \underbrace{\left( \dfrac{\rho_{\gamma', \text{rh}} a_{\text{rh}}^3}{\rho_{\text{vs,f}} a_{\text{f}}^3}  \right)^{3/4}}_{\sim 1} \left( \dfrac{a_{\text{rh}}}{a_{\text{f}}} \right)^{3/4} \sim \left( \dfrac{\Gamma_{\gamma'}}{H_{\text{rh}}} \right)^{1/2} \quad (\mbox{\rm hot HS})}
    \label{eq:D_hot_preRH}
\end{equation}
where, as in the previous section, we used that $\rho_{\rm vs,f} \sim \rho_{\gamma', \text{rh}} (a_{\rm rh}/a_{\rm f})^3$ and the scaling of the Hubble rate during a MD era, $H \propto a^{-3/2}$.

If instead the HS is cold,  the relevant time is when the dark photons are non-relativistic and come to dominate the expansion at $H_{\rm eq}$, 
\begin{equation}
    {D = \dfrac{s_{\text{t,eq}} a_{\text{eq}}^3}{s_{\text{t,f}} a_{\text{f}}^3} \sim \underbrace{\left( \dfrac{\rho_{\gamma', \text{eq}} a_{\text{eq}}^3}{\rho_{\text{vs,f}} a_{\text{f}}^3} \right)^{3/4}}_{\sim 1} \left( \dfrac{\Gamma_{\gamma'}}{H_{\text{eq}}} \right)^{1/2} \sim \left( \dfrac{\rho_{\text{vs, rh}}}{\rho_{\gamma', \text{rh}}} \right) \left( \dfrac{\Gamma_{\gamma'}}{H_{\text{rh}}} \right)^{1/2}\quad (\mbox{\rm cold HS})}
    \label{eq:D_cold_preRH}
\end{equation}
where, for the last identity,  we used that the $\gamma'$ are subdominant, and non-relativistic between $H_{\rm rh}$ and $H_{\rm eq}$, together with $\rho_{\gamma', \text{eq}} \equiv \rho_{\text{vs, eq}}$.
These expressions are similar to the ones in the regime of freeze-out after reheating, see Eqs. \eqref{eq:D_hot_II} and \eqref{eq:D_cold_II}.

\begin{table}[H]
    \centering
    \begin{tabular}{|c|c|}
    \hline
    &\\
       Regime II & $Y_\chi^0 = Y_{\chi, \rm fo}\cdot D \sim\begin{cases}
		\dfrac{\Gamma_{\gamma'}^{1/2}}{\langle \sigma_{\chi\gamma'} v \rangle_{T'} H_{\rm nr}^{1/2} M_P m_\chi} \quad & \quad (\mbox{\rm hot HS}) \\	\dfrac{\xi_{\text{fo}}\Gamma_{\gamma'}^{1/2}}{\langle \sigma_{\chi\gamma'} v \rangle_{T'} H_{\rm eq}^{1/2}M_P m_\chi} \quad &\quad (\mbox{\rm cold HS})   \end{cases}$ 
        \\
        &\\
        \hline
        &\\
      Regime III  &  $Y_\chi^0 = Y_{\chi, \rm fo}\cdot D \sim\begin{cases}
		\dfrac{\Gamma_{\gamma'}^{1/2}}{\langle \sigma_{\chi\gamma'} v \rangle \Gamma_\phi^\chi M_P^{3/2}} \quad &(\text{hot HS}) \\
		\dfrac{\xi_{\rm rh}^{-4}\Gamma_{\gamma'}^{1/2}}{\langle \sigma_{\chi\gamma'} v \rangle \Gamma_\phi^{\text{vs}} M_P^{3/2}} \quad &(\text{cold HS})
	\end{cases}$\\
    &\\
         \hline
    \end{tabular}
    \caption{DM relic yields in the two distinct regimes of thermal production, including entropy dilution from dark photons decays.  Notice that for regime II, we can substitute $H_{\text{eq}}^{1/2}\simeq H_{\rm nr}^{1/2}(\rho_{\gamma',\rm nr}/\rho_{\rm vs, \rm nr}) \simeq H_{\rm nr}^{1/2} \xi_{\rm nr}^4$ with $H_{\rm nr}\simeq m_{\gamma'}^2/\xi_{\text{nr}}^2 M_P$. }
    \label{tab:yields}
\end{table}

\begin{figure}[hbt]
    \centering
    \includegraphics[width=0.99\linewidth]{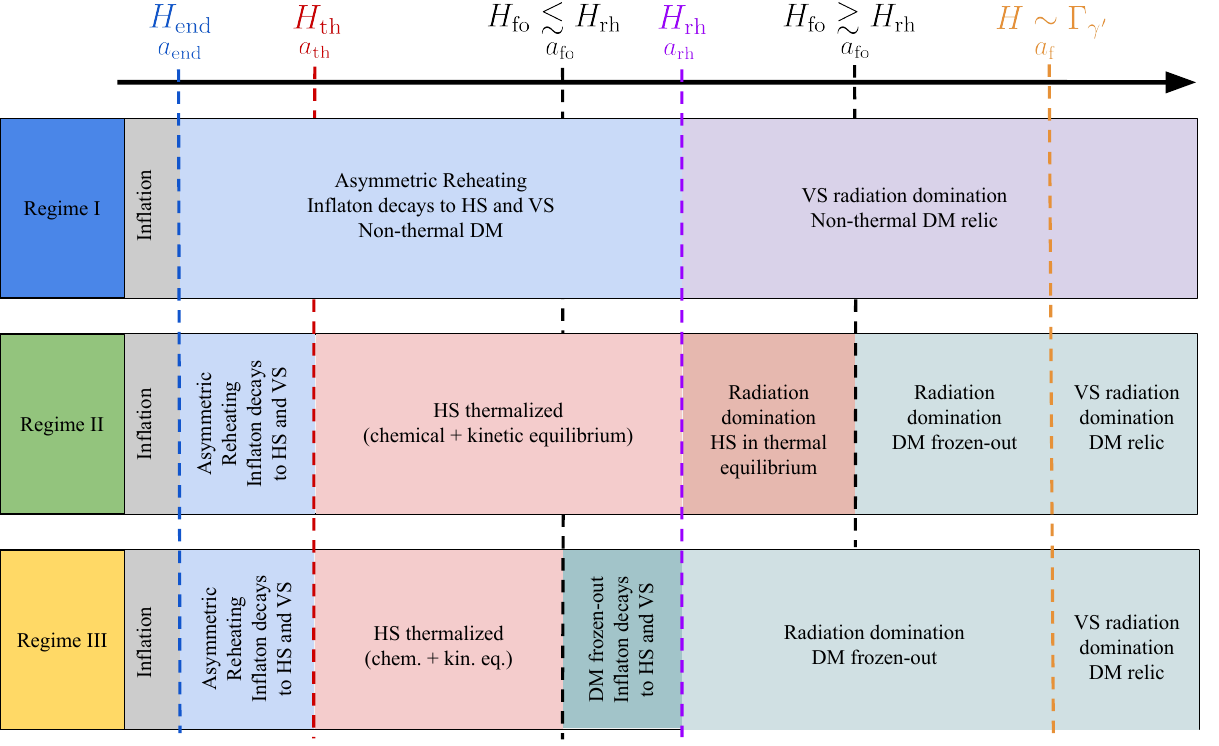}
    \caption{Timeline of the evolution of HS and VS, from the beginning of reheating to the decay of dark photons, for the different regimes of DM production (Regime I, II and III).}
    \label{fig:timeline}
\end{figure}
\subsection{Benchmark dark electron candidates}

We now discuss some benchmark DM candidates, summarizing the different way of setting their relic abundance, depending on the coupling of the HS $(\alpha', y')$, their mass $m_\chi$ as well as the properties of the companion dark photon. In particular, the choice of these parameters determines 1) if and how fast thermalisation is reached within the HS and 2) which regime (I, II, or III) leads to the relic of dark electrons, as determined in section \ref{sec:DMabundance}. We will refer to figures \ref{fig:thermalisation_domain} to delineate the regions in the plane $(\alpha', y')$ at a fixed coupling of inflaton to SM, $y$, for which thermalisation occurs, either before (blue region) or after (yellow) reheating. The regions in which the HS never reaches thermal equilibrium (green) correspond to the same dark electron masses as in figure \ref{fig:thermalisation_domain}. In the case of thermalisation during reheating, we compare $m_\chi$ with $T'_{\rm rh}\sim \max\left[\xi_i^{-1}, 1\right] \sqrt{\Gamma_\phi^\chi M_P}$, to determine whether freeze-out occurs after reheating (regime II), or during reheating (regime III). We then draw lines of iso-relic density of dark electrons for different fixed $m_\chi$, that satisfy the observed DM density \cite{Planck:2018VI}. Such lines continuously link the different regimes of DM production across regions in the plane $(\alpha', y')$.  The results for different dark electron masses, $m_\chi = 1\, \rm GeV,\, 10^{2}\, GeV , 10^{4}\, GeV,$ are shown in figures \ref{fig:iso_relic}, for two different choices of inflaton coupling to SM, $y = 10^{-8},\, 10^{-12}$. In these two panels, we assume negligible entropy dilution from dark photon decays on DM abundance; these effects are taken into account in figures \ref{fig:iso_relic_dilution}. These figures are reminiscent of the ones drawn in \cite{Chu:2011be,Hambye:2019dwd}, in the case of production of dark QED through the kinetic mixing portal.

\begin{figure}[h!]
    \centering
    \includegraphics[width=0.7\linewidth]{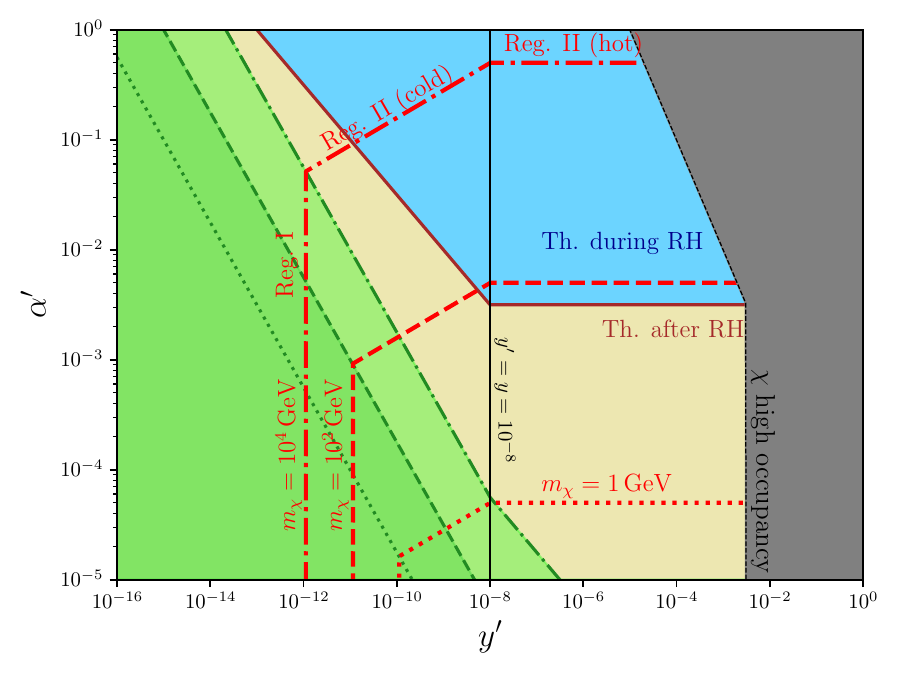}
    \includegraphics[width=0.7\linewidth]{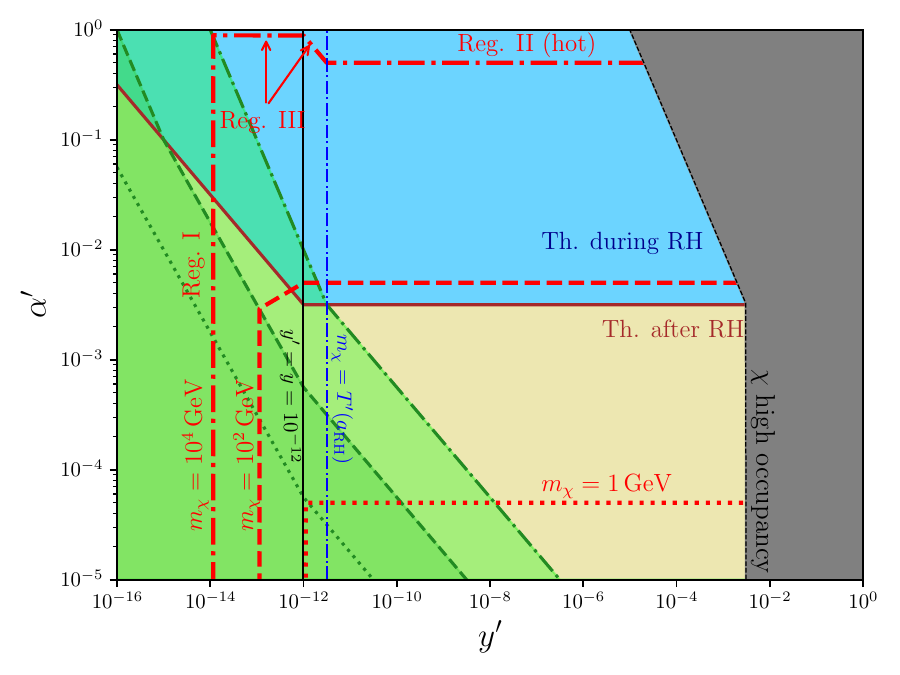}
    \caption{Regimes of DM production associated with the constraint on thermalisation of an {\it abelian} HS, neglecting the effect of entropy dilution from dark photon decays. The constraints are illustrated for a fixed value of $y$ in each panel; for $y' \gtrsim y$, the HS is hot, while for $y' \lesssim y$, the HS is cold. The colored regions correspond to the limits derived in section~\ref{sec:hubble_th} and illustrated in figures \ref{fig:thermalisation_domain}. The red lines correspond to contours of iso-relic density for DM, satisfying the observational constraint $\Omega_\chi h^2= 0.12$, for various DM mass choices. The different regimes of DM production, as discussed in section \ref{sec:DMabundance}, are labeled along the lines for different regions of the parameter space. {\it Upper panel:} $y=10^{-8}$ and $m_\chi = 10^4\, \rm GeV$ (dash-dotted line), $m_\chi = 10^2\, \rm GeV$ (dashed), $m_\chi = 1\, \rm GeV$ (dotted). {\it Lower panel:} $y=10^{-12}$ and $m_\chi = 10^4\, \rm GeV$ (dash-dotted), $m_\chi = 10^2\, \rm GeV$ (dashed), $m_\chi = 1\, \rm GeV$ (dotted). 
    }   
    \label{fig:iso_relic}
\end{figure}

The different features in figures \ref{fig:iso_relic} are the following. 
First, it is evident that for a non-thermal HS (green regions), the dark electron density is fixed solely by inflaton decays, and the DM relic abundance follows Eq.(\ref{eq:Constraint_relic_abundance}), which we called regime I. The relic abundance for each choice of dark electron masses then depends only on $y', y$ and not on $\alpha'$, corresponding to the vertical red lines. The correct relic abundance is obtained from the freeze-in for $y=10^{-8}$, $y'\simeq 10^{-12}, \, 10^{-11}, \, 10^{-10}$ respectively for $m_\chi = 10^{4}\,\rm GeV,\, 10^2\, GeV,\, 1\, GeV$ and for $y=10^{-12}$, $y'= 10^{-14}, \, 10^{-13}, \, 10^{-12} $ respectively for $m_\chi = 10^{4}\,\rm GeV,\, 10^2\, GeV,\, 1\, GeV$. For a given dark electrons mass, they are obviously over-abundant to the right of the corresponding vertical line.

Increasing $\alpha'$, when a given vertical line with freeze-in crosses the boundary of the green regions, they either enter the yellow region (thermalisation of the HS after reheating) or the blue region (respectively, during reheating). If the iso-relic density lines lie in the yellow region, for any mass of dark electron, the relic abundance is set by freeze-out after reheating (regime II). In this situation, one has to distinguish between a cold $(y'<y)$ and a hot $(y'>y)$ HS; for both cases, the DM relic abundances are given by Eq.\eqref{eq:DM_yield_post_RH}. For a cold HS, the abundance scales as $\Omega_\chi h^2\propto \alpha'^{-2}\xi_{\rm fo}$. Assuming no evolution of the temperature ratio, the relic density for fixed $m_\chi$  scales as $\Omega_\chi h^2\propto \alpha'^{-2}\sqrt{y'/y}$. This corresponds to the oblique segments of the red lines in figures \ref{fig:iso_relic} to the left  of the vertical black line, so for $y'<y$. In the case of a hot HS, the abundance of DM is independent of $y', y$, depending solely on $\alpha'$. This case corresponds to the horizontal segments of the red lines. Focusing on the few  benchmarks depicted in the figures, we find that the right relic in regime II of a hot HS is reached for $\alpha' \simeq 0.4, \, 0.004,\, 4\times 10^{-5} $ respectively for $m_\chi = 10^{4}\, \rm GeV,\, 10^{2}\, GeV, \, 1 \, GeV$.

Now, if the constant relic density line lies in the blue region, the regime that leads to the relic abundance depends both on $m_\chi$ and $T'_{\rm rh}$. If $m_\chi<T'_{\rm rh}$, the dark electrons freeze out after reheating (regime II) and their relic is given again by Eq.(\ref{eq:DM_yield_post_RH}). On the contrary, if $m_\chi>T'_{\rm rh}$, freeze-out occurs during reheating while dark electrons are being sourced by inflaton decay (regime III), in which case their final abundance is settled at the end of the reheating era, as given by Eq.(\ref{eq:Y_X_RH_pre_RH_FO}). Such situation occurs for instance for a candidate with $m_\chi = 10^{4}\, \rm GeV$, $y=10^{-12}$, as can be seen in the lower panel of figures \ref{fig:iso_relic} as the red dash-dotted line that lies between the boundary of the green region and the vertical dash-dotted blue line given by $m_\chi = T'_{\rm rh}$. In such situation, once must again distinguish between a cold and a hot HS, for which the final abundance of dark electrons scales as $\Omega_\chi h^2\propto \alpha'^{-2} y^{-1}$ and $\Omega_\chi h^2\propto \alpha'^{-2} y'^{-1}$ respectively. For a cold HS, the relic density is independent of $y'$, resulting in a horizontal segment on the left side of the vertical black line $(y'<y)$, while for hot HS, we see a segment decreasing as $y'$ grows, on the right side of the vertical black line $(y'>y)$.\\

In figures \ref{fig:iso_relic}, it is assumed the dark photon do not produce substantial entropy. For instance, for the case of a hot HS, this occurs if the dark photons thermalize with the VS while they are still relativistic \cite{Coy:2024itg}. We now include the possible impact of entropy dilution from dark photon decays on the final abundance of dark electrons in figure \ref{fig:iso_relic_dilution}. 
\begin{figure}[h!]
    \centering
    \includegraphics[width=0.7\linewidth]{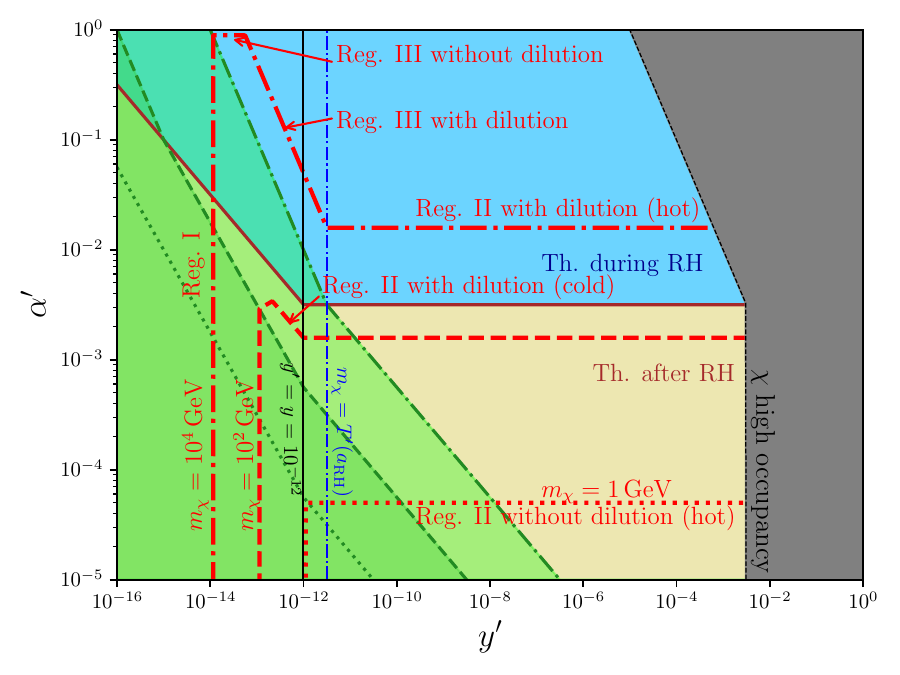}
    \caption{Regimes of DM production associated with the constraint on thermalisation of an {\it abelian} HS, including the effect of dilution via dark photon decays. The constraints are illustrated for a fixed value of $y=10^{-12}$ ; for $y' \gtrsim y$, the HS is hot, while for $y' \lesssim y$, the HS is cold. The colored regions correspond to the limits derived in section~\ref{sec:hubble_th} and illustrated in figures \ref{fig:thermalisation_domain}. The red lines correspond to contours of iso-relic density for DM, satisfying the observational constraint $\Omega_\chi h^2= 0.12$, for various DM mass choices. The different regimes of DM production, as discussed in section \ref{sec:DMabundance}, are labeled along the lines for different regions of the parameter space. The dark photon mass is set such that for each contour $m_{\gamma'}=m_\chi$, the decay width is fixed to $\Gamma_{\gamma'}=10^{-16}\, \rm GeV \sim 10^{9}\, \rm s^{-1}$. The red contours corresponds respectively to $m_\chi = 10^4\, \rm GeV$ (dashdotted), $m_\chi = 10^2\, \rm GeV$ (dashed), $m_\chi = 1\, \rm GeV$ (dotted). \\
    }   
    \label{fig:iso_relic_dilution}
\end{figure}
We consider the same set of parameters, $y=10^{-12}$, $m_\chi = 10^4\, \rm GeV,\, 10^2 \, GeV,\, 1\, GeV$, but fix the dark photon width to $\Gamma_{\gamma'} = 10^{-16}\, \rm GeV = 10^9 \, s^{-1}$ and the dark photon mass to $m_{\gamma'} \lesssim m_\chi$ for each set of iso-relic density lines. The relic from freeze-in for a non-thermal HS (regime I) stays unaffected for this choice of dark photons lifetime, but the impact of dilution is visible both for regimes II and III. For $m_\chi = 1\, \rm GeV$ (dotted), there is no impact of dilution on relic abundance in regime II of the hot HS. However, for $m_\chi = 10^2\, \rm GeV$ (dashed) and $m_\chi = 10^4\, \rm GeV$ (dash-dotted), the observed relic abundance from regime II of a hot HS is obtained at a lower value of HS gauge coupling, respectively for $\alpha' \simeq 0.002$ and $\alpha' \simeq 0.02$, for such dark photons lifetime. This corresponds to the constant dilution factor for the abundance, obtained in Eq.(\ref{eq:D_hot_II}), $D\sim \sqrt{\Gamma_{\gamma'}/H_{\rm nr}}$, which depends only on $\Gamma_{\gamma'}$ and $m_{\gamma'}$.

For a cold HS in regime II, the dilution factor is given by Eq.(\ref{eq:D_cold_II}), $D\sim \sqrt{\Gamma_{\gamma'}/H_{\rm eq}}$.  Taking into account entropy dilution, the relic abundance of DM scales as $\Omega_\chi h^2\propto \alpha'^{-2}y'^{-1}y$, as can be seen for $m_\chi = 10^2\, \rm GeV$ (dashed) in figure \ref{fig:iso_relic_dilution}. The impact of dilution is also strongly visible for regime III, for which the dilution factor for hot and cold HS were obtained in Eq.(\ref{eq:D_hot_preRH}), $D\sim \sqrt{\Gamma_{\gamma'}/H_{\rm rh}}$ and Eq.(\ref{eq:D_cold_preRH}), $D\sim \xi^{-4} \sqrt{\Gamma_{\gamma'}/H_{\rm rh}}$ respectively. By taking into account this dilution factor, we find that both hot and cold HS provides the same final relic abundance for DM, which scales as $\Omega_\chi h^2\propto \alpha'^{-2} y'^{-2}$, as can be seen for $m_\chi = 10^4\, \rm GeV$ (dash-dotted) in figure \ref{fig:iso_relic_dilution}. 

\section{Domain of dark electron candidates}
\label{sec:Constraints_DM}

In the previous section, we have illustrated the different regimes determining the DM relic abundance through some benchmark dark QED candidates. In this final section, we consider them altogether by setting generic constraints on all possible dark electron candidates. These can be reported in the plane $\xi_i= \sqrt{y'/y}$ vs the dark electron mass $m_\chi$. Interpreting $\xi_i$ as the temperature ratio at DM freeze-out, such representation  was called the "domain of thermal DM candidates" in \cite{Coy_2021} (see also \cite{Coy:2024itg} and \cite{Hufnagel:2022aiz}). The logic is to focus on the most extreme DM candidates, meaning the lightest and heaviest admissible candidates for fixed temperature ratio, thus setting the most liberal boundaries on the possible range of DM candidates, so that possible viable DM candidates lie in between. Here, we explore these boundaries for dark QED, taking into account the insights from  production of such HS through inflaton decay. Specifically, we will consider in details 3 new constraints. First, we will impose the requirement that the HS has to be thermalised by the time of DM freeze-out, see section \ref{sec:HS_thermalisation_Domain}. Next, we will take into account the possibility that the temperature ratio at freeze-out may differ from the initial inflation condition $\xi_i = \sqrt{y'/y}$, see section \ref{sec:xi_evolution}. Last but not least, we will study the implications of the inflationary framework on the upper bound on the DM mass as set by unitarity constraints on the DM annihilation cross section, see section \ref{sec:unitarity}. For completeness, we also show how non-thermal candidates fit in the picture of the domain of thermal DM candidates. 

\subsection{Thermalisation}
\label{sec:HS_thermalisation_Domain}

To avoid cluttering of the figures, we first discuss the impact of thermalisation and of the possible evolution of the temperature ratio. We thus concentrate on figure \ref{fig:plot_xi_stability}, which will serve as a canvas for section \ref{sec:TheDomain}. The HS model we consider is simple but still depends on several parameters. So, we have to make specific choices to be able to display constraints in a meaningful way. Our choice requires some explanations as it lies at the core of the logic of the ``domain" \cite{Coy_2021,Coy:2024itg}. A complementary choice of parameters is discussed in Appendix \ref{app:other_constraints}.

The first parameter we fix is the inflaton decay rate. As the ratio $\xi_i$ only depends on the ratio of $y'$ and $y$,  we have introduced a pivot value $y_p$, defined from $\Gamma_\phi \sim (y'^2 + y^2) m_\phi \equiv {y_p^2} m_\phi$, and which is equal to largest of the to Yukawa couplings, $y_p = y'$ for a hot HS and $y_p = y$ for the cold case. We take the value $y_p = 10^{-9}$ for illustrative purpose. Its smallness is chosen so that that reheating is a relatively slow process, leading to cosmological scenarios with late reheating\footnote{We notice that there is an upper bound on $\xi_i>1$  (hot HS), given a value of $y_p = y'$. Indeed, one should require that the temperature of the SM is sufficiently high to ensure BBN, $T \gtrsim \rm T_{\rm BBN}$, at the end of the reheating process, when inflatons have decayed. It implies for hot HS, $\xi_i \lesssim 10^{18}\times y_p = 10^{18}\times y'$. However, as discussed in this section, such high values of the temperature ratio are unstable due to the BBN constraint on dark photon decay.}. From Eqs.\eqref{eq:Tmaxrh} and \eqref{eq:Tprh}, $y_p$ can be expressed in terms of the temperature of the dominant sector at the end of reheating, $T_{\rm rh}$. For $y_p = 10^{-9}$ as in figure \ref{fig:plot_xi_stability}, $T_{\rm rh} \sim y \sqrt{m_\phi M_P} \sim 10^7$ GeV (cold HS) or $T'_{\rm rh} \sim y' \sqrt{m_\phi M_P} \sim 10^7$ GeV (hot HS) for the benchmark value $m_\phi \simeq 3\times10^{13}$ GeV.  In section \ref{sec:unitarity}, we consider different values of $y_p$ and thus of the reheating temperature. 

\begin{figure}[h]
	\centering
	\includegraphics[width=0.7\textwidth]{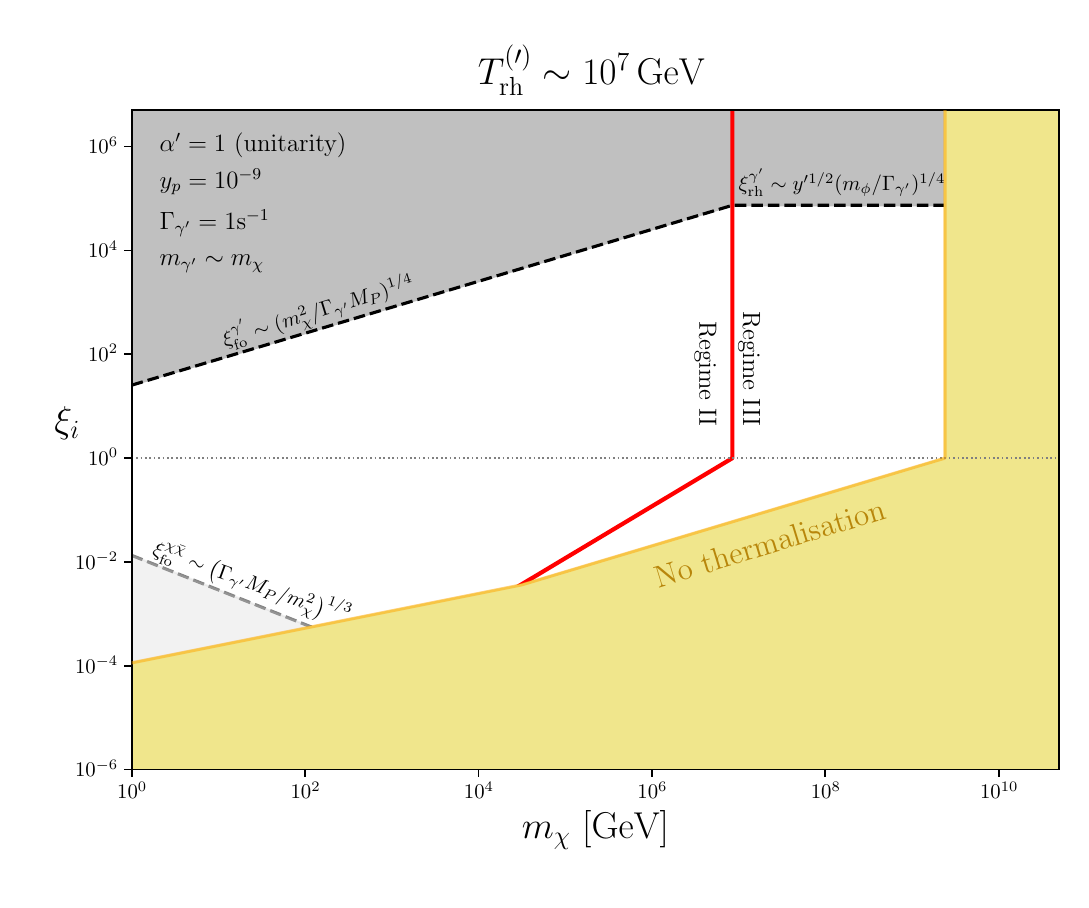}
	\caption{Regions of stable $\xi$ at DM decoupling (white). Points above (below) a dashed line for $\xi_i$ larger (smaller) than one will see $\xi$ evolve vertically until it reaches the white region, at the moment of decoupling. Grey dashed lines indicate the limit imposed by the $\chi \bar{\chi} \leftrightarrow f \bar{f}$ channel, or due to the $\gamma' \leftrightarrow f \bar{f}$ channel. The solid red line shows the separation between pre- and post-reheating freeze-out regimes, and the yellow region contains points for which the HS never thermalises. The dark photon lifetime and mass, as well as the gauge coupling, are maximal (which is the situation considered in section \ref{sec:unitarity}).  }
	\label{fig:plot_xi_stability}
\end{figure}
The next parameter we fix is $\alpha'$. Since $Y_\chi \propto 1/\sigma_{\chi\gamma'}$, an upper bound can be set on the mass of DM particles $\chi$ by considering the largest possible annihilation cross section allowed  by unitarity  \cite{Griest:1989wd}. Consequently, to saturate this bound have we taken $\alpha' = 1$ \cite{Coy_2021,Coy:2024itg} (see section \ref{sec:unitarity}).  For fixed $\alpha'$, the unitarity upper bound on the DM mass can be further increased if there is entropy production between DM freeze-out and today, as this leads to dilution of the DM abundance. Thus, an absolute upper bound on the DM mass is set simultaneously maximizing the value of the entropy dilution factor \cite{Coy:2024itg}. As we have recapped in section~\ref{sec:DMabundance}, the entropy dilution factor depends on the ratio of two time scales, see \cref{eq:D_hot_II,eq:D_cold_II,eq:D_hot_preRH,eq:D_cold_preRH}
and \cite{Coy:2024itg} for further details. For the case of interest here, these are the lifetime of the dark photon $1/\Gamma_{\gamma'}$ on one hand and, on the other hand, the moment when the dark photons become non-relativistic $1/H_{\rm nr}$. As in \cite{Coy:2024itg}, we assume that the dark photon decays at most just before BBN, $\Gamma_{\gamma'} \sim 1\mathrm{s}^{-1}$ and to maximize entropy dilution, we assume that it becomes non-relativistic shortly after dark electron freeze-out, hence the choice $m_{\gamma'} \sim m_\chi$. These parameters are extreme  but, following  the logic of \cite{Coy_2021,Coy:2024itg}, are chosen so as to {\em maximize} the range of possible DM thermal candidates. In more concrete terms,  they  only impact {\em the boundaries} of the figure \ref{fig:plot_xi_stability}. This choice being made, in principle, any realistic thermal DM candidates should lie somewhere  within the (white) region which is left when all generic constraints are applied. 

We can now discuss the constraints from thermalisation of the HS.  The condition for thermalisation has been discussed in section \ref{sec:hubble_th}, both for the case of non-abelian and abelian interactions.  From Eq.\eqref{eq:R_thermalisation_condition_pre_RH_th} we have 
\begin{equation}
	\rho_{\chi, \text{th}} \sim \Gamma_\phi^\chi H_{\text{th}} M_P^2 \sim \alpha'^4 \dfrac{\Gamma_\phi^{\chi 2} M_P^4}{m_\phi^2} \gtrsim m_\chi^4
\end{equation}
or
\begin{equation}
	m_\chi \lesssim \alpha' y' M_P  \quad \rightarrow m_\chi \lesssim \begin{cases}
		\alpha' y_p M_P \quad &(y' \gtrsim y, \text{hot HS}) \\
		\alpha' \xi_i^2 y_p M_P \quad &(y' \lesssim y, \text{cold HS})
	\end{cases}
	\label{eq:thermalisation_limits}
\end{equation}
with $\xi_i = \sqrt{y'/y}$. The yellow shaded  region in figure \ref{fig:plot_xi_stability} corresponds to dark electrons that do not reach thermal equilibrium.   

\bigskip

In this plot, we can also distinguish DM candidates that froze-out after reheating (regime II) or during reheating (regime III), as discussed in section \ref{sec:DMabundance}. If DM particles are still relativistic at the end of reheating, and assuming no  significant sourcing from the SM, their energy density at RH is 
\begin{equation}
	\rho_{\chi, \text{rh}} \sim \dfrac{\Gamma_\phi^\chi}{H_{\text{rh}}} \rho_{\phi, \text{rh}} \sim \Gamma_\phi^\chi \Gamma_\phi M_P^2 
\end{equation}
where $\Gamma_\phi$ is the total inflaton decay rate. The condition $\rho_{\chi, \text{rh}} \gtrsim m_\chi^4$ (regime II) gives
\begin{equation}
	 m_\chi \lesssim \begin{cases}
		y_p (m_\phi M_P)^{1/2} \quad &(y' \gtrsim y  \; ;\;  y_p = y') \\
		\xi_i y_p  (m_\phi M_P)^{1/2} \quad &(y' \lesssim y  \; ;\; y_p = y)
	\end{cases}
    \qquad (\text{regime II})
	\label{eq:pre_post_RH_FO_limit}
\end{equation}

The separation is shown by the red line in the figure \ref{fig:plot_xi_stability}; for $m_\chi$ to the right (left) of this line, freeze-out occurred before (resp. after) reheating. Clearly, this line corresponds to the condition $m_\chi \lesssim T'_{\rm rh}$, which applies for regime II. In regime III, freeze-out occurs before reheating and so at a higher temperature $T'_{\rm fo}\sim m_\chi \gtrsim T'_{\rm rh}$, provided that the condition for thermalisation is satisfied (yellow region). We remind that the figure is built for a fixed reheating temperature {of the dominant sector}, which is the HS for $\xi_i \gtrsim 1$ and the VS for $\xi_i \lesssim 1$.

\subsection{Temperature ratio $T'/T$ at freeze-out}
\label{sec:xi_evolution}

There is some ambiguity in the definition of the temperature ratio $\xi \equiv T'/T$, as it may (and should in the case of a hot HS) evolve along the history of the universe. 
In \cite{Coy_2021} and \cite{Coy:2024itg}, $\xi$ was treated as the temperature ratio at DM freeze-out $\xi_{\rm fo} = \left. T'/T \right\vert_{\rm fo}$, which was treated as free parameter. Here, we use instead the initial ratio $\xi_i \sim \sqrt{y'/y}$ as fixed by the effective coupling of the inflaton to the HS and the VS. Clearly, depending on the model and the scenario, $\xi_i$ may differ from $\xi_{\rm fo}$. To relate the two, is required, not only to  determine the initial conditions for DM freeze-out, but also to evaluate the possible entropy dilution factor. We discuss separately the case of a hot (upper part of figure \ref{fig:plot_xi_stability}) and cold HS (respectively lower part).

\subsubsection*{a. Case of a hot HS}

As shown in section \ref{sec:pre_RH_FO}, in regime III, freeze-out occurs at the end of reheating, so that $\xi_{\rm fo} \sim \xi_{\rm rh}$. Still along regime III, and during reheating, when dark electrons become non-relativistic, their abundance is sourced by the inflaton decay so that
\begin{equation}
	\rho_\chi \sim \sqrt{\dfrac{\Gamma_\phi^\chi m_\chi^2}{\langle \sigma_{\chi\gamma'} v E \rangle_{T'}} \rho_\phi} \quad \text{while} \quad \rho_{\gamma'} \sim \dfrac{\Gamma_\phi^\chi}{H} \rho_\phi \qquad ( \text{hot HS, regime III})
	\label{eq:rho_X_rho_A_HS_dom}
\end{equation}
Meanwhile, the VS is subdominant and its energy density evolves according to
\begin{equation}
	\dot{\rho}_{\text{vs}} + 4 H \rho_{\text{vs}} = \Gamma_\phi^{\text{vs}} \rho_\phi + \langle \sigma_{\chi f} v E \rangle n_\chi^2 + m_{\gamma'} \Gamma_{\gamma'} n_{\gamma'} 
	\label{eq:rho_SM_eq_HS_dom}
\end{equation}
so that
\begin{equation}
	\rho_{\text{vs}} \sim \mbox{\rm max}\left(\dfrac{\Gamma_\phi^{\text{vs}}}{H} \rho_\phi,  \dfrac{\langle \sigma_{\chi f} v E \rangle}{H} n_\chi^2,  \dfrac{m_{\gamma'} \Gamma_{\gamma'}}{H} n_{\gamma'}\right) \quad (\text{hot HS, during reheating})
	\label{eq:rho_SM_sourced}
\end{equation}
depending on which source of heat transfer from the HS to the VS is dominant. 

The first source term corresponds to production through inflaton decay. If it is dominant, then simply
\begin{equation}
	\rho_{\text{vs}} \sim \dfrac{\Gamma_\phi^{\text{vs}}}{\Gamma_\phi^{\text{hs}}} \rho_{\gamma'} \qquad \mbox{\rm and} \qquad \xi_{\rm rh} = \xi_i \sim \sqrt{y'/y}\qquad (\text{VS reheating from $\phi$ decay})
	\label{eq:rho_SM_no_sourcing_pre_RH}
\end{equation}
This situation implies that the temperature ratio at freeze-out can be directly related to the initial inflaton properties. Since the temperature ratio does not change, $T'/T = \xi_i$, we say that the temperature ratio is stable, following the terminology of \cite{Coy:2024itg}.

Production of the VS through inflaton decay is clearly the standard scenario. However, it is conceivable that it is not the dominant channel but that, instead, the VS is produced indirectly, through the HS particles. 
The second source term in Eq. \eqref{eq:rho_SM_sourced} corresponds to annihilation of dark electrons into VS particles (actually, SM fermions). The case of a VS sourcing dominated by this second term during reheating is considered in Appendix \ref{app:other_constraints}. Now, following the logic of \cite{Coy_2021,Coy:2024itg}, the most extreme situation is the one in which the energy is sequestered within the HS as long as possible, corresponding to dark photon decay right before BBN, $\Gamma_{\gamma'} \gtrsim \rm 1s^{-1}$, and thus corresponding to the minimal sourcing of the VS from $\gamma'$ decay. Hence, this corresponds to the largest possible temperature ratio at any given time before dark photon decay, as its decay rate sets the minimum amount of energy that can be injected into the VS. From \eqref{eq:rho_SM_eq_HS_dom}, 
\begin{equation}
	\rho_{\text{vs}} \sim \dfrac{m_{\gamma'} \Gamma_{\gamma'}}{H} n_{\gamma'} \quad \mbox{\rm and}  \quad \xi^{\gamma'}_{\rm rh} \sim \left( \dfrac{H_{\text{rh}}}{\Gamma_{\gamma'}} \right)^{1/4} \sim y_p^{1/2} \left( \dfrac{m_\phi}{\Gamma_{\gamma'}} \right)^{1/4}
	\quad (\text{$\gamma'$ sourcing}; \text{regime III})
	\label{eq:rho_SM_A_sourcing_pre_RH}
\end{equation}
with $H_{\rm rh} \sim \Gamma_\phi$\, and where we replaced $m_{\gamma'}$ with $\rho_{\gamma'}/m_{\gamma'}$, since dark photons are nonrelativistic at the end of reheating in this regime. Any prior value of $\xi_i \gtrsim \xi_{\rm rh}^{\gamma'}$ is erased (in \cite{Coy:2024itg}, it is said that $\xi_i$ is unstable)
by energy injection from the HS to the VS, so this sets an upper bound in the plane $\xi_i-m_\chi$, as shown in figure \ref{fig:plot_xi_stability} (upper right part of the green dashed line).
\bigskip

Moving toward regime II, DM is still relativistic after the end of reheating and redshifts as $a^{-4}$, with $\rho_\chi \sim \rho_{\gamma'} \sim H^2 M_P^2$.
We need to determine $\xi_{\rm fo}$ at DM freeze-out, which now occurs after the end of reheating. 
The standard expectation is again that  $\xi_{\rm fo} = \xi_i$. However, as for regime III, the boundary of the domain is set by considering that the VS is reheated indirectly, through the late decay of the dark photon, so 
\begin{equation}
    \xi_{\rm fo}^{\gamma'} \sim \left( \dfrac{H_{\text{fo}}^{3/2} M_P^{1/2}}{m_{\gamma'} \Gamma_{\gamma'}} \right)^{1/4} \sim \left( \dfrac{m_\chi^3}{m_{\gamma'} \Gamma_{\gamma'} M_P} \right)^{1/4}\quad (\text{$\gamma'$ sourcing}; \text{regime II})
\end{equation}
using $H_{\rm fo} \sim m_\chi^2/M_P$ and $\rho_{\rm vs}$ as in Eq.\eqref{eq:rho_SM_A_sourcing_pre_RH}. {The VS sourcing via $\chi \bar{\chi}$ annihilation in regime III is considered in Appendix \ref{app:other_constraints}.} Again, any $\xi_i \gtrsim \xi_{\rm fo}^{\gamma'}$ is erased through heat injection in the VS, so this sets an upper bound in the upper plane of figure \ref{fig:plot_xi_stability} (upper left part of the black dashed line).
\bigskip

\subsubsection*{b. Case of a cold HS}
For a cold HS, the logic is as in the previous section but reversed, as the VS energy density is dominant, and is driven by inflaton decay during reheating,
\begin{equation}
	\rho_{\text{vs}} \sim \dfrac{\Gamma_\phi^{\rm vs}}{H} \rho_\phi \quad (\text{cold HS})
\end{equation}
while the dark photon energy density is sourced from both inflaton and VS particles. From  \eqref{eq:Full_Boltzmann_system}, 
\begin{equation}
    \dot{\rho}_{\gamma'} + 4 H \rho_{\gamma'} = \Gamma_\phi^{\chi} \rho_\phi + \langle \sigma_{\chi f} v E \rangle_T n_{\rm vs}^2 + \Gamma_{\gamma'} m_{\gamma'} n_{\rm vs}
\end{equation}
and 
\begin{equation}
	\rho_{\gamma'} \sim \mbox{\rm max}\left(\dfrac{\Gamma_\phi^\chi}{H} \rho_\phi, \dfrac{\langle \sigma_{\chi f} v E \rangle_{T}}{H} n_{\text{vs}}^2, \dfrac{\Gamma_{\gamma'} m_{\gamma'} }{H} n_{\text{vs}}\right) \quad (\text{cold HS, during reheating})
	\label{eq:rho_A_sourced_here}
\end{equation}
We consider various possibilities in Appendix \ref{app:other_constraints}, but the most relevant case turns out to be a regime of DM freeze-out after reheating, in which case $\rho_{\rm vs} \sim H^2 M_P^2$. Again, one possibility is that $\xi_{\rm fo} = \xi_i$. Alternatively, the HS may be dominantly heated by production of dark fermions and dark photons from the VS, in which case $\xi_i$ is unstable, $\xi_{\rm fo} \gtrsim \xi_i$. To determine the relevant process, we compare the second and third sources in \eqref{eq:rho_A_sourced_here},
\begin{equation}
	\dfrac{\langle \sigma_{\chi f} v E \rangle_T n_{\text{vs}}}{m_{\gamma'} \Gamma_{\gamma'}} \sim  \dfrac{\varepsilon^2 \alpha\alpha' T^2}{m_{\gamma'}\Gamma_{\gamma'}}
\end{equation}
This ratio is independent of the mixing parameter, $\varepsilon$, and turns out to be  much larger than one at DM freeze-out for the choice of parameters used in figure \ref{fig:plot_xi_stability}. Hence, energy transfer to the HS is driven by annihilation of DM particles and the temperature ratio evolves towards
\begin{equation}
	\xi_{\text{fo}}^{\chi \bar{\chi}} \sim \left. \left( \dfrac{\langle \sigma_{\chi f} v E \rangle_{T} \rho_{\text{vs}}^{1/2}}{H} \right)\right|_{\rm fo}^{1/4} \sim \left( \langle \sigma_{\chi f} v E \rangle_{T} M_P \right)^{1/4}
\end{equation}
Here, the temperature $T_{\text{fo}}$ at which the cross section is averaged contains an implicit dependence on $\xi$, via $T_{\text{fo}} \sim m_\chi/\xi_{\text{fo}}$. Isolating the temperature ratio gives
\begin{equation}
	\xi_{\text{fo}}^{\chi \bar{\chi}} \sim \alpha'^{1/3} \varepsilon^{2/3} \left( \dfrac{M_P}{m_\chi} \right)^{1/3} \quad (\text{cold HS})
\end{equation}
If $\xi_i \lesssim \xi_{\text{fo}}$ this sets a lower bound on the temperature of the HS and then on $\xi_i$ in figure \ref{fig:plot_xi_stability}.
Re-expressing the factor $\varepsilon$ in terms of $\Gamma_{\gamma'}$, $\xi_{\rm fo}$ is depicted by the dashed grey line in the lower part of figure \ref{fig:plot_xi_stability}. 

\subsection{Domain of thermal dark electrons}
\label{sec:TheDomain}

The general frame being set, we finally focus on the unitarity limit for various choices of the reheating temperature, as shown in figures \ref{fig:Unitarity_walls}. For completeness, we also consider the so-called relativistic floor and non-thermal DM candidates.

\subsection*{a. Relativistic floor}
This bound has been introduced in \cite{Hambye:2020lvy} (see also \cite{Coy_2021}). It concerns DM particle candidates that decoupled while being still relativistic, like the Standard Model neutrinos. Under this assumption the DM yield does not depend on the DM mass and temperature ratio at the time of freeze-out and takes a very simple form,
\begin{equation}
    Y_\chi^{\rm dec} \simeq \dfrac{g_\chi T_{\rm dec}^{\prime 3}}{4 g_\ast(T_{\rm dec}) T_{\rm dec}^{\prime 3}} \sim \xi_{\rm dec}^3
\end{equation}
Using the measured DM abundance $m_\chi Y_\chi^0 \simeq 4 \times 10^{-10} \rm GeV$, the locus of such DM candidates is a line in the plane $\xi_i$ vs $m_\chi$, with
\begin{equation}
    m_\chi \gtrsim \xi_{\rm dec}^{-3}\, 10^{-10} \rm GeV
\end{equation}
This is shown as a green solid line in the different panels of figure \ref{fig:Unitarity_walls}. As expected, such DM candidates are only viable for a cold HS. Below the relativistic floor, the particles are subdominant; above, another mechanism, like non-relativistic freeze-out, must set the DM abundance \cite{Hambye:2020lvy,Coy_2021}. 

Focusing on a HS with only dark electrons and dark photons, the relativistic floor could only be reached if dark photons were heavier than dark electrons \cite{Coy_2021}. Here, we have assumed, on the contrary, that the dark photons are lighter than the dark electrons, so the relativistic floor is beyond reach.  Also, we tentatively notice that, unless the dark photon is very long-lived $\Gamma_{\gamma'} \sim 1\, s^{-1}$, the temperature ratio at freeze-out tends to be too large for the relativistic freeze-out to be a viable possibility, see the grey dashed lines in figures  \ref{fig:Unitarity_walls}. We leave a detailed analysis of this regime for possible future work.

\subsection*{b. Unitarity wall}
\label{sec:unitarity}

Moving towards larger couplings, the unitarity wall correspond to DM candidates with the largest annihilation cross section compatible with unitarity \cite{Griest:1989wd}
\begin{equation}
	\sigma_J \leq \dfrac{\pi (2 J + 1)}{p_i^2} \qquad \text{(unitarity bound)}
\end{equation}
where $p_i$ is the modulus of the momentum of one of the ingoing particles and $J$ the orbital angular momentum of the diffusion channel. This bound translates to an upper bound on the DM mass. The twist is to apply the unitarity bound to DM particles within a HS, with temperature $T'$. The most stringent bound for the mass is obtained by taking $J = 0$ \cite{Baldes:2017gzw, Flores:2024sfy}, and we will do so in the rest of this section. In the center-of-mass frame and non-relativistic DM, 
\begin{equation}
	\sigma_0 v\leq \dfrac{4 \pi}{m_\chi^2 v} 
\end{equation}
and the thermally averaged cross section is bounded by \cite{Hambye:2020lvy, Coy_2021}
\begin{equation}
	\langle \sigma v \rangle \leq \dfrac{4 \pi}{m_\chi^2} \dfrac{K_2(2m_\chi/T')}{K_2^2(m_\chi/T')} 
\end{equation}
In the case of large coupling $\alpha'$, $\chi$ particles decouple while being non-relativistic, so we can use the expansion of the modified Bessel function $K_2$ for large arguments, and compute the upper bound on the averaged cross section near freeze-out, with $T'\sim m_\chi$,
\begin{equation}
	\langle \sigma v \rangle \lesssim \dfrac{4 \pi}{m_\chi^2} 
\end{equation}

As for the standard unitarity bound (i.e. for $T' = T$), the bound on the cross section translates into an upper bound on the DM mass. For $T' \neq T$, the bound depends on $T'/T$, but otherwise delineates the domain of thermal DM candidates \cite{Coy_2021}. To set the unitarity wall, we  use the expressions for the relic abundance of DM in each thermalised scenarios (II \& III) discussed in section \ref{sec:DMabundance},  which are to be compared to the measured DM abundance. 
\begin{figure}[h!]
	\centering
	\includegraphics[width=0.49\textwidth]{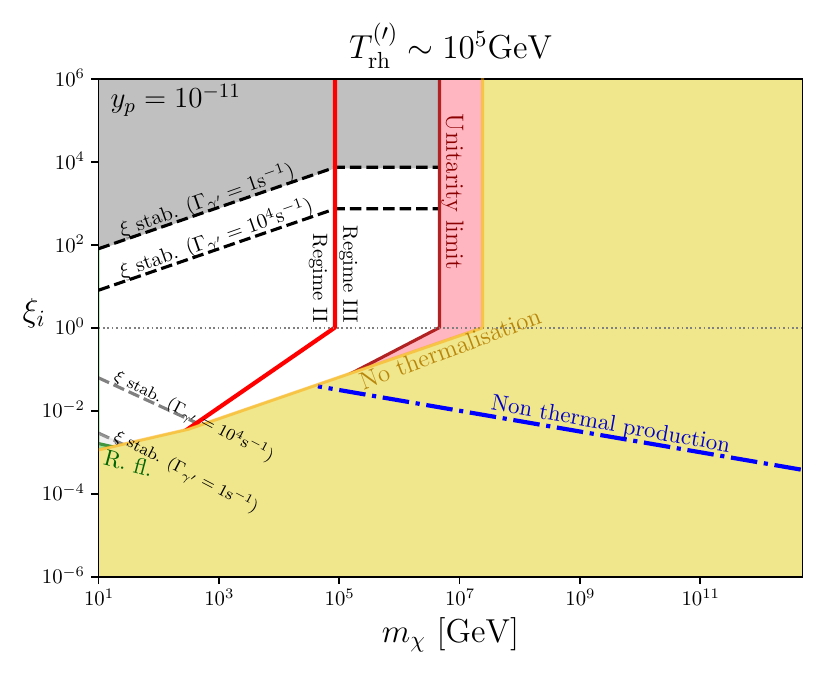}
	\includegraphics[width=0.49\textwidth]{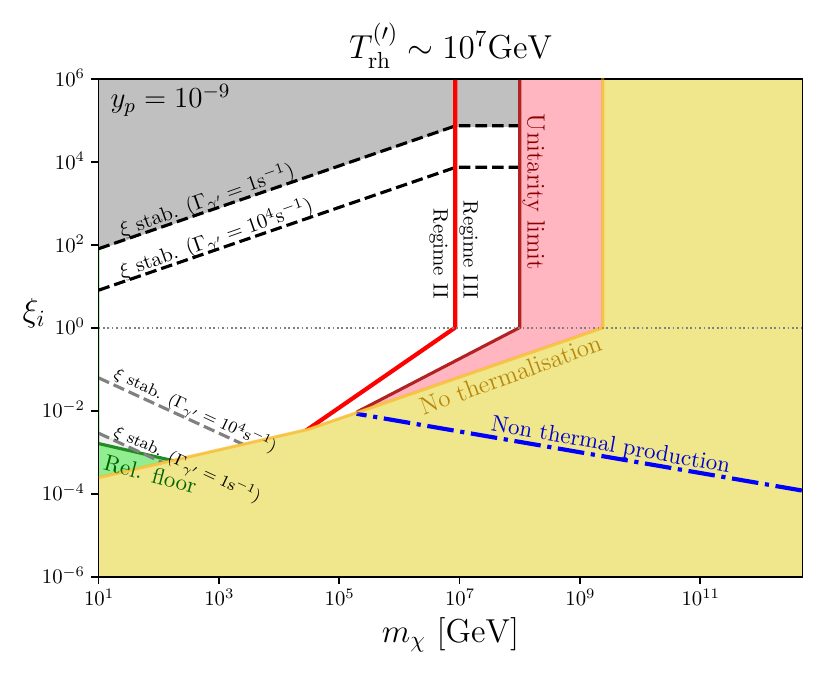}
    \includegraphics[width=0.49\textwidth]{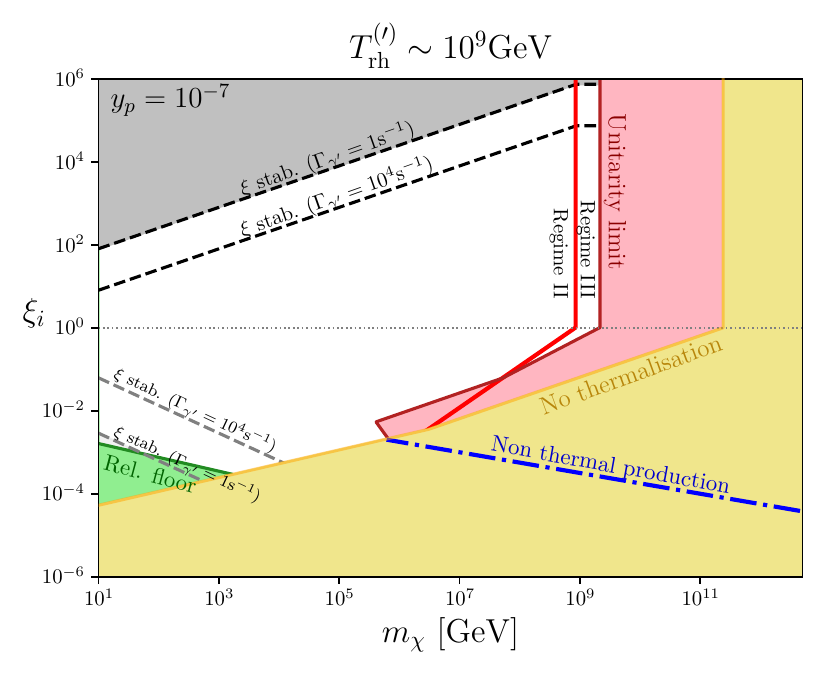}
    \includegraphics[width=0.49\textwidth]{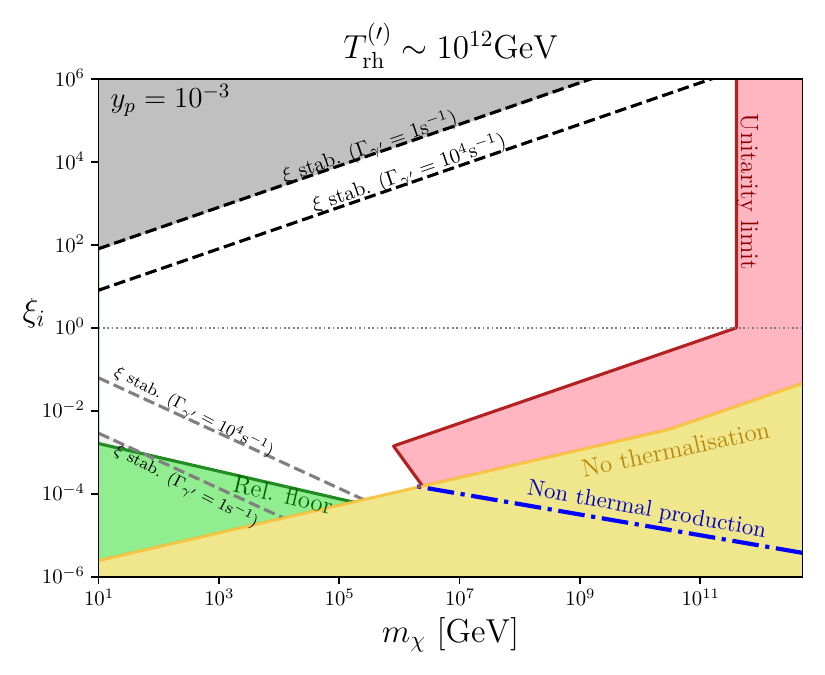}
	\caption{Constraints on the domain of thermal HS DM. Unitarity limits on the DM mass $m_\chi$ are given in red, as a function of the initial temperature ratio $\xi_i \sim \sqrt{y'/y}$, for different values of the pivot Yukawa coupling $y_p$. The red region shows the points for which the thermal HS with such a DM mass is excluded by too large DM abundance, despite the upper bound on annihilation cross-section set by unitarity. Points in the yellow region correspond to situations where the HS does not thermalise before decoupling as derived in section \ref{sec:thermalisation}. The blue dash-dotted line indicates the points that yield the right relic abundance for a non-thermal HS, obtained in section \ref{sec:non_thermal}. Above (below) grey dashed lines at fixed $\Gamma_{\gamma'}$, $\xi$ evolves  vertically until it hits the lines at the moment of DM decoupling for hot (respectively cold) HS.}
	\label{fig:Unitarity_walls}
\end{figure}
As explained in \cite{Coy:2024itg}, the largest unitarity bound on the DM mass occurs when the entropy dilution factor is the largest, which in turn leads to consider dark photons that become non-relativistic soon after DM freeze-out and decay as late as possible.

In the case of hot HS, combining  the DM abundance at freeze-out, Eqs.\eqref{eq:DM_yield_post_RH} and \eqref{eq:Y_X_RH_pre_RH_FO}, with the corresponding entropy dilution factor, Eqs. \eqref{eq:D_hot_II} and \eqref{eq:D_hot_preRH}, gives  for regime II (freeze-out after reheating) and III (resp. during reheating) the following expressions for the DM relic abundance today (see table \ref{tab:yields})
\begin{equation}
	Y_\chi^0 \sim \begin{cases}
		
		\dfrac{\Gamma_{\gamma'}^{1/2}}{\langle \sigma_{\chi\gamma'} v \rangle_{T'} H_{\text{nr}}^{1/2} M_P m_\chi} \sim \dfrac{\Gamma_{\gamma'}^{1/2}}{\alpha'^2 M_P^{1/2}} \quad &(\text{regime II})\\
        \dfrac{\Gamma_{\gamma'}^{1/2}}{\langle \sigma_{\chi\gamma'} v \rangle_{T'} \Gamma_\phi^\chi M_P^{3/2}} \sim \dfrac{m_\chi^2 \Gamma_{\gamma'}^{1/2}}{\alpha'^2 y'^2 m_\phi M_P^{3/2}} \quad &(\text{regime III})
	\end{cases}
	\quad (\mbox{\rm hot HS})
\end{equation}
These expressions show clearly that the largest upper limit  on $m_\chi$ is reached when $\Gamma_{\gamma'}$ is minimal. Taking $\Gamma_{\gamma'} = (0.2 \text{s})^{-1}$ and $\alpha' = 1$ gives the following  upper bound on the dark electron mass, 
\begin{equation}
	m_\chi \lesssim \begin{cases}
		4\times10^{11}\, \text{GeV} \quad &(\text{regime II})\\
        y'^{2/3} \times 10^{14}\, \text{GeV} \quad &(\text{regime III}) \\
	\end{cases}
	\qquad (\mbox{\rm hot HS})
\end{equation}
The largest of these two bounds corresponds to the red vertical lines in figures  \ref{fig:Unitarity_walls} (unitarity limit); regime III (DM freeze-out after reheating) is the only possible scenario if $y' \gtrsim 10^{-5}$.\\

For a cold HS, the relic abundance can be get using Eqs.\eqref{eq:DM_yield_post_RH} and \eqref{eq:Y_X_RH_pre_RH_FO} and the dilution factors \eqref{eq:D_cold_II} and \eqref{eq:D_cold_preRH}, to give (see table \ref{tab:yields})
\begin{equation}
	Y_\chi^0 \sim \begin{cases}
		\dfrac{\xi_{\text{nr}}^{-3} \Gamma_{\gamma'}^{1/2}}{\langle \sigma_{\chi\gamma'} v \rangle_{T'} H_{\text{nr}}^{1/2}M_P m_\chi} \sim \dfrac{\xi_{\text{nr}}^{-2} \Gamma_{\gamma'}^{1/2}}{\alpha'^2M_P^{1/2}} \quad &(\text{regime II})\\
		\dfrac{\xi_{\text{rh}}^{-4} \Gamma_{\gamma'}^{1/2}}{\langle \sigma_{\chi\gamma'} v \rangle_{T'} \Gamma_\phi^{\text{vs}}M_P^{3/2}} \sim \dfrac{m_\chi^2 \Gamma_{\gamma'}^{1/2}}{\alpha'^2 y'^2 m_\phi M_P^{3/2}} \quad &(\text{regime III}) 
	\end{cases}
	 \quad (\mbox{\rm cold HS, entropy dilution})
\end{equation}
Here, we have used the fact that maximal entropy dilution requires that the coupling between the HS and the VS is small (tiny kinetic mixing $\varepsilon$) so that the temperature ratio $\xi_{\text{rh}} =\xi_i$ and can be replaced by $\sqrt{y'/y}$. We replaced $H_{\text{eq}}^{1/2}\simeq H_{\rm nr}^{1/2}(\rho_{\gamma',\rm nr}/\rho_{\rm vs, \rm nr})$ and $H_{\rm nr}\simeq m_\chi^2/\xi_{\text{nr}}^2 M_P$. We also assumed that, for maximal values of the dark photon mass $m_{\gamma'}$, $\xi_{\text{fo}} \approx \xi_{\text{nr}}$, which is reasonable since the two moments are then very close to each other. This leads to the following upper bound on the dark electron mass,
\begin{equation}
	m_\chi \lesssim \begin{cases} \xi_{\text{nr}}^2 \times 4\times 10^{11} \text{GeV} \quad &(\text{regime II})\\
    y'^{2/3} \times 10^{14} \text{GeV} \quad &(\text{regime III}) 
	\end{cases}
	\qquad (\text{cold HS, entropy dilution})
\end{equation}
These bounds correspond to the diagonal branch of the red region in figures \ref{fig:Unitarity_walls}, starting when $\xi_i < 1$. Notice that the limit is not vertical in the $y' < y$ part of figures \ref{fig:Unitarity_walls}, because the initial temperature ratio $\xi_i \sim \sqrt{y'/y}$ is fixed by $\sqrt{y'/y_p}$, with $y_p$ the chosen pivot Yukawa coupling. Hence, $y'^{2/3} \sim y_p^{2/3} \times \xi^{4/3}$ along the lines. 

Last but not least, when entropy production due to dark photon decay is negligible, the DM yield becomes (see table \ref{tab:yields})
\begin{equation}
	Y_\chi^0 \sim \begin{cases}
		\dfrac{\xi_{\text{fo}}}{\langle \sigma_{\chi\gamma'} v \rangle_{T'} M_P m_\chi} \sim \dfrac{\xi_{\text{fo}} m_\chi}{\alpha'^2 M_P} \quad &(\text{regime II})\\
		\dfrac{1}{\langle \sigma_{\chi\gamma'} v \rangle_{T'} (\Gamma_\phi^{\text{vs}})^{1/2} M_P^{3/2}} \sim \dfrac{m_\chi^2}{\alpha'^2 y m_\phi^{1/2} M_P^{3/2}} \quad &(\text{regime III}) 
	\end{cases}\qquad (\mbox{\rm cold HS, no dilution})
\end{equation}
The corresponding upper bound on the DM mass is then,
\begin{equation}
m_\chi \lesssim \begin{cases}
		\xi_{\text{fo}}^{-1/2} \times 3\times10^4 \text{GeV} \quad &(\text{regime II})\\
        
		y^{1/3} \times 2\times 10^8 \text{GeV} \quad &(\text{regime III})
	\end{cases}\qquad (\text{cold HS, no dilution})
    \label{eq:unitarity_wall_constraints}
\end{equation}
This explains the last turn of the red limit in figures \ref{fig:Unitarity_walls}, around $\xi_i \lesssim 10^{-3}$. We emphasise that FO after reheating (regime II) yields the same bounds as those derived in \cite{Coy_2021, Coy:2024itg}, whereas the bounds for regime II freeze-out are new results. To determine which bound is relevant, we can simply use the criterion \eqref{eq:pre_post_RH_FO_limit} to evaluate whether freeze-out takes place before or after the end of reheating. Both relations come naturally together for parameters at the edge of this limit, which is around $y' \sim 10^{-5}$.\\

\subsection*{c. Non-thermal candidates}

Throughout this section, we assumed that the HS has reached thermal equilibrium. Yet, for completeness, and because this aspect was left out of the analysis of \cite{Coy_2021,Coy:2024itg}, we also depict the locus of DM candidates produced through inflaton decay but that do not thermalise, see the dot-dashed blue lines in figures \ref{fig:Unitarity_walls}. The reason this can be done is because we are expressing the vertical axis in terms of the ratio of inflaton effective (Yukawa) couplings to the HS and VS. In general ---but not always, see the grey dashed lines --- this ratio corresponds to the temperature ratio at the time of DM freeze-out, the parametrization used in the previous analysis \cite{Coy_2021,Hufnagel:2022aiz,Coy:2024itg}. The broader, but at the same time, more specific, perspective of the present work allows to also depict the candidates that are produced out-of-thermal equilibrium. We deem of interest to understand how they fit with  the frame of thermal candidates. 

It comes with no surprise that the non-thermal candidate lie in the yellow region, corresponding to HS that could not reach thermal equilibrium. We also notice that the  locus of these non-thermal candidates ends up in the domain of thermal candidates (white region) by crossing simultaneously the unitarity wall. The reason is the following. At the edge of the no-thermalisation regions in figures \ref{fig:Unitarity_walls},
the relations given in Eq.(\ref{eq:R_thermalisation_condition_pre_RH_th}) for regime III  scenarios, and in Eq.(\ref{eq:R_thermalisation_condition_post_RH_cold}) for regime II ones, have to be satisfied. Assuming in addition the unitarity limit value for the gauge coupling provides a one-to-one relation among the parameters $(\xi_i, m_\chi)$,\footnote{We can determine the DM mass for which the non-thermal abundance lines cross the no-thermalisation regions, in figures \ref{fig:Unitarity_walls}, 
by using conditions above together with Eq.(\ref{eq:Constraint_relic_abundance}), to obtain
\begin{eqnarray}
    &&m_\chi \sim y_p^{1/7}\times7\times10^6\, \rm GeV~~~~~~~~~~\text{(regime II)}\\
    && m_\chi \sim y_p^{1/3}\times 2\times 10^8\, \rm GeV~~~~~~~~~\text{(regime III)}
    \label{eq:mass_crossing}
\end{eqnarray}
}
\begin{eqnarray}
    &&y' \approx y^{1/3}\left(\frac{m_\chi^2m_\phi}{M_P^3}\right)^{1/3}\quad \mbox{\rm or}\quad \xi_i^2 \sim y^{-2/3}\left(\frac{m_\chi^2m_\phi}{M_P^3}\right)^{1/3}~~~~~\text{(regime II)}\\
    &&  y' \approx \frac{m_\chi}{M_P} \quad \mbox{\rm or}\quad \xi_i ^2 \sim y^{-1}\frac{m_\chi}{M_P}~~~~~~~~~~~~~~~~~~~~~~~~~~~~~~~\text{(regime III)}
    \label{eq:border_no-thermalisation-regions}
\end{eqnarray}

In figures \ref{fig:Unitarity_walls}, we can see that the unitarity wall and the no-thermalisation region intersect in the regime of low $\xi_i$. At this crossing, the correct relic abundance is obtained for a thermal HS via DM freeze-out, and the effect of dilution due to entropy injection from dark photon decays can be neglected. Being simultaneously on the unitarity wall and on the boundary of the no-thermalisation region requires satisfying both Eq.~(\ref{eq:border_no-thermalisation-regions}) and Eq.~(\ref{eq:unitarity_wall_constraints}). The corresponding mass at this crossing can be shown to coincide with that derived in Eq.~(\ref{eq:mass_crossing}), indicating that it occurs at the same point as the intersection of the non-thermal abundance lines with the no-thermalisation region, as visible in figures~\ref{fig:Unitarity_walls}.

\subsubsection*{d. Impact of dark photon decay on  $\xi_{\rm fo}$}

As discussed in section \ref{sec:xi_evolution}, the temperature ratio at DM freeze-out may evolve through energy transfer from one sector to the other along the process of reheating, and may thus differ from the initial condition $\xi=\xi_i$. In figures \ref{fig:Unitarity_walls}, the largest possible domain in the plane $m_\chi$ vs $\xi_i$ is bounded by the dashed lines, which gives the temperature ratio at DM freeze-out as driven by the dark photon decay channel. The largest possible dark photon lifetime (here, $\Gamma_{\gamma'} \sim 1 \, s^{-1}$) gives the largest possible domain of dark electron candidates. Concretely, for fixed dark electron mass, $\xi_i$ within the grey region (for the case of a hot HS) is driven to the dashed grey temperature ratio by the time of DM freeze-out. We say that such a point is unstable. Conversely, within the white region, $T'/T \sim \xi_i$ at freeze-out, for which we say that the temperature ratio is stable. For faster dark photon decay rate, the energy transfer is more efficient and thus, for the case of a hot HS (cold HS),  the lower (resp. higher) is the temperature ratio at DM freeze-out. The similar feature occurs for a cold HS, now the temperature ratio being driven by energy transfer from the VS to the HS. One may in particular notice  that the relativistic floor tends to be below the $\xi_i$ stability line, except for large reheating temperatures.

\section{Conclusions}
\label{sec:conclusions}

In this work, we have investigated a hidden sector (HS) scenario in which the scalar inflaton field couples to both the visible sector (essentially, the SM) and the HS, with different branching ratios, leading to asymmetric reheating. The HS is taken to be described by dark QED.  The asymmetry in the energy densities of the two sectors sets the stage for a rich and non-trivial thermal (and non-thermal) history that we have investigated in details.

A central aspect of our analysis is the study of the HS thermalisation, which we assume can be handled using a perturbative {approach}. Doing so, we go beyond the commonly adopted instantaneous thermalisation approximation during reheating of a HS. While the problem of thermalisation of a non-abelian sector  has been addressed several times in the literature, in particular in relation with the reheating of the SM, the case of an abelian sector has received much less attention, two notable exceptions being \cite{Garny:2018grs,Arvanitaki:2021qlj}, but from distinct perspectives. Building on the existing literature on the complex problem of thermalisation, we show that a weakly coupled dark QED HS reaches thermodynamical equilibrium  through soft splitting processes which are always affected {by the} Landau-Pomeranchuk-Migdal suppression effect. This can significantly delay the thermalisation of the HS, compared to the case of HS with non-abelian interactions. This delay has important consequences for the further evolution of the HS and visible sectors during and after reheating. We have derived the thermalisation timescales and obtained a simple but, we believe, robust criterion which determines when thermalisation of HS can occur along the process of decay of the inflaton. Doing so, we have  identified several possible outcomes, depending on the parameter space for the HS, as defined by  $\alpha'$ and the mass of the dark electron, $m_\chi$. For the sake of comparison, we provide these both for the cases of abelian and non-abelian interactions, see figure \ref{fig:thermalisation_domain}.

Building on these results, we have analysed the production of DM and identified several distinct regimes. Depending on the strength of the couplings and the thermalisation history of the HS, the dark electrons can be produced through non-thermal processes (i.e freeze-in dominated by inflaton decays), standard freeze-out after reheating of both the HS and VS, or freeze-out during reheating while both sectors are still being sourced by inflaton decays. The latter scenario, in particular, provides a novel realisation for the generation of the DM relic abundance, in which the interplay between annihilations and continuous particle injection from inflaton in the early phase of reheating settles the final DM abundance. For all production mechanisms, we provide both numerical calculations and analytical expressions of the DM abundance at freeze-out. 

We have also examined the role of the dark photon, whose production and subsequent decay can significantly affect the cosmological evolution of both HS and VS. In particular, entropy injection from late decays of a non-relativistic and dominant dark photon can dilute significantly the DM abundance and alter drastically the final DM relic abudance. Taking these effects into account, we have highlighted several benchmark  DM candidates. In particular, in figures \ref{fig:iso_relic} and \ref{fig:iso_relic_dilution}, we provide contours of constant DM relic density in the plane $\alpha'$ vs $m_\chi$, which show how the different ways of DM production are related with each other. This  representation is akin to the ``{\it mesa}" phase diagram  of \cite{Chu:2011be,Hambye:2019dwd}, here applied to the inflaton portal. 

Our results also highlight the importance of consistently taking into account the reheating period after inflation for both sectors, together with the portal interactions between  the HS and the VS, in a fully dynamical framework. In particular, we have shown that for a thermalised HS during reheating, the temperature ratio cannot, in general, be treated as a fixed initial condition, but instead evolves due to the interplay between inflaton decay and the coupling between the hidden and visible sectors arising here from kinetic mixing. The evolution of the temperature ratio during and after reheating depends crucially on the timescale of thermalisation relative to the timescale for the reheating process, as well as on which sector is dominant. This has direct implications for the phenomenology of the model and the generation of the cosmological relics. Indeed, the evolution of the temperature ratio, including entropy injection from dark photon decays, significantly affects the evolution of the HS and the final DM relic abundance. These results are summarised in the plane $\xi_i$ vs $m_\chi$, in which we set boundaries that encompass potentially all possible dark electrons candidates, given generic constraints, in particular the unitarity limit on the DM annihilation cross section, a presentation that has been called the ``domain" \cite{Coy_2021,Hufnagel:2022aiz,Coy:2024itg}, see figures \ref{fig:Unitarity_walls}. Our key new results, taking into account a detailed treatment of the thermalisation of the HS, include the interplay with the case of non-thermal production of dark electrons, the impact of the evolution of the temperature ratio and the role of the reheating temperatures (both $T'$ and $T$) on the domain, in particular on the maximal possible mass for dark electron DM candidates.

 This work concludes a series of studies on HS scenarios featuring a hot HS (in the sense that the HS dominates the early expansion of the universe). Compared to previous analyses (by some of us), the present work adopts a specific but at the same time more holistic perspective by incorporating a concrete production mechanism for the initial state of the HS through asymmetric decay of the inflaton, thereby grounding the scenario in a more complete cosmological picture. While further investigations remain possible, two directions stand out as particularly relevant to us. The first concerns the possibility of freeze-out occurring while the dark electrons are still relativistic (akin to the SM neutrinos), the so-called relativistic floor regime, which would require revisiting the underlying assumptions we made in this work and thus requires a new analysis \cite{Hambye:2020lvy,Coy_2021,Coy:2024itg,Henrich:2025sli}. The second concerns the thermalisation of an abelian dark sector, which, to the best of our knowledge, remains incompletely addressed in the cosmological literature. In particular, it would be valuable to assess whether the simple thermalisation criteria we have derived are truly robust. A minor aspect, in our opinion, is that we have neglected quantum statistics effects, focusing on generic features. These may however have a strong impact on specific candidates. We leave these issues and other questions for future work.

\section*{Acknowledgments}

We thank Nicolás Bernal, Marcos A.G. Garcia, Mathias Garny, and Mathieu Gross for useful discussions. 
SC is supported by a Humboldt Research Fellowship of the Alexander von Humboldt Foundation. This work was made possible by Institut Pascal at Université Paris-Saclay with the support of the program “Investissements d’avenir” ANR-11-IDEX-0003-01, the P2I axis of the Graduate School of Physics of Université Paris-Saclay, as well as IJCLab, CEA, IAS, OSUPS, and APPEC. 

\appendix

\section{Cross sections and rates}
\label{app:average_cross_sections_and_decays}

\subsubsection*{Inflaton  decay into 2 dark photons}
Inflaton decay  into  a pair of dark photons 
arises at one-loop through the exchange of dark electrons, see \eqref{eq:loop_decay}. From \cite{Shifman:1979eb, Marciano:2011gm}, 
\begin{equation}
    \Gamma_{\phi}^{\gamma'} = \dfrac{\alpha'^2 y'^2}{256 \pi^3} \dfrac{m_\phi^3}{m_\chi^2} \left| F(\beta_\chi) \right|^2
\end{equation}
with $\beta_\chi = 4 m_\chi^2/m_\phi^2$ and 
\begin{equation}
    F(\beta_\chi) = -2 \beta_\chi \left[ 1 + \dfrac{1}{4} (\beta_\chi - 1) \left( \ln \dfrac{1+\sqrt{1-\beta_\chi}}{1-\sqrt{1-\beta_\chi}} - i \pi \right)^2 \right] 
\end{equation}
with
\begin{eqnarray} 
    &|F(\beta_\chi)|^2& \stackrel{\beta \to +\infty}{\longrightarrow} \frac{16}{9}\ , \\
    \nonumber
    &|F(\beta_\chi)|^2& \stackrel{\beta \to 0^+}{\sim} \frac{1}{4} \beta_\chi^2 \ln^4\left(\frac{4}{\beta_\chi}\right) \left[ 1 + \frac{2(\pi^2 - 4)}{\ln^2(4/\beta_\chi)} + \frac{(\pi^2 + 4)^2}{\ln^4(4/\beta_\chi)} \right]
\end{eqnarray}

\subsubsection*{Dark photon decay}
The rate $\Gamma_{\gamma'} \equiv \Gamma_{\gamma' \rightarrow f \bar{f}}$ into SM fermions is given in general by\cite{Berger:2016vxi, Coy:2024itg}:
\begin{equation}
    \Gamma_{\gamma' \rightarrow f \bar{f}} = \dfrac{N_c c_W^2 Q_f^2 \alpha}{3} m_{\gamma'} \varepsilon^2 \left( 1 + \dfrac{2 m_f^2}{m_{\gamma'}^2} \right) \sqrt{1 - \dfrac{4 m_f^2}{m_{\gamma'}^2}}
\end{equation}
where $N_c$ is the number of colors for species $f$. Summing over every possible outgoing species, one needs to be careful when it comes to the hadron part, for which we use the data extracted from $e^+ e^- \rightarrow \text{hadrons}$ rate \cite{Workman:2022ynf, Berger:2016vxi, Coy:2024itg},
\begin{equation}
    \Gamma_{\gamma' \rightarrow \rm hadrons} = \Gamma_{\gamma' \rightarrow \mu^+ \mu^-} \times \mathcal{R}(m_{\gamma'})
\end{equation}
with the ratio $\mathcal{R}$ defined as
\begin{equation}
    \mathcal{R} = \dfrac{\sigma(e^+ e^- \rightarrow \text{hadrons})}{\sigma(e^+ e^- \rightarrow \mu^+ \mu^- )}
\end{equation}
Hence
\begin{equation}
    \Gamma_{\gamma'} = \sum_\ell \Gamma_{\gamma' \rightarrow \ell \bar{\ell}} + \Gamma_{\gamma' \rightarrow \text{hadrons}}
\end{equation}

\subsubsection*{Dark matter annihilation into dark photons}
See \eqref{Fig:DE_interactions}. The cross section $\sigma_{\chi \gamma'} \equiv  \sigma_{\chi \bar{\chi} \rightarrow \gamma' \gamma'}$ is given by\cite{Chu:2011be, Coy:2024itg}
\begin{equation}
    \sigma_{\chi \bar{\chi} \rightarrow \gamma' \gamma'}(s) = \dfrac{4\pi \alpha'^2}{s} \left[ \dfrac{2 s^2 + 8 m_\chi^2 s - 16 m_\chi^4}{s (s - 4 m_\chi^2)} \tanh^{-1} \left( \sqrt{1 - \dfrac{4 m_\chi^2}{s}} \right) - \dfrac{s + 4 m_\chi^2}{\sqrt{s (s - 4 m_\chi^2)}} \right]
    \label{eq:sigma_XX_AA}
\end{equation}
assuming that the the dark photon mass is much smaller than that of the dark electron (see \cite{Aboubrahim:2021ycj} for a full expression with a non-zero $m_{\gamma'}$).

\subsubsection*{Dark matter annihilation into fermions}
See diagrams \eqref{Fig:inflaton_portal} and  \eqref{Fig:DE_interactions}). Through the inflaton, the cross section $\sigma_{\chi f}^\phi \equiv \sigma_{\chi \bar{\chi} \rightarrow f \bar{f}}^\phi$ is given by
\begin{equation}
    \sigma_{\chi \bar{\chi} \rightarrow f \bar{f}}^\phi(s) = \dfrac{(y y')^2 s}{16 \pi (s-m_\phi^2)^2} \sqrt{1 - \dfrac{4 m_\chi^2}{s}} \left( 1-\dfrac{4 m_f^2}{s} \right)^{3/2}
\end{equation}
Through the dark photon, the cross section $\sigma_{\chi f}^\epsilon \equiv \sigma_{\chi \bar{\chi} \rightarrow f \bar{f}}^{\gamma'}$ is
\begin{equation}
    \sigma_{\chi \bar{\chi} \rightarrow f \bar{f}}^{\gamma'}(s) = \dfrac{4 \pi}{3} \alpha \alpha' \varepsilon^2 \dfrac{\left(1 + \dfrac{2 m_f^2}{s} \right) \left(1 + \dfrac{2 m_\chi^2}{s} \right)}{s \left[ \left( \dfrac{m_{\gamma'}^2}{s} - 1 \right) + \dfrac{m_{\gamma'}^2 \Gamma_{\gamma'}^2}{s^2} \right]} \sqrt{\dfrac{1 - \dfrac{3 m_f^2}{s}}{1 - \dfrac{4 m_\chi^2}{s}}}
\end{equation}
\subsubsection*{Averaged cross sections}
In Boltzmann equations, the cross sections are averaged on the distribution of incoming particles (we neglect quantum statistics effects). If the distribution is thermal, we rely on the standard results from \cite{Gondolo:1990dk}. For $2-2$ processes, 
\begin{equation}
    \langle \sigma v \rangle_T = \dfrac{1}{8 m^4 T K_2^2(m/T)} \int_{4 m^2}^\infty \sigma(s) \ (s-4m^2) \sqrt{s} K_1(\sqrt{s}/T) ds \quad (\text{thermal distribution})
    \label{eq:Thermally_averaged_cross_section}
\end{equation}
with  $m$ the mass of incoming particles and $K_n$ are modified Bessel functions of the second kind. For energy exchanges, we need $\langle \sigma v E \rangle$. Following the same steps as in \cite{Gondolo:1990dk}, we have obtained the following expression
\begin{equation}
    \langle \sigma v E \rangle_T = \dfrac{1}{16 m^4 T K_2^2(m/T)} \int_{4 m^2}^\infty \sigma(s) \ (s-4m^2) s K_2(\sqrt{s}/T) ds \quad (\text{thermal distribution})
\end{equation}

When incoming particles have not reach thermal equilibrium (a situation corresponding to early stages of reheating, see section \ref{sec:Thermalisation_HS}) and follow the distribution given in \eqref{eq:distribution_nonthermal}, the averaged cross section is given by \cite{Garcia:2018wtq}
\begin{equation}
    \langle \sigma v \rangle_{NT} = \dfrac{18}{m_\phi^3} \int_0^{m_\phi^2} \sigma(s) \ \sqrt{s} \left[ \ln \dfrac{m_\phi + \sqrt{m_\phi^2-s}}{\sqrt{s}} - \dfrac{\sqrt{m_\phi^2-s}}{m_\phi} \right] ds \quad (\text{non-thermal distribution})
\end{equation}
In numerical solutions to the Boltzmann equations Eqs (\ref{eq:Full_Boltzmann_system}), as illustrated in figures \ref{fig:Boltzmann_illustrations_post_RH}-\ref{fig:Boltzmann_illustrations_pre_RH}, we have numerically evaluated these integrals with the relevant distribution, to obtain the appropriate averaged cross sections. 

\bigskip

\section{Landau-Pomeranchuk-Migdal effect}

\label{app:LPM}

A slight complication of the problem of thermalisation  is that key processes, like bremsstrahlung, are affected by the so-called  Landau-Pomeranchuk-Migdal (LPM) effect \cite{Baier:2000sb,Arnold:2002zm,Kurkela:2011ti}. 
It is instructive to compare the cases of abelian (A) and non-abelian (NA) gauge interactions, even though our focus is on a HS with abelian interactions only. For generality, we keep factors of the dark fermion density $n_\chi$ explicit in this Appendix, and replace them by their expression in terms of cosmological production only in the bulk of the text, see section \ref{sec:hubble_th}.

The gist of the LPM effect is as follows. A gauge boson (daughter particle) emitted in a soft collision is typically collinear with its parent particle, so it takes a time $t_{\rm form}$
before the two particles can be considered fully separated. In a medium, the parent and, in the case of NA interactions, the daughter particle may undergo multiple scatterings during 
$t_{\rm form}$. As a result, one must sum over several radiation amplitudes, which interfere destructively. Physically, multiple scatterings cause the particles transverse momentum to undergo a random walk, so its trajectory is not well defined during the formation time and the resulting phase decoherence leads to destructive interference and so a suppression of the bremsstrahlung rate \cite{Landau:1953um,Migdal:1956tc}. In the context of reheating, the LPM effect can cause a delay in thermalisation \cite{Kurkela:2011ti,Mukaida:2015ria}. A proper treatment  involves using in-medium effective theories \cite{Arnold:2002zm}. Here, we restrict ourselves to a simple estimate of 
$t_{\rm form}$
 following \cite{Mukaida:2015ria}, as our aim is  to have  a practical condition for determining when thermalisation of the HS becomes effective.\footnote{A related problem is the energy loss of high-energy particles injected into a medium that is already in thermal equilibrium (see, e.g. \cite{Mukaida:2022bbo,Drees:2022vvn} which are excellent introductions to the topic) in a cosmological setup.}

The  timescale  $t_{\rm form}$ can be estimated as follows. We set  $k$
as the momentum of the daughter particle, 
$p$ that of the parent particle, and $\theta \ll 1$ the angle between their spatial momenta. Typically, though not necessarily, $k \ll p$. Denoting by $k_\perp$ and $p_\perp$ their transverse momentum, one has $\theta \sim k_\perp/k \sim p_\perp/p$. 
One distinguishes two situations, depending on whether the daughter or the parent particle interacts more frequently with the medium.  A key  point (easily overlooked with a schematic approach, as here) is that the dominant contribution to transverse-momentum diffusion comes from many soft, small-angle scatterings rather than from relatively rare hard scatterings. This is  precisely the regime in which the LPM effect arises, with the formation time $t_{\rm form}$ overlapping with a sequence of soft in-medium interactions. The net effect of these many soft collisions is that of an accumulated diffusion, with transverse momentum growing with the square root of the number of scatterings within a given time interval $N_{\rm scat}(t)$. Focusing on the parent particle, 
 \begin{equation}
 p_\perp(t) \sim \sqrt{N_{\rm scat}(t)} \Delta p
 \end{equation}
 where $\Delta p$ denotes the typical momentum transfer per collision. The  precise value of $\Delta p$ is not critical because the number of scatterings in a time $t$ is given by \footnote{In the literature dedicated to the LPM effect, diffusion is usually expressed in terms of the momentum-broadening coefficient defined as \cite{Arnold:2002zm,Kurkela:2011ti}, 
 \begin{equation}
     \hat q = {d \langle p_\perp^2\rangle\over dt} \sim \Gamma_{2-2}^s \Delta p^2\,.
 \end{equation}}
 \begin{equation}
     N_{\rm scat}(t) = t/t_{\rm scat} \sim \Gamma^s_{2-2} t  \sim \alpha'^2 {n_\chi\over\Delta p^2} \, t\quad \quad \mbox{\rm so that} \quad p_\perp \sim \sqrt{\alpha'^2\, n_\chi \,t}
     \label{eq:diffuse}
 \end{equation}
In the case of non-abelian interactions, the same considerations apply to the transverse momentum $k_\perp$ of a non-abelian gauge boson, which can also undergo many soft, t-channel scatterings. An abelian gauge boson, however, can only have relatively rare hard collisions --essentially $s/u$ channel Compton scatterings-- and so its transverse momentum does not diffuse. \\

\subsubsection*{Abelian interactions}

\noindent When the daughter particle splits from its parent particle,  $k_\perp/k \sim p_\perp/p \sim\theta$, from projection of the gauge boson momentum along the direction of the fermion. As the fermion transverse momentum diffuses, the gauge boson inherits that change proportionally to its own energy fraction $k/p$, $k_\perp(t) \sim k \,p_\perp(t)/p$. To estimate $t_{\rm form}$, we request that the particle diffuses over a distance of the order of its transverse size, $v_\perp t_{\rm form} \gtrsim 1/k_\perp$, during $t_{\rm form}$. With $v_\perp = k_\perp/k$, this gives $t_{\rm form} \sim k/ k_\perp^2 \sim p^2/k\, p_\perp^2$. Using \eqref{eq:diffuse} leads to 
\begin{equation}
    t_{\rm form} \sim \sqrt{p^2\over  \alpha'^2 n_\chi\, k} 
\end{equation} 
Rewriting this in term of the rate for soft $2-2$ collisions, see Eq.\eqref{eq:soft2-2_abelian} in section \ref{sec:thermalisation}, 
\begin{equation}
\label{eq:soft2-2_abelian_here}
    \Gamma^s_{2-2} \approx {\alpha'^2\over \bar m^2} n_\chi 
\end{equation}
with cut-off scale by the in-medium Debye mass or the dark photon mass,
\begin{equation}
    \bar m = \mbox{\rm max}(m_D,m_{\gamma'}) \sim \mbox{\rm max}(\alpha' n_\chi/m_\phi,m_{\gamma'})
 \end{equation}
 For the former, the rate is constant, $\Gamma^s_{2-2} \sim \alpha' m_\phi$. In either cases, 
the LPM effect takes the form of a simple suppression factor,
\begin{equation}
  t_{\rm form} \sim \left\{\begin{array}{cc} \sqrt{k^A_{\rm LPM}\over k}/\Gamma_{2-2}^s & \quad k \lesssim k^A_{\rm LPM}\\
 \\ {1/\Gamma_{2-2}^s} & \quad k \gtrsim k^A_{\rm LPM}
  \end{array}\right.
\end{equation}
where the pivot momentum $k^A_{\rm LPM}$ is given by
\begin{equation}
\label{eq:pivo_abelian}
    k^A_{\rm LPM}\sim \left\{\begin{array}{cc} {p^2 m^2_\phi/n_\chi} & \quad m_D \gtrsim m_{\gamma'}\\
    \\
   \frac{\alpha'^2 p^2 n_\chi\, }{m_{\gamma'}^4} & \quad m_{\gamma'} \gtrsim m_D
    \end{array}\right.
\end{equation}
So the rate for bremsstrahlung, $\Gamma_{2-3} \sim \alpha'/t_{\rm form}$,  is estimated to be of order
\begin{equation}
\label{eq:23abelian}
  \Gamma_{2\rightarrow 3}  \sim \left\{\begin{array}{cc} \alpha' \, \Gamma_{2-2}^s\sqrt{k\over k^A_{\rm LPM}} & \quad k \lesssim k^A_{\rm LPM}\\
 \\ \alpha'\, \Gamma_{2-2}^s & \quad k \gtrsim k^A_{\rm LPM}
  \end{array}\right. 
\end{equation}
\\

For $p\sim m_\phi$, and $m_{\gamma'} \lesssim m_D$, $k_{\rm LPM}^{\rm A} \sim m_\phi^4/n_\chi \gg m_\phi$ is always larger than the initial energy of the parent particle. Thus, the LPM effect is always relevant and  the emission of soft gauge bosons, $k\ll p$, is comparatively suppressed and the daughter and parent particle tend to share the same energy $k \sim p$.  Once dark photons with 
$k \sim p$ begin to be effectively produced, they  in turn can create dark fermion pairs with momentum $\sim p/2$, through
\begin{center}
\scalebox{1.}{
\begin{tikzpicture}[baseline={-0.1cm}]
  \begin{feynman}[every blob={/tikz/fill=gray!30,/tikz/inner sep=2pt}]
    \vertex (i1) at (-1.5, 0.5) {\(\gamma'\)};
    \vertex (i2) at (-1.5,-0.5) {\(\chi\)};
    \vertex (c1) at (0,0.5);
    \vertex (c2) at (0.6,-0.5);
    \vertex (f1) at (0.6,0.5) ;
    \vertex (f3) at (1.8, 0.5) {\(\chi\)}; 
    \vertex (f4) at (1.8, 1.2) {\(\bar{\chi}\)};
    \vertex (f2) at (1.8,-0.5) {\(\chi\)};
    \diagram* {
      (i1) -- [boson] (c1),
        (i2) -- [fermion] (c2),
      (c1) -- [fermion] (f1),
      (f1) -- [boson] (c2),
      (f1) -- [fermion] (f3),
        (f4) -- [fermion] (c1),
      (c2) -- [fermion] (f2) }; \end{feynman}
\end{tikzpicture}}
\end{center}
These newly produced particles can in turn radiate dark photons, with a rate $\Gamma_{2-3}$ similar to \eqref{eq:23abelian}. This leads to a cascade of processes, in which the energy is degraded at each step by a factor ${\cal O}(2)$, ultimately driving the HS toward thermalisation. Consequently, the bottleneck for thermalisation of dark QED is the emission of the first dark photons, with a timescale sets by Eq.\eqref{eq:23abelian} for $m_{\gamma'} \lesssim m_D$,
\begin{equation}
\label{eq:abelianth}
    t_{\rm th}^{\rm A} \sim 1/\Gamma_{2\rightarrow 3} \sim {1\over \alpha'^2 m_\phi} \left({m_\phi^3\over n_\chi}\right)^{1/2}\qquad \mbox{\rm (abelian thermalisation)}
\end{equation} 
where we set $k \sim p \sim m_\phi$.\\

If  $m_{\gamma'}\gtrsim m_D$, the rate for soft-collision decreases with the density $n_\chi$. So, one expects that the LPM becomes less operative, as $k_{\rm LPM}^A \propto n_\chi$ too in that case. Taking $k \sim k_{\rm LPM}^A\sim p\sim m_\phi \sim$,  gives
\begin{equation}
    m_{\gamma'}^2 \gtrsim \alpha' (m_\phi n_\chi)^{1/2} \sim \alpha' (\Gamma_\phi^\chi H M_P^2)^{1/2} \sim \alpha' T'^2
\end{equation}
where we used Eq.\eqref{eq:nchifromphi}, section \ref{subsec:kinetic}, for $n_\chi$ and \eqref{eq:Tmaxrh}, section \ref{sec:framework}, to estimate $T'$. This amounts to say that the LPM suppression becomes irrelevant when dark photon mass becomes larger than the thermal Debye mass at thermalisation. 
In that case, the splitting rate is not suppressed by the LPM effect, and the rate for thermalisation becomes
\begin{equation}
\label{eq:timethA}
    t_{\rm th}^A \sim \frac{1}{\Gamma_{2-3}} \sim \frac{1}{\alpha' \Gamma_{2-2}^s}\sim \frac{1}{\alpha'^3m_\phi} \left(\frac{m_\phi^3}{n_\chi}\right)\left(\frac{m_{\gamma'}}{m_\phi}\right)^2
\end{equation}
which, by construction is larger than \eqref{eq:abelianth} and also increases faster as the density decreases. In the bulk of the article, we assume that the dark photon mass is always less than the Debye mass.

\subsubsection*{Non-abelian interactions}

For the sake of comparison, we now discuss the LPM effect in the case of non-abelian interactions. For NA interactions, the gauge boson that is produced through bremsstrahlung can itself undergo soft, t-channel scatterings in the medium. Consequently, its transverse momentum evolves as $k_\perp = \sqrt{\alpha^{\prime 2} n_\chi\, t}$, using the same notation for $\alpha'$ as in the abelian case. Repeating the logic as in the previous section, the timescale for separation is 
\begin{equation}
    t_{\rm form} \sim \sqrt{k\over \alpha'^2 n_\chi} 
\end{equation}
which can be expressed as
\begin{equation}
  t_{\rm form} \sim \left\{\begin{array}{cc} {1\over \Gamma_{2-2}^s} \sqrt{k \over k^{\rm NA}_{\rm LPM}} & \quad k \gtrsim k^{\rm NA}_{\rm LPM}\\
 \\ {1/\Gamma_{2-2}^s} & \quad k \lesssim k^{\rm NA}_{\rm LPM}
  \end{array}\right.
\end{equation}
Here, the pivot momentum  is given by
\begin{equation}
\label{eq:kLPMNA}
    k^{\rm NA}_{\rm LPM} \sim {n_\chi/m^2_\phi} \qquad \text{(non-abelian interactions)}
\end{equation}
The rate for bremsstrahlung of non-abelian gauge bosons is thus estimated to be given by 
\begin{equation}
\Gamma_{2\rightarrow 3}^{\rm NA} \sim \left\{
\begin{array}{cc}
    \alpha'\, \Gamma_{2-2}^s & \quad  k \lesssim k^{\rm NA}_{\rm LPM} \\
    \\
    \alpha' \,\Gamma_{2-2}^s \sqrt{k^{\rm NA}_{\rm LPM}/k}  &\quad  k \gtrsim k^{\rm NA}_{\rm LPM}\end{array}\right.
    \label{eq:23nonabelian}
\end{equation}

There are key differences with respect to abelian interactions. First, in non-abelian theories the LPM effect primarily suppresses the emission of relatively hard gauge bosons. This is a consequence of gauge boson self-interactions, which trigger a cascade whereby soft gauge bosons are produced copiously and subsequently populate momentum space from low momenta upward. These self-interacting gauge particles provide a form of positive feedback: the larger the gauge boson density, the more efficiently additional radiation is generated, driving the system toward equilibrium much more rapidly than in the abelian case. This mechanism, known as bottom-up thermalisation \cite{Baier:2000sb}, has been studied extensively in a cosmological context in~\cite{Kurkela:2011ti,Harigaya:2013vwa,Mukaida:2015ria}. In practice, this effect amounts to progressively replacing the fermion number density $n_\chi$ in Eq.~\eqref{eq:kLPMNA} by the density of gauge bosons. In particular, as shown in \cite{Harigaya:2013vwa}, the thermalisation timescale for non-abelian particles is parametrically given by
\begin{equation}
\label{eq:timethNA}
t_{\rm th}^{\rm NA}\sim \frac{1}{\alpha'^2 m_\phi}
\left(\frac{m_\phi^3}{n_\chi}\right)^{3/8}
\qquad \text{(non-abelian thermalisation)}
\end{equation}
As discussed in section \ref{sec:hubble_th}, these results imply that, all other things being the same, thermalisation is more efficient for non-abelian interactions. 

\section{More on $T'/T$ vs $\xi_i$}
\label{app:other_constraints}

In section \ref{sec:xi_evolution}, we have set boundaries on the domain of thermal dark QED candidates making a specific choice of parameters. Those were chosen so as to determine the largest possible domain of mass for thermal DM candidates. In figure \ref{fig:plot_xi_stability}, we fixed  the lifetime of the dark photon to be the longest possible, and took also the dark photon to be as heavy as possible. This led to maximize the entropy production after DM freeze-out, and so to determine the largest possible DM mass consistent with the unitarity bound. For completeness, we  provide here figure \ref{fig:plot_xi_stability_app}. Here, instead, we consider a light dark photon, with a fixed mixing parameter, $\epsilon$. This allows us to illustrate the evolution of the temperature ratio as function of $\epsilon$ depending on the possible source of heating of the VS. See section \ref{sec:xi_evolution} for details.
\begin{figure}[H]
	\centering
	\includegraphics[width=0.7\textwidth]{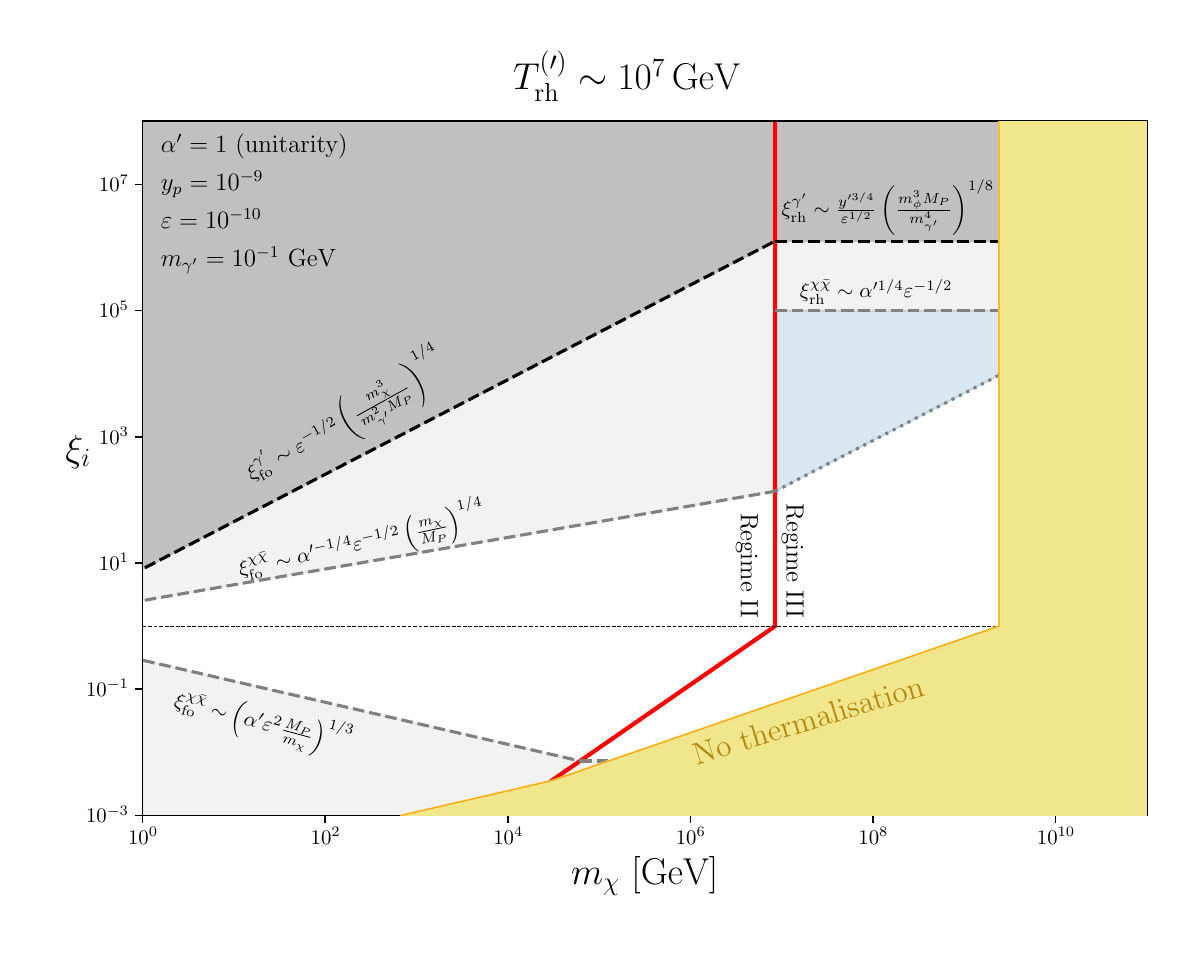}
	\caption{Regions of stability of $\xi$ at DM decoupling (white) for fixed dark photon mass and mixing parameter. Points above (below) a dashed line for $\xi_i$ larger (smaller) than one will see $\xi$ evolve vertically until it reaches the white region, at the moment of decoupling. Grey dashed lines indicate the limit imposed by $\chi \bar{\chi} \leftrightarrow f \bar{f}$ process, while dark dashed lines show the one due to the $\gamma' \leftrightarrow f \bar{f}$.}
	\label{fig:plot_xi_stability_app}
\end{figure}

As in fig \ref{fig:plot_xi_stability}, this figure shows the regions above which the temperature ratio is driven to the value $\xi_{\rm rh}$ at the end of reheating, either from $\chi\bar \chi$ annihilation ($\xi_{\rm fo}^{\chi\bar\chi}$, light dashed line) or through the decay of the dark photon  ($\xi^{\gamma'}_{\rm fo}$, dark dashed line). Below these lines (white region) $T'/T  \sim \xi_i$, and this is thus the initial condition as set by inflation, which is stable. Conversely, in the shaded regions above (below) the limits, the temperature ratio is lower (larger) at reheating time compared to the initial value $\xi_i$. For a low dark photon mass $m_{\gamma'} = 10^{-1}~\rm GeV$, dark electron annihilation turns out to be always the dominant process that drives the evolution of the temperature ratio. 
As in section \ref{sec:xi_evolution}, we need to distinguish the case of a cold HS from that of hot HS as well as the regime II (freeze-out after reheating) and regime III (freeze-out during reheating). By solving the Boltzmann equations Eqs (\ref{eq:Full_Boltzmann_system}) in each of the relevant situation, following the same steps as in section \ref{sec:xi_evolution}, we find the following parametric solutions for the different boundaries of the regions in figure \ref{fig:plot_xi_stability_app}, 
\begin{equation}
	\xi_{\text{fo}}^{\chi\bar\chi} \sim \alpha'^{-1/4} \varepsilon^{-1/2} \left( \dfrac{m_\chi}{M_P} \right)^{1/4} \quad ( \text{hot HS, regime II})
\end{equation} 
\begin{equation}
	\xi_{\text{rh}}^{\chi\bar \chi} \sim \alpha'^{1/4} \varepsilon^{-1/2}  
	\quad ( \text{hot HS, regime III})
\end{equation}

\begin{equation}
	\xi_{\text{fo}}^{\chi \bar{\chi}} \sim \alpha'^{1/3} \varepsilon^{2/3} \left( \dfrac{M_P}{m_\chi} \right)^{1/3} \quad (\text{cold HS, regime II})
\end{equation}
and
\begin{equation}
	\xi_{\text{rh}}^{\chi \bar{\chi}} \sim \begin{cases}
		\alpha'^{1/4} \varepsilon^{1/2} \left( \dfrac{M_P}{T_{\text{rh}}} \right)^{1/4} \quad (T_{\text{rh}} \gg m_\chi) \\
		\alpha'^{1/4} \varepsilon^{1/2} \left( \dfrac{M_P}{m_\chi} \right)^{1/4} \quad (T_{\text{rh}} \ll m_\chi)
	\end{cases}
	\quad ( \text{cold HS, regime III})
\end{equation}

%%%%%%%%%%%%%%%%%%%%%
\bibliographystyle{JHEP}
\bibliography{biblio}
%%%%%%%%%%%%%%%%%%

\end{document}